\documentclass[11pt]{article}

\usepackage{import}

\usepackage{amsmath,amsfonts,amstext,amssymb,mathrsfs,mathtools}
\usepackage{graphicx}
\usepackage{comment}
\usepackage{feynmp-auto}
\usepackage{color}
\usepackage{cancel}
\usepackage{multicol}
\usepackage{multirow}
\usepackage{graphicx}
\usepackage[sorting=none,style=numeric-comp]{biblatex}
\bibliography{references}
\usepackage{tikz}
\usepackage{wrapfig}
\usepackage{geometry}

\newcommand{\mathds}[1]{\mathbbm{#1}}

\usepackage[english]{babel}
\usepackage{csquotes}
\usepackage{subcaption}
\usepackage{scrextend}

\usepackage{empheq}
\usetikzlibrary{calc}

\usetikzlibrary{decorations.pathmorphing}
\usetikzlibrary{shapes,arrows,fit,positioning}
\usepackage{bm}
\usepackage{bbm}
\usetikzlibrary{calc}
\usepackage{float}

\usepackage{resizegather}

\newcommand{\secref}[1]{Section \ref{#1}}
\newcommand{\appref}[1]{Appendix \ref{#1}}

\newcommand{\figref}[1]{Figure \ref{#1}}

\newcommand{\ed}{\text{d}}

\newcommand{\mbh}{M_{\text{BH}}}
\newcommand{\mpl}{M_{\text{Pl}}}

\newcommand{\lpl}{\ell_{\text{Pl}}}

\newcommand{\gap}{\text{ }}
\newcommand{\sgap}{\gap\gap}
\newcommand{\mgap}{\sgap\sgap}
\newcommand{\lgap}{\gap\gap\gap\gap\gap\gap}
\newcommand{\p}{\partial}

\newcommand{\proj}{\text{p}}

\newcommand{\rfs}{R_0}
\newcommand{\flatgrav}{\mathfrak{h}}

\newcommand{\fdfield}[1]{\mathbf{#1}}

\newcommand{\prop}{\mathbf{P}}

\usepackage{jheppubmod}

\usepackage[labelfont={bf}, width={.9\textwidth}, font={small}]{caption}

\begin{document}

\title{All elastic amplitudes in the (black hole) eikonal phase}
\author{Nico Groenenboom}
\emailAdd{n.groenenboom@uu.nl}
\affiliation{Institute for Theoretical Physics and Center for Extreme Matter and Emergent Phenomena, Utrecht University, 3508 TD Utrecht, The Netherlands.}
\date{\today}

\abstract{In this article we calculate the eikonal scattering amplitude for an arbitrary number of in- and out-particles, using covariant quantization in a spherical harmonics basis on the Schwarzschild background. We extend prior results to resummation over all partial waves, restoring contributions from transverse separation and correctly taking into account the particle masses in the pole structure. We consider leading order interactions mediated by scalar-scalar-graviton vertices and scalar electrodynamics. We perform our calculations in the black hole eikonal phase. The $2\to 2$ eikonal amplitude is measured by the transverse Green's function $(-\Delta_{\Omega}+a)G(\Omega,\Omega')=\delta^{(2)}(\Omega-\Omega')$. As a consistency check, we use our formalism in flat space to find an exact match with the known flat space eikonal $2\to 2$ amplitude in literature. We then extend the eikonal amplitude to arbitrarily many particles for the first time in both flat space and on the black hole background. We show that the black hole amplitude matches the black hole S-matrix as derived by 't Hooft. We conclude that this amplitude provides the most general elastic contribution one can achieve in the eikonal phase.}

\maketitle

\clearpage

\addtocontents{toc}{\protect\setcounter{tocdepth}{2}}

\section{Introduction}
The Schwarzschild spacetime was already discovered in 1916 and provided the first classical solution to Einstein's field equations. Discovered as a spherically symmetric solution characterized only by its mass $\mbh$, it was later understood to have a radius from which nothing can escape, the event horizon $R=2G \mbh$ . Decades later, Bekenstein showed that a black hole of mass $\mbh$ has non-zero entropy $S\sim \frac{R^2}{G}$ violating the classical picture, and suggesting that black holes contain a large amount of information \cite{Bekenstein:1973ur}. Additionally, in the same decade Hawking showed that they slowly evaporate, emitting low energetic radiation at a very low temperature \cite{Hawking:1974rv,HawkingParticleCreation}. He argued that semi-classical black hole physics is well approximated by free quantum fields on the curved Schwarzschild background. This approximation results in thermal Hawking radiation that contains no information. Thus, the large entropy seemingly evaporates into nothing, leading to a violation of unitarity and the information paradox.\\
Since the discoveries of Hawking the question of black hole unitarity has been an active field of research. There are many different perspectives and proposals to tackle this problem \cite{HarlowLectures}, among which predominantly AdS/CFT, which generally involves free quantum fields in accordance with Hawking's picture. On the contrary, 't Hooft argued the exact opposite \cite{tHooft1996}, that interactions between ingoing particles and outgoing Hawking radiation strongly affect the entropy of the outgoing radiation and thus unitarity. The proposal is called the S-matrix proposal, where the ingoing radiation is related to the outgoing radiation by an S-matrix
\begin{align}
    |\text{out}\rangle=S|\text{in}\rangle
\end{align}
that should be unitarity $SS^{\dagger}=1$ to resolve the information paradox. In particular, he proposed that gravitational interaction must dominate this interaction as it becomes the strongest coupling near the horizon, leading to an equation for the S-matrix using semi-classical methods and quantum mechanics \cite{tHooft1996,Stephens1993} (which we shortly summarise in \secref{sec:intro background}). Further investigations have since been made, notably extensions to other theories and deeper analysis of the S-matrix \cite{Veneziano2008Wosiek,Veneziano2012,Veneziano2004alone,Veneziano2008Wosiek2,Ciafaloni2008Colferai}. \\
\\
The S-matrix derived by 't Hooft involves a quantum mechanical interpretation of the external states as momentum distributions of an arbitrary number of particles. This strongly clouds the possibility for inelastic interactions, which demand a field theory to be well understood. In the past years we have developed exactly such a toolbox for scattering of particles near the Schwarzschild horizon \cite{Toolbox}. While we replicated 't Hooft's S-matrix for each mode in a decoupled partial wave basis by considering an eikonal resummation \cite{ShortPaper,LongPaper}, we show in this article that the resummation over coupled partial wave sheds new light on the correct field theoretic interpretation of the complete S-matrix of 't Hooft. We recently included electromagnetic interactions \cite{SQEDpaper} showing indeed that gravitational interactions dominate. The eikonal resummation involves a resummation over ladder diagrams: There are two conserved lines of matter fields that exchange an arbitrary amount of interaction bosons, as performed by \cite{LevySucher}. The resulting combinatorics results in the interaction bosons becoming soft (vanishing momentum). It was shown that the eikonal resummation is leading for certain field theories \cite{Tiktopoulos1971,Eichten1971}, which we are forced to assume. We have also calculated a particular set of inelastic diagrams \cite{2to2N}, which show an exponential decay similar to the exponential thermal factor of Hawking radiation, and a time delay of order page time. \\
\\
In this article, we use the methods developed in \cite{Toolbox} to calculate all possible elastic scattering amplitudes within a certain regime of phase space. In \secref{chap: scalar fields} we shortly summarize the important results of \cite{Toolbox} that we need for this article with additional remarks for a stronger foundation. In order to be able to do these calculations we work only in the proposed black hole eikonal phase:
\begin{align}\label{eqn:bheikonal}
E \, \mbh \gg \mpl^2 \, .
\end{align}
where $E$ is the scattering energy for any two particles. The field theory in \cite{Toolbox} involves a covariant graviton interaction between scalar fields, expanded into a spherical harmonics basis. This expansion into harmonics is well-known \cite{ReggeWheeler,MartelPoisson,KalloshRahman,KalloshSpherical,Zerilli1970,Vishveshwara1970,Nagar2005}. The spherical harmonics basis has the strong advantage to simplify the action using the spherical background symmetry, however it introduced a large amount of different modes to consider, and in particular an infinite summation over angular momentum modes $\ell m$ in each interaction vertex. Since 't Hooft's S-matrix involved decoupled states \cite{tHooft2016}, we enforced a decoupling limit in \cite{ShortPaper,LongPaper,SQEDpaper} that removes this summation by always fixing one particle to be at $\ell=0$. In this article we remove the decoupling altogether and calculate all amplitudes including a full resummation over partial waves within the black hole eikonal phase. We may then consider the external states in the position basis instead, interpreting them as single particles localized at different angles $\Omega$ on the two-sphere. We thus extend the results of \cite{ShortPaper,LongPaper,SQEDpaper} to include new interpretations and calculations.\\
\\
In \secref{chap: eikonal} we first extend the familiar eikonal resummation of graphs to include all partial waves, removing the minimal coupling, and show that the resulting $2\to 2$ scattering is of a different form to both our previous results and 't Hooft's S-matrix. We find an identical structure with eikonal amplitudes in literature, and are in particular able to exactly reproduce the flat space eikonal amplitude derived in \cite{KabatOrtiz}. In \secref{chap: MN to MN} we extend the familiar black hole eikonal ladder diagrams, to a many-particle generalization calculating a $K\to K$ diagram for an arbitrary amount of particles $K$. It is natural to write this new proposed diagram as $N+M\to N+M$, splitting $K=N+M$ into $N$ particles falling into the black hole and $M$ going out of it (which is crucially distinct from entering and exiting the Feynman diagram), that all interact eikonally. Within the black hole eikonal phase this diagram may be calculated exactly, and we can show that it agrees with 't Hooft's S-matrix, providing the complete field theoretic generalization and ensuring functionality of the toolbox.\\
\\
Finally, we make some remarks on the regimes we work with. As mentioned we work in the black hole eikonal phase 
$E \, \mbh \gg \mpl^2 $. For large semi-classical black holes, this condition is easily satisfied even with low scattering energies. For an earth mass black hole (with $R_s \sim 1 cm \gg \lpl$), \eqref{eqn:bheikonal} implies that $s \gg 10^{-64} \mpl^2$. This shows that the eikonal phase on black holes is satisfied easily, and ensures that our results are valid for any standard model particle. Based on intuition one would expect from the eikonal approximation in flat space that trans-Planckian physics is required $s\gg \mpl^2$ \cite{Almheiri2012, Almheiri2013, Marolf2013, PolchinskiChaos2015}, but the background black hole ensures to regulate this. Scale issues only become important when the black hole size becomes extremely small, which would be a highly unstable regime. Additionally the gravitational interaction is determined by a coupling
\begin{align}\label{eqn:approx}
    \gamma=\frac{\mpl}{\mbh}
\end{align}
which is incredibly small. This ensure that the theory is valid up to a number of particle $N\sim \frac{1}{\gamma}\sim \frac{\mbh}{\mpl}$ which is incredibly large. \\
In the remainder of this introduction we provide a short summary of literature results that are especially relevant for this paper. In particular we outline the semi-classical S-matrix derived by 't Hooft \cite{tHooft1996} that we aim to reproduce in this article. We refer the reader to \cite{Toolbox} for more commentary and discussion on the field theory itself.

\subsection{Schwarzschild many-particle S-matrix}
\label{sec:intro background}
\begin{figure}[h!]
    \centering
    \includegraphics[width=0.7\textwidth]{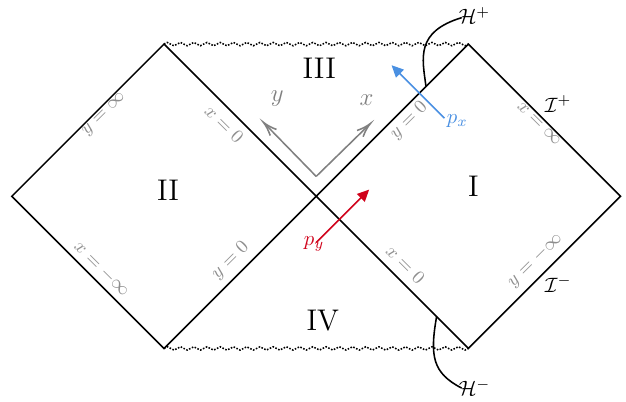}
    \caption{The Penrose diagram for the maximally extended Schwarzschild black hole. The four different regions are labelled in the convention of 't Hooft. The direction of the coordinates $x,y$ and the definition of the horizons are shown, as well as the conventional notation for null infinity. The momentum direction is orthogonal to the coordinate direction because of the off-diagonal metric.}
    \label{fig:penrose}
\end{figure}

Since we want to perform scattering on a black hole background, we need to specify the metric in a chosen set of coordinates. As mentioned most of the work is done in Kruskal-Szekeres coordinates. The main reason for this choice of coordinates is that it describes the entirety of the Schwarzschild Spacetime, and it is regular on the horizon. This last property is important for us to be able to define a stable field theory. We will employ coordinates $x,y$ such that 
\begin{align}
    \eta_{ab}&=\begin{pmatrix}
    0 & -1 \\ -1 & 0
    \end{pmatrix},\\
    \gamma_{AB}&=\begin{pmatrix}
    1 & 0 \\ 0 & \sin^2\theta
    \end{pmatrix},
\end{align}
in terms of which the full metric is given by
\begin{align}
    \ed s^2=f(r)\eta_{ab}\ed x^a \ed x^b+r^2\gamma_{AB}\ed x^A\ed x^B.
\end{align}
The coordinates are related to the original Schwarzschild coordinates by
\begin{align}
    xy&=2R^2\left(1-\frac{r}{R}\right)e^{\frac{r}{R}-1},\\
    x/y&=\text{sgn}\left(1-\frac{r}{R}\right)e^{2\tau} \lgap\lgap \tau=\frac{t}{2R},\\
    f(r)&=\frac{R}{r}e^{1-\frac{r}{R}},
\end{align}
where $R$ is the Schwarzschild radius and $\mu=1/R$ the inverse Schwarzschild radius. In \figref{fig:penrose} a visual representation of the spacetime with the coordinates direction has been shown. Remark that region I has $xy<0$ so that actually $x>0$ and $y<0$. More details on the background and conventions are in \appref{chap: background}.
We first shortly review 't Hooft's shockwave analysis in the case of a charged particle \cite{tHooft:1996rdg} propagating in the background of a Schwarzschild black hole, and how it leads to a semi-classical S-matrix. We first calculate the back-reaction of a highly boosted charged shockwave on a probe test particle \cite{Dray:1984ha,DrayShockwave}. The gravitational back-reaction of the shock leaves an imprint on the gravitational field experienced by the probe. The probe then experiences geodesics that are shifted across the null surface traced out by the shockwave. For a particle with momentum $p_{\text{in}}$ at location $x_{0}$ and a point on the sphere $\Omega_{0}$ the metric is given by \cite{Hooft:2016itl, Betzios2016}
\begin{align}
\label{eq:gravshockwave}
\mathrm{d}s^{2} ~ &= ~ - 2 f \mathrm{d} x \Big(\mathrm{d}y - p_{\text{in}}\delta\left(x \right) \lambda_{1}\left(\Omega, \Omega_{0}\right) \mathrm{d} x \Big) + r^2 \mathrm{d}\Omega^{2} \, ,\\
\label{eqn:lambda1}
	\lambda^{\ell m}_{1} ~ &= ~ \dfrac{8 \pi G}{\ell^{2} + \ell + 1} \, .
\end{align}
Outside of the location of the source shock, a probe particle experiences the background Schwarzschild solution. At the location of the source $\delta\left(x \right)$, however, a probe particle experiences an instantaneous shock. In analogy to the gravitational back-reaction discussed above, an electrodynamical shockwave leaves an imprint on a charged particle. This extension was performed in \cite{SQEDpaper} and resulted in a shift given by
\begin{align}\label{eqn:lambda2}
	\lambda^{\ell m}_{2} ~ = ~ -\dfrac{q_{\text{in}}}{\ell^{2} + \ell} \, .
\end{align}
Using the shockwave solution we can write down a quantum-mechanical S-matrix using semi-classical methods. The aim is to calculate the S-matrix for the wavefunction of a charged particle in the presence of a gravitationally back-reacting charged shockwave. To this end, let us first begin by writing the wavefunction of a charged particle in said eigenbasis as $\psi \left(p_{\text{in}}, q_{\text{in}}\right) ~ = ~ \langle \psi | p_{\text{in}}, q_{\text{in}} \rangle$. 
The resulting S-matrix is then given by \cite{SQEDpaper,tHooft1996}
\begin{align}
\label{eq:22semiclas}
	S\left(p_\text{in}, q_{\text{in}} ; p_\text{out}, q_{\text{out}}\right) ~ = ~ \exp\left( i \, \lambda_{1} \, p_{\text{in}} \, p_\text{out} + i \lambda_{2} \, q_{\text{in}} q_{\text{out}}\right) \, .
\end{align}
This shows that the semi-classical S-matrix for two particles is a simple complex exponent, where the $\lambda_i=\lambda_i(\Omega_1,\Omega_2)$ is a transverse Green's function whose value depends on the transverse separation of the two particles. This type of identical contribution for the graviton and photon was schematically predicted by 't Hooft in \cite{HOOFT1987dominance,tHooft1996}, without the Green's functions. \\
\\
We would like to generalise the previous results to the case of many particles in order to then take a continuum limit to describe a distribution of particles on the horizon. Since quantum mechanics does not allow for particle production, we may safely assume that the number of incoming and outgoing particles is equal and large; we call the number of incoming and outgoing particles $N_{\text{in}}$ and $N_{\text{out}}$ respectively. We will label the $i^{\text{th}}$ incoming particles by its longitudinal position $x_{i}$, angular position on the horizon $\Omega_{i}$, momentum $p^{i}_{\text{in}}$ and charge $q^i_{\text{in}}$such that $i \in N_{\text{in}}$. Similarly, outgoing particles are labelled by $y_{j}, \Omega_{j}, p^{j}_{\text{out}}$, $q^j_{\text{out}}$ and $j \in N_{\text{out}}$. Assuming that there is no more than one particle at each angular position on the horizon, the basis of states may be written as
\begin{align}
    | p_{\text{in, tot}} , q_{\text{in, tot}} \rangle ~ = ~ \bigotimes_{i} |p^i_{\text{in}} , q^{i}_{\text{in}} \rangle \quad \text{and} \quad |p_{\text{out, tot}} , q_{\text{out, tot}} \rangle ~ = ~ \bigotimes_{j} | p^j_{\text{out}} , q^{j}_{\text{out, tot}} \rangle \, ,
\end{align}
where we assumed a factorised Hilbert space because all parallel moving particles are independent. The resulting S-matrix in terms of these states is given by \cite{SQEDpaper}
\begin{align}
	S_{\text{tot}} ~ = ~ \exp\left(i \lambda^{ij}_{1} p^{i}_{\text{in}} p^{j}_{\text{out}} + i \lambda^{ij}_{2} q^{i}_{\text{in}} q^{j}_{\text{out}} \right) \, ,
\end{align}
where a sum over all in and out particles is implicit. The continuum limit $N_{\text{in}} = N_{\text{out}} \rightarrow \infty$, where the positions of particles may be described by distributions $x\left(\Omega\right)$ and $y\left(\Omega\right)$, is now easy to achieve. We first promote the momenta and charges to be distributions as smooth functions of the sphere coordinates and then replace the sum over in and out particles with integrals over the sphere coordinates as
\begin{align}\label{eqn:shockwaveQMSmatrix}
	S_{\text{tot}} ~ &= ~ \exp\left[ i \int \mathrm{d} \Omega \, \mathrm{d} \Omega' \left(\lambda_{1} \left(\Omega, \Omega' \right) p_{\text{in}} \left(\Omega\right) p_{\text{out}} \left(\Omega'\right) + \lambda_{2} \left(\Omega, \Omega' \right) q_{\text{in}} \left(\Omega\right) q_{\text{out}} \left(\Omega' \right) \right) \right] \nonumber \\
	&= ~ \exp\left[ i \left(\dfrac{8 \pi G \, p_{\text{in}} p_{\text{out}}}{\ell^{2} + \ell + 1} - \dfrac{q_{\text{in}} q_{\text{out}}}{\ell \left(\ell + 1\right)} \right) \right] \, ,
\end{align}
where we expanded the expression in partial waves in the second line and substituted for $\lambda_{1}$ and $\lambda_{2}$ using \eqref{eqn:lambda1} and \eqref{eqn:lambda2}. Of course, the momentum and charge distributions are also expanded in spherical harmonics, but their partial wave indices have been suppressed. Originally our aim was to re-derive the S-matrix above using field theoretic methods within $2\to 2$ scattering. While we found the same equation in \cite{LongPaper,ShortPaper,SQEDpaper}, closer inspection showed that the interpretation is different, and the correct $2\to 2$ S-matrix was found in \secref{chap: eikonal} which indeed matches \eqref{eq:22semiclas}, and the correct generalization of \eqref{eqn:shockwaveQMSmatrix} was instead found in \secref{chap: MN to MN}.

\subsection{Flat space \texorpdfstring{$2\to 2$}{2->2} S-matrix}
\label{sec:eikonalintro}
\label{sec:minkEikonal}
An analogous shockwave for massless particles exists on the Minkowski metric, called the Aichelburg-Sexl metric \cite{aichelburg1971gravitational}, further researched in \cite{Gray2021,Cristofoli2020}. We will write only the gravitational part for the flat space calculations. For a particle moving with energy $E$ in the $\hat{z}-$direction the shockwave metric takes the form
\begin{align}
\label{eq:flatspaceshockwave}
\mathrm{d}s^{2} ~ &= ~ - \ed t^2+\ed x^2+ \ed y^2+\ed z^2+4 E G\delta\left(t-z\right) \log|x^{\perp}-x^{\perp}_0|\bigr(\ed t-\ed z\Big)^2 \, ,
\end{align}
where $|x^{\perp}-x^{\perp}_0|^2=(x-x_0)^2+(y-y_0)^2$. The semi-classical S-matrix is given by
\begin{align}
	S\left(p_\text{in} ; p_\text{out}\right) ~ = ~ \exp\left( - 8iG  \, E_{\text{in}} \, E_\text{out} \log|x^{\perp}_{\text{in}}-x^{\perp}_{\text{out}}| \, \right) \, .
\end{align}
This S-matrix still depends on $x^{\perp}$, so it is partly in momentum space (in the $t,z$ coordinates) and partly in position (in $x,y$). We can write down the full momentum space S-matrix as the Fourier transform
\begin{align}
\label{eq:semiclasSmatrix}
    S=\int\ed^2 x^{\perp} ~ e^{i p^{\perp}\cdot (x^{\perp}-x^{\perp}_0)} ~ S\left(p_\text{in} ; p_\text{out}\right),
\end{align}
which can be solved to find
\begin{align}
\label{eq:flatspacesmatrix}
S ~ = ~  \dfrac{\pi \Gamma\left(1 - i Gs\right)}{\Gamma\left(iGs\right)} \left(\dfrac{4}{k_{\perp}^2}\right)^{1-iGs} \, ,
\end{align}
as was derived by 't Hooft in \cite{HOOFT1987dominance}, where $s=4 \, E_{\text{in}} \, E_\text{out}$. This S-matrix has also been derived through field theory by means of the eikonal resummation. In the flat space eikonal limit, elastic forward scattering of massive scalar particles can be calculated exactly \cite{HOOFT1987dominance, KabatOrtiz,VerlindeVerlinde}, with further research by \cite{IyerWill,AdamoGravitonScattering,Muzinich1987,Ciafaloni2018Colferai,Mogull2020,Fabbrichesi1993,Adamo:2021rfq}. On flat space the eikonal phase demands trans-Planckian energies, which for small impact parameters should lead to black hole production \cite{Giddings2007,Eardley2002,Banks1999,Addazi2016} (so large impact parameters are required). This was shown to be fundamentally different for black hole eikonal scattering due to the emergent mass scale $\frac{1}{R}$. The eikonal amplitude has also been calculated on an AdS background \cite{Parnachev2020,Cornalba2006,Cornalba2007,Cornalba2006two,Cornalba2007two,Lam2018}, and within celestial holography \cite{RaclariuEikonalCFT}. \\
Of the four external particles, the two ingoing ones are taken to carry momenta $p_1$ and $p_2$ while the outgoing momenta are labelled by $p_3$ and $p_4$. The Mandelstam variables of interest are 
\begin{align}
    s ~ \coloneqq ~ - \left(p_1 + p_2\right)^2 \qquad \text{and} \qquad t ~ \coloneqq ~ - \left(p_1 - p_3\right)^2 \, ,
\end{align}
and we focus on the eikonal limit $s \gg t$. Moreover, to avoid large transverse momentum transfer, the impact parameter is taken to be large; in flat space, the only available scales to compare the impact parameter with are the Planck length, i.e. $b\gg \lpl$, and the scale associated to the centre of mass energy of the collisions, i.e. $b \gg G_{N} \sqrt{s}$. Therefore, the two scattering particles maintain most of their momentum in the scattering direction which we call longitudinal, i.e. $p_1^{\parallel}\approx p_2^{\parallel}$. The two particles do exchange a small amount of momentum in the transverse directions, such that $p_1^{\perp}\neq p_2^{\perp}$. Nevertheless, for all particles, we take $p_i^{\parallel}\gg p_i^{\perp}$. In this limit the amplitude involves a resummation only over ladder diagrams, and the result for massless scalars is given by \cite{KabatOrtiz}
\begin{align}
\label{eq:amplitudeMinkowskicalc2}
i \mathcal{M} ~ = ~ 2s\int\ed^2 x_{\perp} \gap e^{- i q_{\perp} \cdot x_{\perp}} \bigr(e^{-2iG s\log (\tilde{\mu}x_{\perp})} - 1\bigr) \, .
\end{align}
where $\tilde{\mu}$ is an infrared regulator for the graviton. This equation clearly matches \eqref{eq:semiclasSmatrix} up to an overall kinematical factor $2s$ and the $-1$ free field contribution. Indeed solving the integral gives
\begin{align}
\label{eq:amplitudeMinkowski}
i\mathcal{M} ~ = ~ \dfrac{ i \kappa^2 s^2}{-t} \dfrac{\Gamma\left(1 - i\alpha(s)\right)}{\Gamma\left(1 + i \alpha(s)\right)} \left(\dfrac{4\tilde{\mu}^2}{-t}\right)^{-iGs} \, .
\end{align}
This amplitude contains all power of $G$, but is valid only to leading order in $s$. Therefore, the approximation gets better with ultra-high energy scattering. When $\tilde{\mu}=1$, this is equal to \eqref{eq:flatspacesmatrix}, up to a conventional prefactor. Thus the scattering amplitude matches the semi-classical scattering matrix derived by 't Hooft in \cite{HOOFT1987dominance} based on a first quantised description of shockwaves on an Aichelburg-Sexl metric \cite{aichelburg1971gravitational}.\\
\\
We expected to be able to reconstruct the black hole semi-classical S-matrix using field theory as well. Previous attempts in \cite{ShortPaper,LongPaper,SQEDpaper} appear to give the correct result in a harmonics basis, however the interpretation of the external states is different. \eqref{eqn:shockwaveQMSmatrix} holds for a distribution of many particles, while the previous papers correlate two particles only. In this paper we extend the eikonal analysis to an arbitrary number of particles.

\clearpage

\section{Near-horizon field theory}
\label{sec:field theory}
Here we provide a short summary of the field theory developed in \cite{Toolbox} and \cite{SQEDpaper}, with some new insights. Using this field theory we will calculate the amplitudes in the next sections.

\subsection{Scalar fields}
\label{chap: scalar fields}
As matter content we will consider scalar fields, both complex and real. For a complex scalar field we find the following action in spherical harmonics \cite{SQEDpaper}
\begin{align}
\label{eq:scalaraction}
S ~ &= ~ -\sum\limits_{\ell m}\int\ed^2  x ~  \,\phi_{\ell m} \left(-\p^2+ \frac{f(r)\ell(\ell+1)}{r^2}+\frac{1}{r}\p^2 r\right)\bar{\phi}_{\ell m} \,, 
\end{align}
where all remaining contractions are made with the flat metric $\eta_{ab}$. The scalar field expansion is defined by
\begin{align}
    \phi(x^{\mu})=\sum\limits_{\ell m} \frac{\phi_{\ell m}(x^a)}{r}Y_{\ell m}(x^a).
\end{align}
The action thus becomes of a Klein-Gordon form with a mass-potential $M_{\ell}^2$ defined above. We now seek to approximate near the horizon, which yields different results in different coordinates. In our coordinates assuming $x^2\approx 0$ we find
\begin{align}
S ~ &= ~ -\sum\limits_{\ell m}\int\ed^2  x ~  \,\phi_{\ell m} \left(-\p^2+\mu^2\lambda\right)\bar{\phi}_{\ell m} \, ,
\end{align}
where we used the inverse radius $\mu=\frac{1}{R}$ as effective mass, and defined $\lambda=\ell^2+\ell+1$ as shorthand notation for the angular momentum contribution. Since the mass does not depend on $r$ any more we can Fourier transform, resulting in
\begin{align}
    S ~ &= ~ -\sum\limits_{\ell m}\int\frac{\ed^2p}{(2\pi)^2} ~  \,\phi_{\ell m}(p) \left(p^2+ \mu^2\lambda\right)\bar{\phi}_{\ell m}(p).
\end{align}
For a real scalar the action is identical up to a prefactor $1/2$. The interactions are governed by spin 1 and 2 gauge fields, specifically a U(1) coupled gauge field to the complex scalar field, and graviton perturbations for both scalar fields. 

\subsection{Electromagnetism: Spin-1}
\label{chap: EM}

\begin{figure}[h!]
\centering
\includegraphics[width=\textwidth]{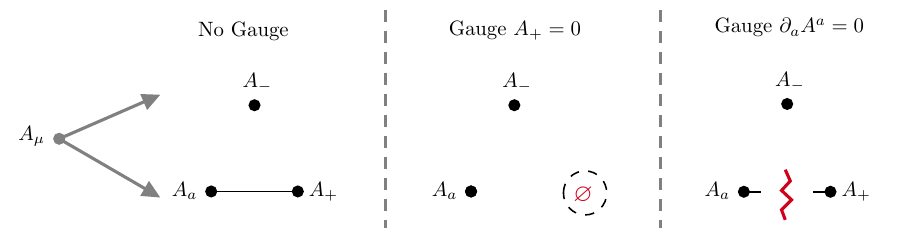}
\caption{\label{fig:EMgauges} An illustration of the different harmonic modes that split off from $A_{\mu}$, denoted by black dots. In general the equal parity modes have interactions indicated by the solid line. In the first gauge choice $A_+$ is removed explicitly (replaced by an empty set), leaving only the two decoupled modes. In the last gauge choice all 3 modes remain, however the gauge choice breaks the coupling in the even sector.}
\end{figure}

\label{sec:gaugeaction}
In this subsection we shall perform the splitting of the metric to derive the relevant propagators for the electromagnetic gauge field $A_{\mu}$. Because the field is of spin 1, this is a lot more involved than the scalar. The spherical harmonics expansion for a spin-1 field splits the four degrees off freedom of $A_{\mu}$ into three degrees of freedom $A_a,A_+,A_-$. In order to find the propagators we need to fix gauge explicitly. We will use the same gauge as in \cite{SQEDpaper} and propose a new gauge. While the results in the end are the same, the new gauge is mathematically more elegant, and requires fewer approximations. We will first consider the quadratic actions. The gauge field action is defined by
\begin{align}
    S_{EM}=-\frac{1}{4}\int\ed^4 x ~ \sqrt{-g} ~ F_{\mu\nu}F^{\mu\nu},
\end{align}
where $F_{\mu\nu}=\p_{\mu}A_{\nu}-\p_{\nu}A_{\mu}$ where we may use partial derivatives instead of covariant derivatives because of the antisymmetry (and the torsion-free background). Recall that the gauge field obeys a symmetry
\begin{align}
    A_{\mu}\to A_{\mu}+\p_{\mu}\lambda,
\end{align}
that leaves the action invariant, for any local scalar parameter $\lambda$. We will need to fix the gauge later, however we shall delay this choice until after applying the metric and spherical harmonics expansion. As outlined in \appref{app: spherical harmonics} a spin-1 field can be expanded as
\begin{align}
    A_a&=\sum\limits_{\ell m}\fdfield{A}_a^{\ell m}(x^a) Y_{\ell m}(x^A),\\
    A_A&= \sum\limits_{\ell m}\fdfield{A}_+^{\ell m}(x^a) \p_A Y_{\ell m}(x^A)-\sum\limits_{\ell m} \fdfield{A}_-^{\ell m}(x^a)\epsilon_{A}^{\mgap B}\p_B Y_{\ell m}(x^A),
\end{align}
where we will use the shorthand notation $\eta_{A,\ell m}^+=\p_A Y_{\ell m}(x^A),\eta_{A,\ell m}^-=-\epsilon_{A}^{\mgap B}\p_B Y_{\ell m}(x^A)$, the minus sign is a convention without loss of generality. Henceforth,, we will omit the dependencies on $x^a,x^A$, and remark that $\epsilon_{AB}$ is by our definition raised and lowered with the two-sphere metric $\gamma_{AB}$ only. The modes $\fdfield{A}_a^{\ell m},\fdfield{A}_+^{\ell m}$ we will call the \textit{even parity} modes since their eigenfunctions $\eta_{A,\ell m}^+$ remain the same under parity transformations $x^A\to - x^A$, whereas for $\fdfield{A}_-^{\ell m}$ the eigenfunction obtains a minus sign under the same parity transformation, hence called the \textit{odd parity mode}. We can already argue from the underlying spherical symmetry, that any coupling between the odd an even modes must vanish. Supposing that a term with one even- and one odd-parity existed in the action, then the action would change sign under a parity transformation. For this reason such terms cannot exist in any spherical background. This decoupling was shown explicitly in \cite{SQEDpaper}. As will be argued in \secref{chap: eikonalized theory} we may neglect the odd parity modes in the eikonal limit, so we will ignore them for this subsection for brevity as well.\\
\\
In harmonics we find that the gauge transformation has the following form:
\begin{align}
    \delta \fdfield{A}_a^{\ell m}&=\p_a \lambda^{\ell m},\\
    \label{eq:gaugetransf}
    \delta \fdfield{A}_+^{\ell m}&= \lambda^{\ell m}.
\end{align}
The gauge transformation acts on the even modes only, in an expected way with a derivative on the vector-mode $\fdfield{A}_a$. However the scalar mode $\fdfield{A}_+$ changing with the gauge parameter without any derivative. This already hints that the easiest gauge-choice is to remove the scalar mode altogether $\fdfield{A}_+=0$, which is a valid gauge choice except for $\ell=0$, since $\fdfield{A}_+$ does not exist at $\ell=0$. The calculation in this gauge has been done in \cite{SQEDpaper}. The field must be redefined by a Weyl transformation
\begin{align}
    \fdfield{A}_a&=\frac{\sqrt{f}}{r}A_a,
\end{align}
and $\fdfield{A}_-=A_-$ unchanged. This gives the following quadratic action
\begin{align}
    S_{even}&=-\frac{1}{2}\sum\limits_{\ell m}\int\ed^2 x  ~A^a\left(\eta_{ab}q^2-q_aq_b+\frac{\ell(\ell+1)}{R^2}\eta_{ab}\right)A^b,
\end{align}
where we used the horizon approximation $x^2=0$ and shockwave approximation $x^a A_a=0$ as discussed in \cite{SQEDpaper}. For $\ell=0$ an extra condition is needed. The shockwave approximation appears to be quite strong, however we may propose a different gauge where it is not needed. Let us define a lightcone harmonic gauge by $\p_a \fdfield{A}^a=0$ and $\fdfield{A}_+=A_+$ unchanged. In that case we find
\begin{align}
    S_{even}&=-\frac{1}{2}\sum\limits_{\ell m}\int\ed^2 x  ~\left( A^a (-q)\eta_{ab}\left(q^2+\frac{\ell(\ell+1)}{R^2}\right)A^b(q)+\ell(\ell+1)A_+(-q)q^2 A_+(q)\right).
\end{align}
The upshot is that to derive the action above only the horizon approximation $x^2=0$ is needed and it is valid for all $\ell$, although this has been traded for an additional term for $A_+$. This shows, however, that a method avoiding the shockwave approximation is possible. More importantly: In the soft limit $q\to 0$ both actions do coincide, while we would have found a mismatch between both gauges had we not applied the shockwave approximation \footnote{Without the shockwave approximation the mass term for the $A_+=0$ gauge obtains a shift $\ell(\ell+1)\to \ell(\ell+1)+1$, while the mass term in the lightcone harmonic gauge remains the same.}. This shows that the horizon approximation may only be consistent when paired with the shockwave approximation.

\subsubsection{Interactions}
\label{sec:gaugeinteractions}
The interactions we consider are those sourced by a complex matter current $J_{\mu}$ that is classicaly conserved. This gives the following interaction term:
\begin{align}
    S=iq \int\ed^4x ~ \sqrt{-g}~ A^{\mu} \left(\phi \p_{\mu}\bar{\phi}-\bar{\phi}\p_{\mu}\phi\right).
\end{align}
Writing all fields in harmonics, including their relevant normalization factors, gives
\begin{align}
    S&=i\mu q \sum\limits_{\{\ell m\}}CL[1,2,3]\int\ed^2 x  ~  A_{\ell_1 m_1}^a \left(\phi_{\ell_2 m_2}\p_a \bar{\phi}_{\ell_3m_3}-\bar{\phi}_{\ell_3m_3}\p_a \phi_{\ell_2m_2}\right)\nonumber \\
    \label{eq: even gauge interactions}
    &+i\mu Q \sum\limits_{\{\ell m\}}\left(\mu^2\ell_2(\ell_2+1)-\mu^2\ell_3(\ell_3+1)\right)CL[1,2,3]\int\ed^2 x  ~A^+_{\ell_1 m_1} \phi_{\ell_2 m_2} \bar{\phi}_{\ell_3m_3},
\end{align}
where $\{\ell m\}$ is used to denote summation over $\ell_1m_1,\ell_2m_2$ and $\ell_3m_3$ at the same time, and we recognized the definitions of the $CL$ functions as defined in \appref{app: spherical harmonics}. These received a shorthand notation $CL[i,j,k]=CL[\ell_im_i, \ell_jm_j,\ell_k m_k]$ for brevity. Recognize that $\mu^2\ell_3(\ell_3+1)-\mu^2\ell_2(\ell_2+1)$ is precisely the difference between the on-shell masses for the scalar field, indicating how combining both the $A_a$ and $A_+$ vertex for on-shell scalars returns the Ward identity. In principle because we are considering scalar electrodynamics there is also a quartic coupling $A_{\mu}A{^\mu}\phi \bar{\phi}$. We will neglect these couplings in the eikonal limit, as was shown in \cite{SQEDpaper}.

\subsection{Gravity: Spin-2}
\label{chap: gravity}

\begin{figure}[h!]
    \centering
    
\includegraphics[width=\textwidth]{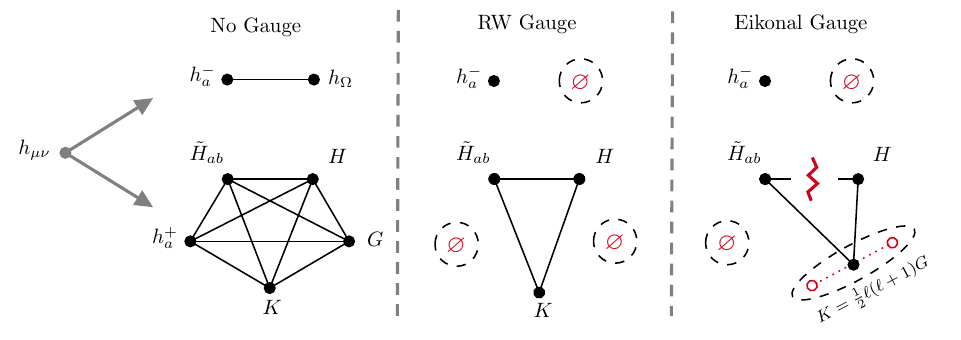}
    \caption{An illustration of the different harmonic modes (denoted by black dots) and gauges for gravity. Here $\tilde{\fdfield{H}}_{ab}$ is the traceless version of $\fdfield{H}_{ab}$. In both the even and odd sector there are many couplings between the modes. In the Regge-Wheeler (RW) gauge three modes are explicitly set to zero (replaced by empty sets in the image), also removing many couplings. In the eikonal gauge only two modes are removed, however the even scalar modes are combined into one, and the coupling between $\tilde{H}_{ab}$ and $H$ is broken, effectively reducing in less couplings than the RW gauge.}
    \label{fig:grav gauges}
\end{figure}

In this subsection we write down the actions for the gravitational field. Additionally, we must apply the spherical harmonics expansion, which for a spin-2 field splits the 10 degrees of freedom of $h_{\mu\nu}$ over 6 modes $H_{ab},h^+_a,K,G ,h^-_a,h_{\Omega}$. This subsection is based on \cite{LongPaper,Toolbox}. For our spin-2 interactions we will consider linearized gravitons around the metric ansatz. We start from the Einstein-Hilbert action
\begin{align}
    S=\frac{1}{16\pi G}\int\ed^4 x~\sqrt{-g}~R.
\end{align}
Metric fluctuations are defined in the background field method about the Schwarzschild background as $\bar{g}_{\mu\nu} = g_{\mu\nu} + \kappa h_{\mu\nu}$ where $\kappa^2=8\pi G$. Since the Schwarzschild metric $g_{\mu\nu}$ is a vacuum solution of the equations of motion, the on-shell action and the variation of it to linear order in $h_{\mu\nu}$ vanish. In the soft limit, therefore, the path integral is dominated by quadratic terms in $h_{\mu\nu}$. We expand the graviton field in harmonics as well. The even parity modes are given by
\begin{align}
    \nonumber h^+_{ab}&=\fdfield{H}_{ab}Y_{\ell}^m ,\nonumber \\
    h^+_{aA}=h^+_{Aa}&= \fdfield{h}_a^+\eta^+_{A,\ell m} ,\label{eq: gravitoneven index}\\
    \nonumber h^+_{AB}&=\fdfield{K} g_{AB}Y_{\ell}^m+r^2\fdfield{G}\tilde{\nabla}_{(A}\eta^+_{B),\ell m}.\nonumber
\end{align}
Here $\tilde{\nabla}_A$ is a covariant derivative involving the two-sphere metric $\gamma_{AB}$ only. The odd parity harmonics are given by
\begin{align}
    \nonumber h^-_{ab}&=0 ,\nonumber\\
    h^-_{aA}=h^-_{Aa}&= \fdfield{h}_a^-\eta^-_{A,\ell m} ,\label{eq: gravitonodd index}\\
    \nonumber h^-_{AB}&=\fdfield{h}_{\Omega}\tilde{\nabla}_{(A}\eta^-_{B),\ell m}. \nonumber
\end{align}
The definition of even and odd is similar to spin 1 determined by the action of a parity transformation. Since the background is spherically symmetric and we expect parity invariance of the action we may expect any couplings between a single odd and a single even field to vanish. This was shown explicitly for a specific gauge in \cite{LongPaper}. Since the graviton field transforms under gauge transformations
\begin{align}
    h_{\mu\nu}\to h_{\mu\nu}+\nabla_{\mu}\xi_{\nu}+\nabla_{\nu}\xi_{\mu}
\end{align}
we must fix gauge in order to define invertible quadratic actions. Similar to the gauge field, ss will be argued in \secref{chap: eikonalized theory} we may neglect the odd parity modes in the eikonal limit, so we will ignore them for this subsection for brevity as well.\\
\\
The most logical choice is again to set explicit modes to vanish: $h_a^{+}=0,h_{\Omega}=0,G=0$. This is the original Regge-Wheeler gauge as originally used in \cite{ReggeWheeler,MartelPoisson} and by us in \cite{LongPaper,ShortPaper,2to2N,Toolbox}. We also redefine the fields for appropriate normalization of the kinetic terms in the action:
\begin{align}
    \nonumber h^+_{ab}&=\frac{f(r)}{r}H_{ab}Y_{\ell}^m,\\
    h^+_{aA}=h^+_{Aa}&= 0,\\
    \nonumber h^+_{AB}&=\frac{1}{r}K g_{AB}Y_{\ell}^m.
\end{align}
The resulting quadratic action is given by \cite{LongPaper,Toolbox}
\begin{align}
\label{eq:evenactionfinal}
S ~ = ~ \dfrac{1}{4} \int\ed^2 k \biggr(H^{ab}\Delta^{-1}_{abcd}H^{cd} + H^{ab} \Delta^{-1}_{L,ab} K + K\Delta^{-1}_{R,ab} H^{ab} + K\Delta^{-1}K\biggr) \, ,
\end{align}
where
\begin{align}
\Delta^{-1} ~ &= ~ k^2 + \mu^2 \, , \\
\Delta^{-1}_{ab} ~ &= ~ \eta_{ab} \left(k^2+\dfrac{1}{2} \mu^2 \ell(\ell+1)\right) - k_a k_b \, , \\
\Delta^{-1}_{abcd} ~ &= ~ \dfrac{\mu^2 \left(\ell^2+\ell+1\right)}{2} \bigr(\eta_{ab}\eta_{cd}-\eta_{a(c}\eta_{d)b}\bigr) \, . 
\end{align}
Here the shockwave approximation $x_a H^{ab}=0$ and horizon approximation $x^2=0$ were applied.

\paragraph{Eikonal gauge:} For this paper we observe the existence of a different possible gauge inspired by the interaction vertex. The gauge choice $K=\frac{1}{2}\ell(\ell+1)G$ appears to provide the most optimal vertex when neglecting angular momenta, hence the name eikonal gauge. In this gauge the quadratic operators become
\begin{align}
\Delta^{-1} ~ &=  -\ell(\ell+1)\frac{\ell^2+\ell-2}{4}\left(k^2+\mu^2\right)~ \, ,  \\
\Delta^{-1}_{ab} ~ &= ~ \mu^2\ell(\ell+1)\frac{\ell^2+\ell-2}{4} \eta_{ab} , \\
\Delta^{-1}_{abcd} ~ &=\dfrac{1}{2}\mu^2 \left(\ell^2+\ell+1\right) \bigr(\eta_{ab}\eta_{cd}-\eta_{a(c}\eta_{d)b}\bigr) \, ,
\end{align}
valid for $\ell>1$ only, although we may simply use the same conditions as the Regge-Wheeler gauge to extend to $\ell=0,1$. The eikonal gauge resolves a subtlety in the derivation of the quadratic operators: In the Regge-Wheeler gauge there was an antisymmetry in the operators that had to be resolved by undoing part of the Weyl transformation (Section 4.1 of \cite{LongPaper}), while this antisymmetry was never present in the eikonal gauge, ensuring full consistency with the approximations. Finally, the tensor-scalar coupling $\Delta^{-1}_{ab}\sim \eta_{ab}$ has metric tensor structure. This indicates that the traceless tensor $\tilde{H}_{ab}=H_{ab}-\tfrac{1}{2}\eta_{ab} H$ completely decouples from the trace $H$ and scalar $G$, providing a simpler structure for the interactions.

\subsubsection{Interactions}
\label{sec: gravity interactions}
The interactions terms are given by the linear interaction with the graviton
\begin{align}
    S_{int}=-\kappa\int\ed^4 x ~ \frac{\delta S_M}{\delta \bar{g}^{\mu\nu}(x)}\biggr\rvert_{\bar{g}=g} h^{\mu\nu}(x)= \frac{\kappa}{2}\int\ed^4 x ~ h^{\mu\nu}(x) T_{\mu\nu},
\end{align}
where $T_{\mu\nu}$ is the stress-energy tensor. We neglect higher order graviton interactions. In principle from the 4D theory both scalars fields and the gauge field contribute to the stress-energy. We will ignore the gauge-graviton coupling, because of the additional complexity to the field theory, and focus on the scalar couplings. Additionally we may assume this coupling to give sub-leading effects in the eikonal limit, although this remains to be proven on the black hole background. Observe that the difference between the complex and real scalar field is only a factor of 2, and symmetrization over indices, so if we have one the other is easily transcribed. Splitting all indices, and recognizing the definitions in \appref{app: spherical harmonics}, we find for the real scalar
\begin{align}\label{eq: even graviton interaction}
    S&=\frac{\mu\kappa}{2}\sum\limits_{\{\ell m\}}CL[1,2,3]\int\ed^2  x ~ \left(\tilde{H}_{\ell_1 m_1}^{ab}-\left(K_{\ell_1m_1}-\tfrac{1}{2}\ell_1(\ell_1+1) G_{\ell_1m_1}\right) \eta^{ab}\right)\p_a \phi_{\ell_2 m_2}\p_b \phi_{\ell_3m_3}\nonumber \\
    &-\frac{\mu\kappa}{4}\sum\limits_{\{\ell m\}}\mu^2CL_+[2,3;1]\int\ed^2  x ~ H_{\ell_1m_1}\phi_{\ell_2m_2}\phi_{\ell_3m_3}\nonumber \\
    &+\frac{\mu\kappa}{2}\sum\limits_{\{\ell m\}}\int\ed^2  x h^{a,+}_{\ell_1m_1}\left(\phi_{\ell_3m_3}\p_a\phi_{\ell_2m_2}\mu^2CL_+[1,3;2]+(2\leftrightarrow 3)\right)\nonumber \\
    &+\frac{\kappa}{2}\sum\limits_{\{\ell m\}}\mu^2CL_G[2,3;1]\int\ed^2  x G_{\ell_1m_1} \phi_{\ell_2m_2}\phi_{\ell_3m_3}
\end{align},
where further insertion of the identities in \appref{app: spherical harmonics} is possible but at this moment not fruitful.

\subsection{Eikonalised Theory}
\label{chap: eikonalized theory}
So far we have derived the actions near the horizon, however we are only interested in performing calculations in the eikonal phase $s\gg \gamma \mpl$. As argued in \cite{LongPaper} this allows us to simplify all interactions by keeping only leading order terms $\sim  s$. Since the Mandelstam variable may only emerge from the lightcone momenta $p_a$, we may neglect all transverse momenta $\p_A$ in the vertex. In the propagators we refrain from doing so when possible to avoid changing the pole structure, however in the vertex all corrections are automatically polynomial. For the interactions we may thus set $\p_A\to 0$ or $\p_a \gg \frac{\sqrt{\ell(\ell+1)}}{\mu}$. The gauge field interaction in \eqref{eq: even gauge interactions} becomes
\begin{align}
    S&=iq\mu \sum\limits_{\{\ell m\}}CL[\ell_1m_1,\ell_2m_2,\ell_3m_3]\int\ed^2 x  ~ \ A_{\ell_1 m_1}^a \left(\phi_{\ell_2 m_2}\p_a \bar{\phi}_{\ell_3m_3}-\bar{\phi}_{\ell_3m_3}\p_a \phi_{\ell_2m_2}\right).
\end{align}
We observe that the angular modes $A_{\pm}$ drop out entirely, and only the longitudinal mode contributes to eikonal scattering. The gravitational interaction after setting $\p_A\to 0$ is simplified immensely, \eqref{eq: even graviton interaction} becomes only
\begin{align}
    S&=\mu\sqrt{\kappa}\sum\limits_{\{\ell m\}}CL[\{\ell m\}]\int\ed^2  x ~ \left(\tilde{H}_{\ell_1 m_1}^{ab}-\left(K_{\ell_1m_1}-\tfrac{1}{2}\ell_1(\ell_1+1) G_{\ell_1m_1}\right) \eta^{ab}\right)\p_a \phi_{\ell_2 m_2}\p_b \bar{\phi}_{\ell_3m_3},
\end{align}
where the contribution from the odd modes vanish entirely. We see that in the eikonal limit almost all terms drop out already before gauge fixing, where the vertex depends only on a specific linear combination of graviton modes. The field $h_a^+$ and the trace $H$ decouple completely from the scalars. It can now clearly be observed that the gauge $K=\frac{\ell(\ell+1)}{2}G$ is also interesting, which is why we developed the eikonal gauge.

\subsubsection{Eikonalised gauge fields}
Because certain field components do not contribute to the vertices, we may integrate them out of the theory as a whole. For the gauge field this simply means integrating $A_+$ and $A_-$ out, however since in both gauges these fields decouple, we may simply ignore them, and assume only $A_a$ to exist. The eikonalised theories for gravity are more involved. The interaction vertex in Kruskal-Szekeres coordinates however only couples to a very specific linear combination of the field modes. We redefine these into an effective coupling field
\begin{align}
    \mathfrak{h}_{ab}&=\tilde{H}_{ab}-\left(K-\tfrac{1}{2}\ell(\ell+1) G\right)\eta_{ab}
\end{align}
to reduce the amount of vertex couplings to a single field. For the eikonal gauge this simply means integrating out the scalar modes $H,K$. Since these are already decoupled, we immediately find
\begin{align}
    S_{eik}&=\frac{1}{4}\int\ed^2 x ~ \mathfrak{h}^{ab}~\mathcal{P}^{-1}_{abcd} ~ ~\mathfrak{h}^{cd},\\
    \prop_{abcd}&= -\frac{1}{\mu^2\lambda}T_{abcd},
\end{align}
where the propagator does not yet contain the necessary symmetry factors. Here $T_{abcd}=\eta_{ab}\eta_{cd}-\eta_{ac}\eta_{bd}-\eta_{ad}\eta_{bd}$ is the traceless identity tensor. For the Regge-Wheeler gauge the process is more involved. This was done in \cite{Toolbox} and gives
\begin{align}
S ~ = ~ \sum_{\ell m} \dfrac{1}{4} \int \dfrac{\mathrm{d}^{2} p}{ \left(2\pi\right)^{2}} \biggr(\mathfrak{h}_{ab} \left(\hat{\prop}^{-1}\right)^{abcd} \mathfrak{h}_{cd} + \hat{K} \hat{\prop}^{-1} \hat{K}\biggr) \, .
\end{align}
where
\begin{align}
\hat{\prop}_{abcd} ~ &= -\frac{1}{\mu^2\lambda}T_{abcd}+\prop_K(\eta_{ab}+\tilde{\proj}_{ab})(\eta_{cd}+\tilde{\proj},_{cd})~\\
\hat{\mathcal{P}} ~ &= \frac{4}{\mu^2\lambda}~ \, ,
\end{align}
where $\tilde{\proj}_{ab}$ is the traceless version of $\proj_{ab}$. The second propagator is surprisingly simple, however since we are free to integrate out the $\hat{K}$ field we pay no further attention to it. The first propagator resembles the one we originally had before the field transformation, in particular the soft part in front has become traceless without further addition. Writing out the terms we see that $\hat{\prop}_{abcd}$ contains terms quadratic in $k_a$. This seems to indicate problematic momentum behaviour. However we want to note that similar higher order momentum behaviour is present in any massive bosonic theory with non-zero spin, indeed $\proj_{ab}\sim \frac{k_ak_b}{\mu^2\lambda}$. We have also seen that the we did not have this problem in the eikonal gauge at all, in fact there the propagator is given only by the soft term $\frac{1}{\mu^2\lambda}T_{abcd}$. This indicates that the behaviour in terms of $k_ak_b$ is gauge-dependent, and we expect that the extra terms in this gauge do not affect any physics\footnote{We want to stress that this only holds for the eikonal limit. In general we do not expect gauge-dependence in any physical observable using the approximations above, however when not working in the eikonal limit this means for consistency we must also include all existing modes and vertex contributions, and calculate all possible diagrams. We expect that consistently taking all contributions into account ensures gauge invariance in general, whereas in our field theory we only find gauge invariance when working in the eikonal limit consistently}.\\
\\
These expressions are strictly speaking valid only for the multipole modes $\ell>1$. However since in the case of $\ell=1,0$ we have $K=0$, the procedure above simplifies a lot: The only effect is the propagator becoming traceless. Thus for $\ell=1,0$ the propagator is given by
\begin{align}
    \prop_{abcd}=-\frac{1}{\mu^2\lambda}T_{abcd},
\end{align}
which coincides with the eikonal gauge. This shows that working in the eikonal gauge is easier in the eikonal limit, the propagator simplifies immensely \textit{and} holds for all $\ell$.

\subsubsection{Eikonalized Feynman rules}
\label{sec: feynman rules}
\begin{figure}[h!]
    \centering
\includegraphics[width=0.6\textwidth]{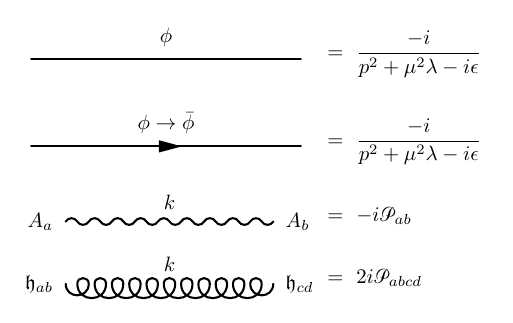}
    \caption{The necessary propagators in the black hole eikonal phase. For all different spin fields only a single relevant mode survives.}
    \label{fig:bh props}
\end{figure}
\begin{figure}[h!]
    \centering
    \includegraphics[width=0.8\textwidth]{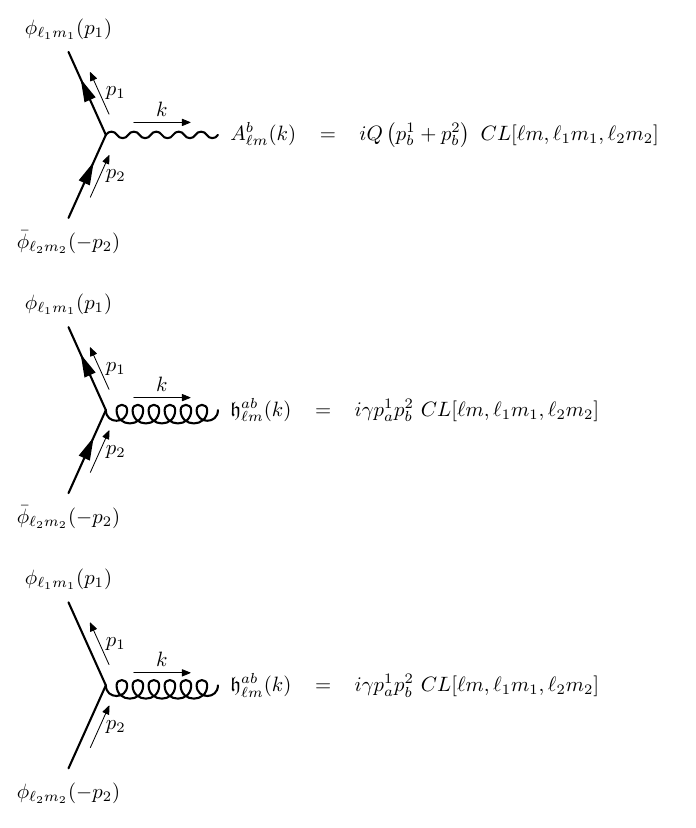}
    \caption{The interaction vertices for the different fields. The gauge field only interacts with the complex scalar, whereas the graviton interacts with both in an identical fashion. One could symmetrize the graviton vertices over the indices, however the fact that the graviton propagator is symmetric automatically takes care of this.}
    \label{fig:bh inters}
\end{figure}

With the results of the previous section, we can now formally write down the Feynman rules, including any factors that were in the action and so far ignored. The only remaining propagators are shown in \figref{fig:bh props}. The first propagator describes the real scalar, the second one the complex scalar. The expressions for the propagators depend on the gauge, and are given by
\begin{align}
    A_+&=0\text{ gauge} & &\mathcal{P}_{ab}= \frac{1 }{k^2+\mu^2(\lambda-1)-i\epsilon}\left(\eta_{ab}+\frac{k_ak_b}{\mu^2(     \lambda-1)}\right),\\
    \p_a \fdfield{A}^a&=0\text{ gauge} &&\mathcal{P}_{ab}= \frac{\eta_{ab} }{k^2+\mu^2(\lambda-1)-i\epsilon},
\end{align}
for the photon and 
\begin{align}
    &\text{RW gauge} && \mathcal{P}_{abcd}~=~-\frac{1}{\mu^2\lambda}T_{abcd}+\prop_K(k)(\eta_{ab}+\tilde{\proj}_{ab})(\eta_{cd}+\tilde{\proj}_{cd}),\\
    &\text{Eikonal gauge}&&\mathcal{P}_{abcd}~=     -~\frac{1}{\mu^2\lambda}T_{abcd},
\end{align}
for the graviton. Here
\begin{align}
    \prop_K&=-\frac{\lambda}{\lambda-2}\frac{1}{k^2+\mu^2\left(\lambda-\frac{\lambda-1}{\lambda-2}\right)-i\epsilon }\\
    \tilde{\proj}_{ab}&=-\frac{2}{\mu^2\lambda}\left(k_ak_b-\tfrac{1}{2}\eta_{ab}k^2\right).
\end{align}
We stress that the expressions above in the first gauges for both fields are only valid for $\ell>1$ or $\ell\ge 1$ for the graviton and photon respectively. For the special cases $\ell=1,0$ instead the expressions in the second gauges must be used, which hold for arbitrary $\ell$. The interactions are shown in \figref{fig:bh inters}, where we defined new coupling constants $\gamma=\mu\kappa$, which is dimensionless, and an effective charge $Q=\mu q$ . Note that the direction of momentum is important for the sign of $p_a$ in the vertex.

\subsubsection{Flat space analogue}
\label{sec: flat space analogue}
\begin{figure}[h!]
    \centering
    \includegraphics[width=0.6\textwidth]{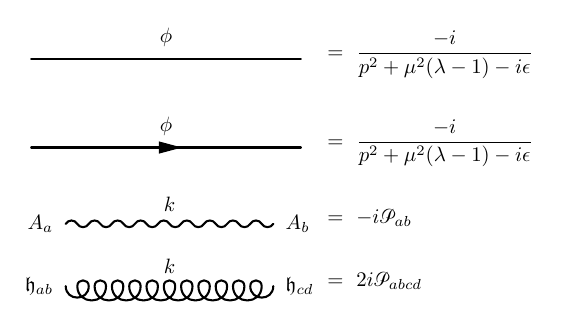}
    \caption{The field propagators for flat space on a fixed radial shell of radius $\rfs=1/\mu$. The structure is identical to the black hole, with minor changes in the mass terms.}
    \label{fig:flat props}
\end{figure}

For comparison with literature it would be nice to have a flat space version and interpretation of the Feynman rules in \secref{sec: feynman rules}. The analogous thing to do would be to set $r=\infty$ to look at future and past null infinity. Taking this strict limit is not effective, however we can choose to fix $r=\rfs$ some constant value, which we assume to be much larger than any other scale present, but keep written as a regulator. Because we are in flat space we use coordinates $(t,r,\theta,\phi)$ and the metric is given by $f(r)=1$. The actions on flat space in harmonics become instead
\begin{subequations}
\begin{align}
    &\text{complex scalar field}: & &S=- \int\ed^2 x ~ \phi\left(-\p^2+\frac{\ell(\ell+1)}{\rfs^2}\right)\bar{\phi},\\
    &\text{real scalar field}: & &S=-\frac{1}{2} \int\ed^2 x ~ \phi\left(-\p^2+\frac{\ell(\ell+1)}{\rfs^2}\right)\phi,\\
    &\text{gauge field}: && S=-\frac{1}{2} \int\ed^2 x ~ A^a\left(-\p^2+\frac{\ell(\ell+1)}{\rfs^2}\right)\eta_{ab}A^b,\\
    &\text{graviton  mode}: && S=-  \dfrac{1}{4} \int \ed^2 x \mathfrak{h}^{ab}\frac{\ell(\ell+1)}{4\rfs^2}T_{abcd} \mathfrak{h}^{cd},
\end{align}
\end{subequations}
where we only kept the modes relevant for the eikonal limit, and work in the eikonal gauge for the graviton mode and the lightcone harmonic $\p_aA^a=0$ gauge for the gauge field. All actions contain implicit summation over $\ell,m$. What we observe is that the propagators are all almost identical to the Kruskal-Szekeres one, upon identifying $\mu\leftrightarrow\frac{1}{\rfs}$ and changing the contribution of $\ell$ at specific locations. The interactions do not contain any potentials, and thus we can immediately find the flat space interactions using the identification $\mu\leftrightarrow\frac{1}{\rfs}$ only. The set of flat space propagators is given in \figref{fig:flat props}.

The first propagator describes the real scalar, the second one the complex scalar. The expressions for the propagators depend on the gauge, and are given by
\begin{align}
    \p_a A^a=0\text{ gauge:} &&\mathcal{P}_{ab}= \frac{\eta_{ab} }{q^2+\mu^2(\lambda-1)-i\epsilon},
\end{align}
for the photon and 
\begin{align}
    &\text{Eikonal gauge:}&&\mathcal{P}_{abcd}~=     ~-\frac{1}{\mu^2(\lambda-1)}T_{abcd},
\end{align}
for the graviton. The vertices are completely identical to the black hole case upon writing $\mu=1/\rfs$ and changing the interpretation of the momenta and coordinates. A comment is in order on the $\ell=0$ behaviour. While in these gauges the propagators are supposed to be regular, the masses vanish, which leads to a pole for the graviton. Indeed the graviton modes $\flatgrav^{00}_{ab}$ appear to vanish exactly on flat space. These must be disregarded from the theory to ensure the summations are finite.\\
Furthermore it is important to note that this subsection describes a significantly physically distinct system from the black hole horizon. While the equations look similar, this is because of our choice of coordinates and definitions so that the horizon calculations resemble flat space for ease of calculations. However the black hole horizon rules are defined for Kruskal-Szekeres coordinates, so the momenta of the particles are defined differently than in flat space, and on a different part of the Penrose diagram. Because the coordinates are related exponentially $x\sim e^{t-r^*},y\sim e^{t+r^*}$ the black hole momenta are also exponentially scaled versions of the flat space momenta (where $r=r^*$).

\clearpage

\section{Tree level amplitude}
\label{chap: tree}

In this section we will first investigate in detail the behaviour of harmonics scattering at tree level. In particular we will concern our-self with calculating the amplitude corresponding to
\begin{align}
    \langle a(p_4,\Omega_4)a(p_3,\Omega_3)a^{\dagger}(p_2,\Omega_2)a^{\dagger}(p_1,\Omega_1)\rangle,
\end{align}
where the operators are to be understood as particles moving with lightcone momenta $p_i$ inserted at a specific angle $\Omega_i$ on the sphere. These angles correspond to positions of insertion; the angular momentum of these states is undefined. All of the external momenta now simply obey the original asymptotic massless condition $p_i^2=0$. This section will focus on the calculation of the tree-level amplitude only, and understanding its behaviour and kinematics. The next section performs the perturbatively exact eikonal resummation.

\subsection{Resummation over partial waves}
The amplitude above depending on angle, can be constructed from the partial waves one by resumming over spherical harmonics appropriately:
\begin{align}
   a(p_i,\Omega_i)    &=\sum\limits_{\ell_i m_i}a_{\ell_i m_i}(p^{\ell_i m_i}_i)Y_{\ell_i m_i}(\Omega_i),
\end{align}
so we will focus on calculating $\langle a_{\ell_4 m_4}(p^{\ell_4 m_4}_4)a_{\ell_3 m_3}(p^{\ell_3 m_3}_3)a^{\dagger}_{\ell_2 m_2}(p^{\ell_2 m_2}_2)a^{\dagger}_{\ell_1 m_1}(p^{\ell_1 m_1}_1)\rangle$. Notice that these momenta are defined in order to satisfy the equation of motion. The original scalar fields are massless and so $p_i^2=0$. Thus we define
\begin{align}
    p_1=(p_{1x},0), \lgap p_2=(0,p_{2y}).
\end{align}
For the partial wave momenta this then implies
\begin{align}
    p^{\ell m}_1=\left(p_{1x},\frac{\mu^2\lambda_1}{2p_{1x}}\right), \lgap p^{\ell m}_2=\left(\frac{\mu^2\lambda_2}{2p_{2y}},p_{2y}\right).
\end{align}
This will be the definition used throughout the calculation. For brevity throughout the calculation we will denote $p^{\ell_i m_i}_i=\bar{p}_i$ since the $\ell_i m_i$ are untouched until the end. $p^{\ell m}_3$ and $p^{\ell m}_4$ are defined analogously.

\subsubsection{Kinematics}
It is interesting to first investigate the possible kinematics if all scalar particles have different masses $p_i^2=-m_i^2=-\mu^2\lambda_i$. First we define momentum exchange as
\begin{align}
    q=p_1\epsilon_1+p_2\epsilon_2,
\end{align}
where $\epsilon_1,\epsilon_2$ are coefficients to be determined. Surprisingly, demanding all particles to be on-shell gives only two exact solutions for $\epsilon_1,\epsilon_2$ because the phase-space is two-dimensional. The full expression for the solution for $\epsilon_1,\epsilon_2$ is very large and of little importance, for brevity we write these solutions in the limit of large $s$ as
\begin{align}
    q^-&=\frac{m_2^2-m_4^2}{s}p_1+\frac{m_3^2-m_1^2}{s}p_2,\\
    q^+&=p_2-p_1+\frac{m_3^2-m_2^2}{s}p_1+\frac{m_1^2-m_4^2}{s}p_2.
\end{align}
Clearly the second case corresponds to the case of large momentum transfer, and so for this paper we will always consider the small momentum transfer $q=q^-$ in the eikonal limit. We may use this form of $q$ explicitly if desired: the phase-space restricts $q$ to be exactly equal.\\
A useful notation is as follows:
\begin{align}
    q&=\frac{\mu^2}{s}\biggr((\lambda_2-\lambda_4)p_1+(\lambda_1-\lambda_3)p_2\biggr),\\
    \label{eq:momentum exchange}
    t&=-\frac{\mu^4}{4s}(\lambda_1-\lambda_3)(\lambda_2-\lambda_4).
\end{align}
We will return to the explicit form of the exchange later. Again because the momentum space is two-dimensional, only two solutions were available, of which one leading. This reinforces the expectation that the eikonal limit gives the physical results we are interested in for black holes; there is only one other solution possible which is immediately strongly sub-leading.

\subsection{Scattering process: Gravity}
\begin{figure}[h!]
    \centering
    \includegraphics[width=0.35\textwidth]{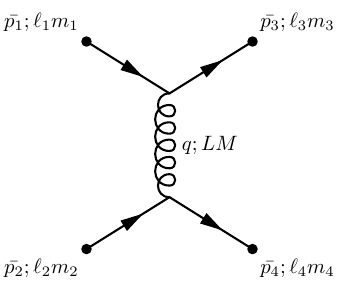}
    \caption{The leading order tree-level diagram in the small $t$ limit. All particles carry different $\ell m$, and in principle the graviton interaction may carry any $L M$ that satisfies angular momentum conservation.}
    \label{fig:resum tree level}
\end{figure}
We will calculate the result for gravitational interactions in the large $s$-limit. The $t-$channel tree level diagram is given in \figref{fig:resum tree level}. The other two possible configurations give sub-leading contributions, so that the leading order contribution is given by:
\begin{align}
    i\mathcal{M}_T=\sum\limits_{LM}\frac{i\gamma^2s^2}{\mu^2\lambda_{\ell}}CL[\ell_1m_1,\ell_3,m_3;LM]CL[\ell_2m_2,\ell_4,m_4;LM].
\end{align}
This amplitude was calculated for the complex scalar denoted in the diagram, but the result is identical for the real scalar, and for complex particles or antiparticles. The amplitude is summed over all possible internal graviton angular momenta that do not violate conservation of momentum, utilizing the $CL$-functions. Using the definition of the coefficients an alternative way of writing this is as
\begin{align}
    i\mathcal{M}_T=\int\ed\Omega\ed\bar{\Omega}\sum\limits_{LM}\frac{i\gamma^2s^2}{\mu^2\lambda_{L}}Y_{\ell_1m_1}(\Omega)Y_{\ell_3m_3}(\Omega)Y_{LM}(\Omega)Y_{\ell_2m_2}(\bar{\Omega})Y_{\ell_4m_4}(\bar{\Omega})Y_{LM}(\bar{\Omega}).
\end{align}
The four harmonics that depend on the external particles can now be isolated and grouped into an initial value contribution, thus called
\begin{align}
       Y_{IV}&(\Omega,\bar{\Omega}) := Y_{\ell_{1} m_{1}}(\Omega) Y_{\ell_{3}m_{3} }(\Omega) Y_{\ell_{2} m_{2}}(\bar{\Omega})Y_{\ell_4m_4}(\bar{\Omega}).
\end{align}
Then finally we can write the tree-level amplitude more compactly as
\begin{align}
    i\mathcal{M}_T=2s\frac{i\gamma^2s}{2\mu^2}\int\ed\Omega\ed\bar{\Omega}Y_{IV}(\Omega,\bar{\Omega})G_1(\Omega,\bar{\Omega}),
\end{align}
where the Green's function $G_a(\Omega,\bar{\Omega})$ is defined by  
\begin{align}
    G_a(\Omega,\bar{\Omega})=\sum\limits_{LM}\frac{1}{L^2+L+a}Y_{L M}(\Omega)Y_{L M}(\bar{\Omega}).
\end{align}
The amplitude has been written in a suggestive way: The factor of $2s$ has been kept separate since it corresponds to the phase space volume of the in-state. The Green's function is well-defined for all values of $a$ except $a=0$, which we will treat separately in the next section.\\
\\
By extension of our knowledge of the eikonal summation, we would expect the eikonal amplitude to be given by the exponent of the tree level amplitude, with the phase space measure subtracted:
\begin{align}
    i\mathcal{M}_{eik}\stackrel{?}{=}2s\text{Exp}\left(\frac{i\gamma^2s}{2\mu^2}\int\ed\Omega\ed\bar{\Omega}Y_{IV}(\Omega,\bar{\Omega})G_1(\Omega,\bar{\Omega})\right).
\end{align}
This would match well with the amplitude of 't Hooft \cite{tHooft1996}, however we shall see that doing the entire calculation correctly does not place the integrals over angles in the exponent. Instead the calculation by 't Hooft does not to correspond to a $2\to 2$ eikonal amplitude, but instead many particles interacting in a very specific way. This will be discussed in \secref{chap: MN to MN}.

\begin{figure}[h!]
    \centering
    \includegraphics[width=0.6\textwidth]{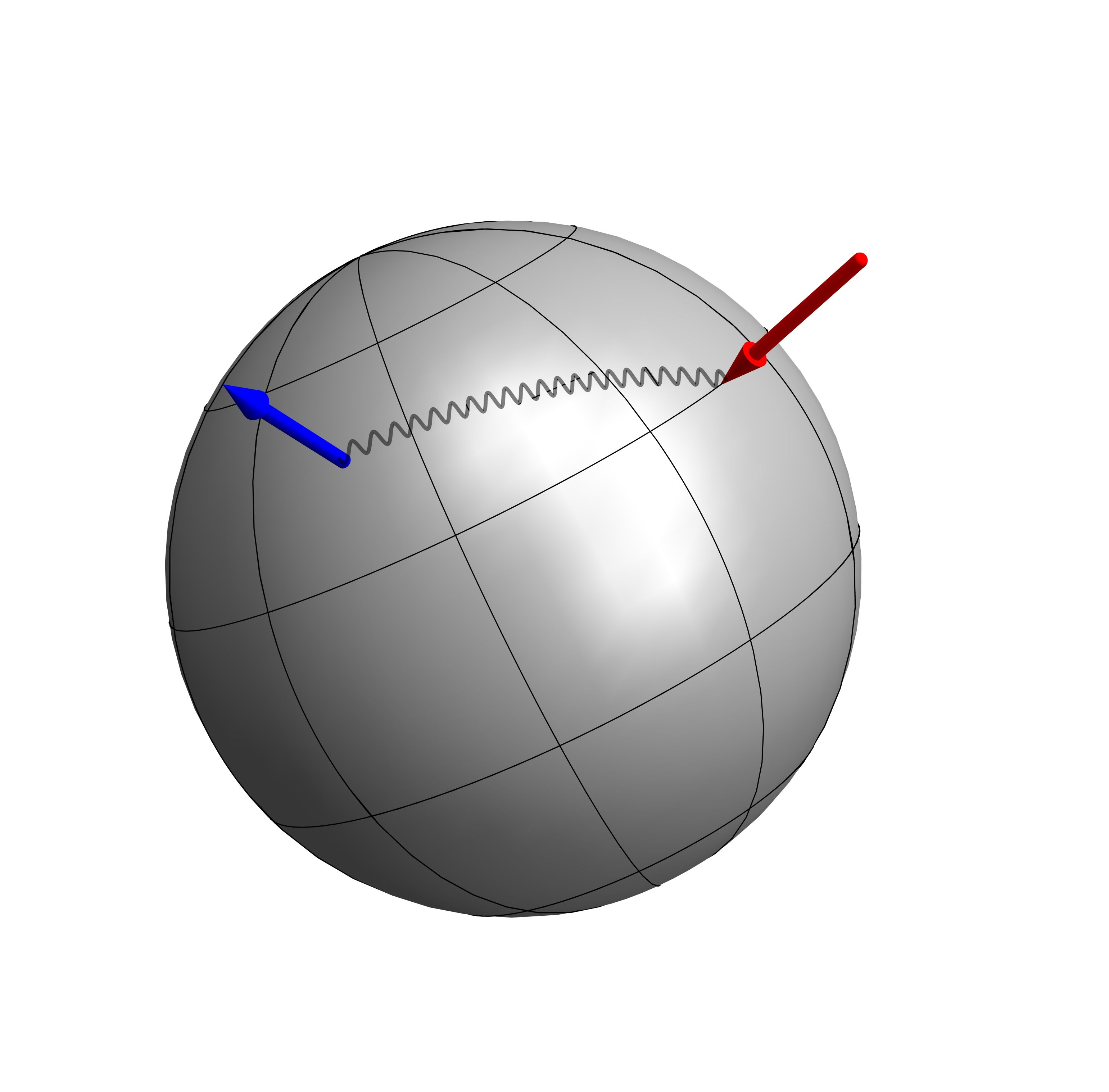}
    \caption{An illustration of tree-level scattering on the horizon in the angular basis. One particle enters the black hole, and one exits the black hole. They interact with a single gauge field that lives on the horizon.}
    \label{fig:tree level scattering}
\end{figure}

We may resum all harmonics to write down the amplitude in terms of angles instead. A graphical illustration of the interpretation in this amplitude has been given in \figref{fig:tree level scattering}. Since the sum of two spherical harmonics quickly gives a delta function we find
\begin{align}
    i\mathcal{M}=2s~ \delta^{(2)}(\Omega_1-\Omega_3)\delta^{(2)}(\Omega_2-\Omega_4)\frac{i\gamma^2s}{2\pi\mu^2} G_1(\Omega_1,\Omega_2).
\end{align}
So two effects can clearly be observed: First and foremost, the in-and-out-particles must share the same angles. This conservation law follows from the eikonal limit: If two particles have small momentum exchange, their paths will hardly deviate. This is projected onto a delta function: The particles on the top line and bottom line keep moving in the same direction. More importantly, compared to \cite{ShortPaper,LongPaper}, a measure of transverse separation is present $G_a(\Omega_1,\Omega_2)$. On closer inspection we can identify it to be the Green's function of the spherical Laplacian:
\begin{align}
    (-\Delta_{\Omega}+a)G_a(\Omega,\bar{\Omega})=\delta^{(2)}(\Omega,\bar{\Omega}).
\end{align}
In \secref{sec:greens func} we shall look at this function more closely for arbitrary $a$, its consequences will be discussed in \secref{chap: eikonal}.

\subsection{Scattering process: Electromagnetism}
\begin{figure}[h!]
    \centering
    \includegraphics[width=0.35\textwidth]{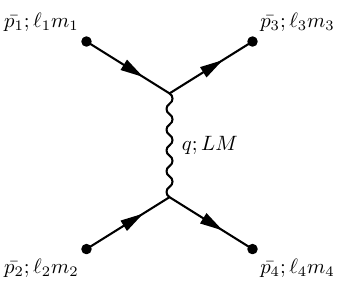}
    \caption{The leading order tree-level diagram for scalar electrodynamics. The only difference compared to the graviton interaction is the different internal propagator, with corresponding vertices.}
    \label{fig:resum tree level gauge}
\end{figure}
The case of electromagnetism is largely similar. The diagram is given in \figref{fig:resum tree level gauge}. The leading contribution is given by
\begin{align}
    i\mathcal{M}_T=-2s\frac{iQ^2}{2\mu^2}\int\ed\Omega\ed\bar{\Omega}Y_{IV}(\Omega,\bar{\Omega})G_0(\Omega,\bar{\Omega}),
\end{align}
which contains the problematic Green's function, because for $a=0$ the $\ell=0$ mode diverges. The obvious modification is to exclude the $\ell=0$ mode:
\begin{align}
    G_0(\Omega,\bar{\Omega})&=\sum\limits_{L>0}\frac{1}{L^2+L}Y_{L M}(\Omega)Y_{L M}(\bar{\Omega}).
\end{align}
However in the field theory this mode was present and would lead to an obvious divergence. It appears that for $\ell=0$ we run into the familiar infrared divergence for massless particles, that was avoided for the graviton. Of course this would be regulated by a term of the form $\frac{1}{q^2}$, however then we are ignoring the fact would likely still be a contribution from sub-leading horizon terms $\mathcal{O}(x^2)$ that contribute larger than $q^2$ to the mass. Instead we resort to a different solution.\\
\\
This effect on the Green's function was also observed by 't Hooft in \cite{tHooft1996}. The $\ell=0$ mode contributes to an overall net charge present in the electromagnetic interaction, and the straightforward resolution is to add a cancelling charge to the defining equation
\begin{align}
 -\Delta_{\Omega}G_0(\Omega,\bar{\Omega})=\delta^{(2)}(\Omega,\bar{\Omega})-\frac{1}{4\pi}.
\end{align}
A motivation on why this is no problem to do, is given in the next section.
\begin{figure}[h!]
    \centering
    \includegraphics[width=0.35\textwidth]{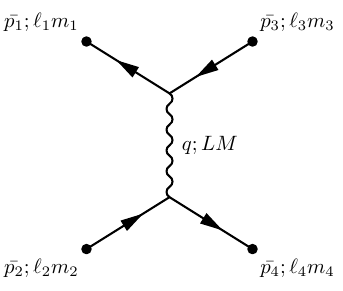}
    \caption{For electrodynamics the charge arrows may go in different directions, so long as overall charge is conserved. This diagram is another possible tree-level diagram. Note that for gravity in principle we could also consider these diagrams, but the result is trivially identical.}
    \label{fig:resum tree level gauge oppo}
\end{figure}
First for electromagnetism we must also consider the diagram with the charge flow for one particle in opposite direction, as shown in \figref{fig:resum tree level gauge oppo}. The resulting amplitude for this diagram is identical up to a sign
\begin{align}
    i\mathcal{M}_T=\sum\limits_{LM}\frac{iQ^2 s}{\mu^2(\lambda-1)}CL[\ell_1m_1,\ell_3,m_3;LM]CL[\ell_2m_2,\ell_4,m_4;LM],
\end{align}
which is directly explained by the fact that the sign of the momentum in the vertex Feynman rule is linked to the direction of charge. This reasoning will also extend to loops: Reversing the charge arrow just adds a factor of $-1$ for each vertex. Of course we can also draw the diagram with both arrows in opposite direction, but then the signs will become positive again. Using the notation $Q_{\text{in}}=\pm Q,Q_{\text{out}}=\pm  Q$ to account for this difference of charge sign for the respective particles, the four possible diagrams can be summed up as 
\begin{align}
    i\mathcal{M}_T=-\sum\limits_{LM}\frac{iQ_{\text{in}}Q_{\text{out}} s}{\mu^2(\lambda-1)}CL[\ell_1m_1,\ell_3,m_3;LM]CL[\ell_2m_2,\ell_4,m_4;LM].
\end{align}

\subsection{The Green's function \texorpdfstring{$G_a$}{Ga}}
\label{sec:greens func}

\begin{figure}[h!]
\centering
\includegraphics[width=0.6\textwidth]{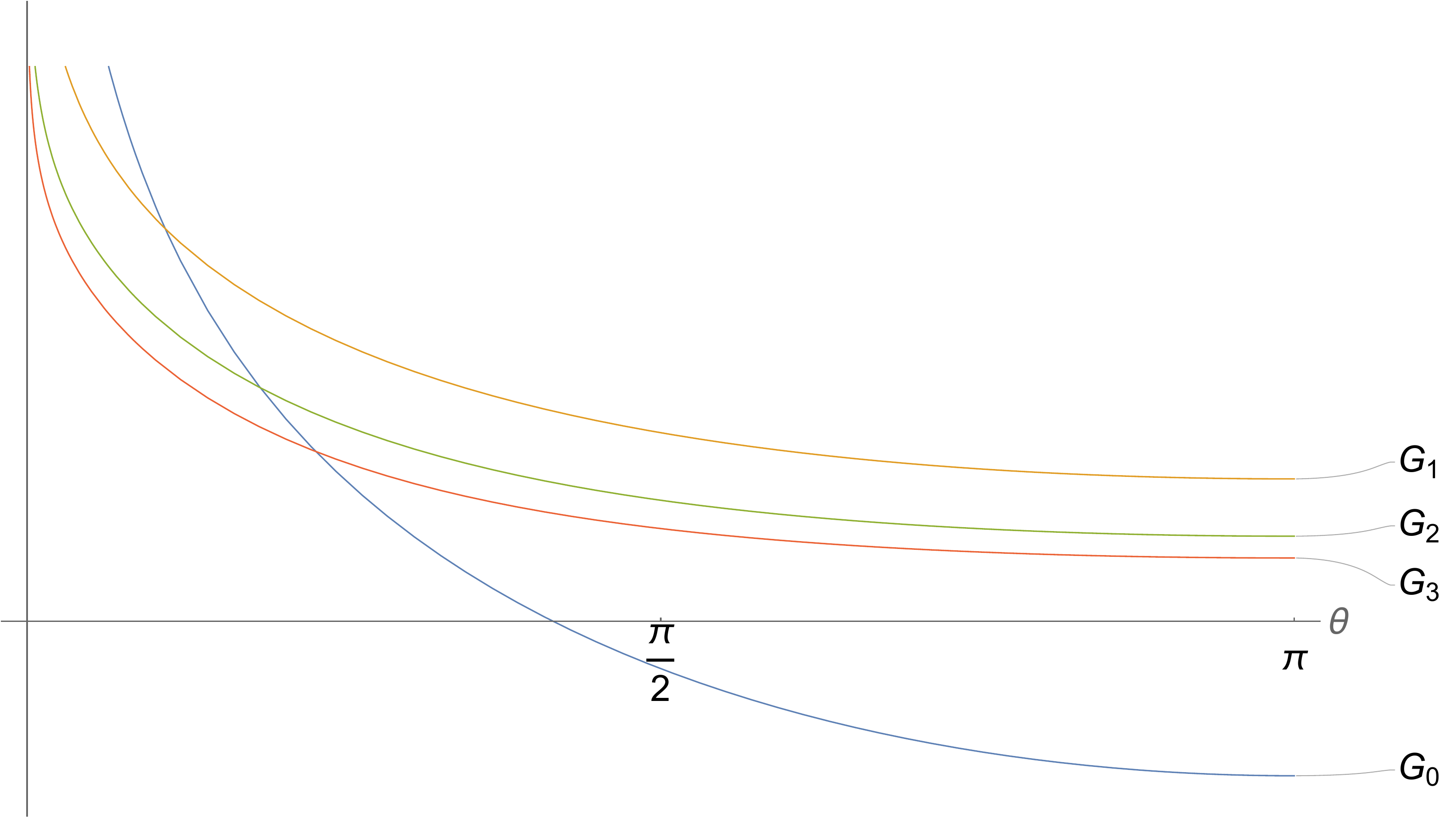}
\caption{\label{fig:greensfunc} A sketch of the Green's function for different values of $a$.}
\end{figure}

In this section we will calculate the general Green's function determined by
\begin{align}
    (-\Delta_{\Omega}+a)G_a(\Omega,\bar{\Omega})=\delta^{(2)}(\Omega,\bar{\Omega}).
\end{align}
The Green's function of the spherical Laplacian is generally not unique, namely it can be changed by homogeneous solutions. To investigate this for arbitrary $a$ the homogeneous equation in spherical harmonics is given by:
\begin{align}
    (\ell^2+\ell+a)G_a^{\ell m}=0.
\end{align}
There is only a non-trivial solution if $\ell^2+\ell+a=0$. This means that for any $a>0$ there is no homogeneous solution, for the special case $a=0$ of the spherical Laplacian these are constants. In the case where $a$ is a negative integer, there are other possible solutions, but in general we will keep $a$ positive. Thus we may conclude that any Green's function $G_a(\Omega,\bar{\Omega})$ for $a>0$ is unique, whereas $G_0(\Omega,\bar{\Omega})$ is not unique, but may be changed by a constant. This constant may be anything, in particular we can subtract the problematic $\ell=0$-piece $\lim\limits_{a\to 0}\frac{1}{a}Y_{00}^2=\frac{1}{4\pi}\lim\limits_{a\to 0} \frac{1}{a}$ in a regulated fashion. This is analogous to subtracting the $-\frac{1}{4\pi}$ in the equation as done before, and shows that this corresponds to a specific choice for Green's function, determined by a zero-net charge boundary condition.  \\
\\
An explicit form of the Green's function can be found in terms of an Appel hypergeometric function
\begin{align}
    G_a(\cos\theta)&=\frac{1}{4\pi a}\text{Re}\biggr[(\tfrac{1}{2}+i\alpha)F_1\left(\tfrac{1}{2}-i\alpha,\tfrac{1}{2},\tfrac{1}{2},\tfrac{3}{2}-i\alpha,e^{-i\theta},e^{i\theta}\right)\biggr],
\end{align}
where $\alpha=\sqrt{a-\tfrac{1}{4}}$.This equation is only valid when $a^2>1/4$ in which case $\alpha>0$ and real. The extension to $a^2<1/4$ is however directly obtained by setting $\alpha\to i\beta$. The $a=0$ pole is still present, since we did not exclude the $\ell=0$ mode here. Truncating the $\ell=0$ result from the function above instead gives a well-defined finite result for $G_0$. An explicit solution for $a=0$ is known, and depends on the boundary condition at $G_0(-1)$. Our value is uniquely determined by the spherical harmonic summation excluding $\ell=0$, giving
\begin{align}
    G_0(\cos\theta)=-\frac{1}{4\pi}\log(\tfrac{1}{2}-\tfrac{1}{2}\cos\theta)-\frac{1}{4\pi},
\end{align}
which numerically can be checked to be identical to the Appel hypergeometric function definition upon subtracting $\frac{1}{4\pi a}$ and taking the $a\to 0$ limit. Graphically the Green's function is shown in \figref{fig:greensfunc}. 
Towards $\theta=0$ there is an obvious divergence, indicating that Planckian effects must be taken into account. It is tempting to solve the Green's function in this limit by approximating the sphere near the north pole as a 2D plane, however this appears to give a mismatch. The angle-dependent part is an exact match, but the constants do not match when compared numerically. Apparently the metric-approximated method neglects certain factors coming from the global normalization.\\
Numerically we find a solution for $G_a(\cos\theta)$ for up to order $\mathcal{O}(\cos\theta-1)$ given by
\begin{align}
    G_a(\cos\theta)&\approx-\frac{1}{4\pi}\log\left(\frac{1}{2}-\frac{1}{2}\cos\theta\right)-\frac{\gamma_E}{4\pi}-\frac{1}{2\pi}\text{Re}[\psi(\tfrac{1}{2}-i\alpha)],\nonumber\\
    &\approx G_0(\cos\theta)-\frac{\gamma_E-1}{4\pi}-\frac{1}{2\pi}\text{Re}[\psi(\tfrac{1}{2}-i\alpha)].
\end{align}
Thus clearly the leading behaviour is divergent, specifically logarithmically. Here $\gamma_E$ is the Euler-Mascheroni constant. We can identify the $\theta-$dependent part as $G_0(\cos\theta)$.

\clearpage

\section{Eikonal resummation}
\label{chap: eikonal}

\begin{figure}[h!]
    \centering
\includegraphics[width=\textwidth]{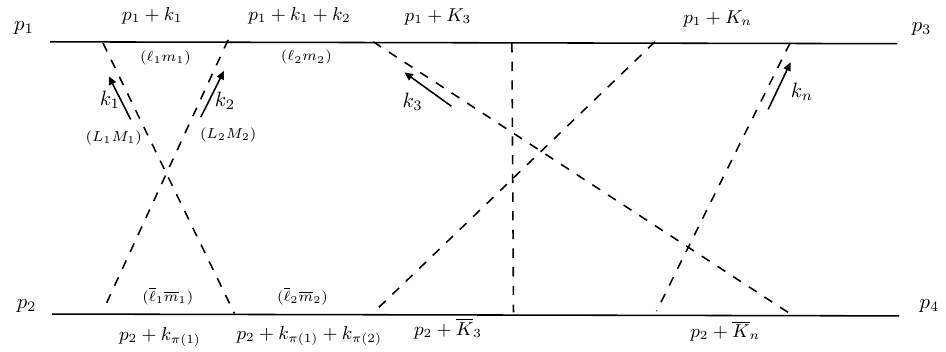}
\caption{\label{fig: general ladder} The ladder-diagram for a given order $n$, drawn schematically for visibility. The dashed lines are either graviton or photon exchanges. The scalars are then real or complex. All used parametrizations are added.}
\end{figure}

In this section we generalize the tree-level diagram to an the eikonal summation of ladder diagrams. This is same eikonal approximation that was done in \cite{LongPaper} but extended to include $\ell m's$ and particle effective masses. This will increase the need for careful bookkeeping of all new factors. The resulting S-matrix is given in \eqref{eq:resummed smatrix}. The notation we use for a typical ladder diagram is given in \figref{fig: general ladder}.

This is a shortened version of what was done by \cite{LevySucher}, where instead we choose to always parametrize the order of the graviton legs by the order at which they hit the $\bar{p}_1$ row. The way they hit the $\bar{p}_2$ row is described by any permutation $\pi$ such that to preserve generality we must sum over all possible $\pi$. So far the effect of momentum conservation still has to be applied. The following definitions will be useful:
\begin{align}
    K_i&=\sum\limits_{j=1}^i k_j, & \bar{K}_i&=\sum\limits_{j=1}^i k_{\pi(j)},\\
    I_i&=\frac{1}{(\bar{p}_1+K_i)^2+\mu^2\lambda_{\ell_i}-i\epsilon}, & \bar{I}_i&=\frac{1}{(\bar{p}_2-\bar{K}_i)^2+\mu^2\lambda_{\bar{\ell}_{\pi(i)}}-i\epsilon},
\end{align}
then the loop amplitude can in total be written as
\begin{align}
    i\mathcal{M}_{n-1}&= i^{2n}(-i)^{n-1}(-i)^{n-1}(i)^n\gamma^{2n}\sum\limits_{\pi}\prod\limits_{i=1}^{n-1}\left(\sum\limits_{\ell_i\ell_i}\sum\limits_{\bar{\ell}_i\bar{\ell}_i}\right)\prod\limits_{i=1}^n\left(\sum\limits_{L_iM_i}\int\frac{\ed ^2k_i}{(2\pi^2)}\right)\nonumber\\
    &\times(2\pi)^2\delta^{(2)}(q-K_{n})\times \prod\limits_{i=1}^{n-1}\left(I_i\bar{I}_i\right)\prod\limits_{i=1}^{n}(\bar{p}_1+K_{i-1})^{a_i}(\bar{p}_1+K_{i})^{b_i}\nonumber\\
    &\times \mathcal{P}^{L_iM_i}_{a_{i-1}b_{i}c_{\pi(i-1)}d_{\pi(i)}}(k_i)(\bar{p}_2-\bar{K}_{i-1})^{c_{\pi(i-1)}}(\bar{p}_2-\bar{K}_{i})^{d_{\pi(i)}}\nonumber\\
    &\times \prod\limits_{i=1}^n CL(\ell_{i-1} m_{i-1},\ell_{i}m_i,L_iM_i)CL(\bar{\ell}_{i-1}\bar{m}_{i-1},\bar{\ell}_{i}\bar{m}_{i},L_{\pi(i)}M_{\pi(i)}).
\end{align}
This is a large but exact expression. We now seek to simplify this as much as possible. The calculation for gravitational interaction is outlined in the next section, the result for scalar electrodynamics is added in the end. The bulk of the calculation is outlined in \appref{app: calculations for resum}, here we provide a short summary.

\subsection{Black hole eikonal resummation}
The first step is approximate the matter propagators  for small $K$ similar to \cite{LevySucher}
\begin{align}
    \frac{1}{(\bar{p}_1+K_i)^2+\mu^2\lambda_{\ell_i}-i\epsilon}\approx \frac{1}{2\bar{p}_1\cdot K_i+\mu^2(\lambda_{\ell_i}-\lambda_1)-i\epsilon}.
\end{align}
The presence of the masses is a problem: The combinatorics for the eikonal ladder do not simplify \textit{and} the harmonics can not be resummed over all $\ell m$. We can not neglect the mass term because $K\sim \frac{\mu}{\sqrt{s}}$ as we saw in \secref{chap: tree}. The trick to proceed is to redefine the loop momenta $k_i\to \tilde{k}_i+b_i$, in order to exactly remove the mass terms by incorporating the mass exchange at each vertex in the graviton momenta. Indeed making this choice gives to highest order in $s$ that
\begin{align}
    b_i&=\frac{\mu^2}{s}\left(-p_2 (\lambda_{\bar{\ell}_{\pi(i)}}-\lambda_{\bar{\ell}_{\pi(i-1)}})+p_2 (\lambda_{\ell_i}-\lambda_{\ell_{i-1}})\right),
\end{align}
which is precisely of the form of $q_-$ in \secref{chap: tree}. Defining the summed up momentum as $B_i=\sum\limits_{j=1}^i b_j$ we can also see
\begin{align}
    B_n=\frac{\mu^2}{s}\left(p_1 (\lambda_{2}-\lambda_{4})+p_2 (\lambda_{3}-\lambda_{1})\right)=q_-
\end{align}
so in total indeed we shift the graviton momenta precisely by the momentum exchange, incorporating it exactly while at the same time getting rid of the mass terms. This means that the extension to massive scalars is trivial. So far we did not yet consider the fact that the $\lambda$'s are summed over, meaning that in principle we are working with arbitrary masses. Then we may have given the scalars an additional $4D$ mass term $m_i^2+\mu^2\lambda_i$ and this would not change the calculation. Of course the $4D$ mass does change the interpretation of the external states, but the eikonal scattering behaviour is unmodified (so long as $s\gg m_i^2$ is satisfied for all $m_i^2$).

\subsubsection{Small momentum exchange approximation} 
\begin{figure}[h!]
    \centering
    \includegraphics[width=0.6\textwidth]{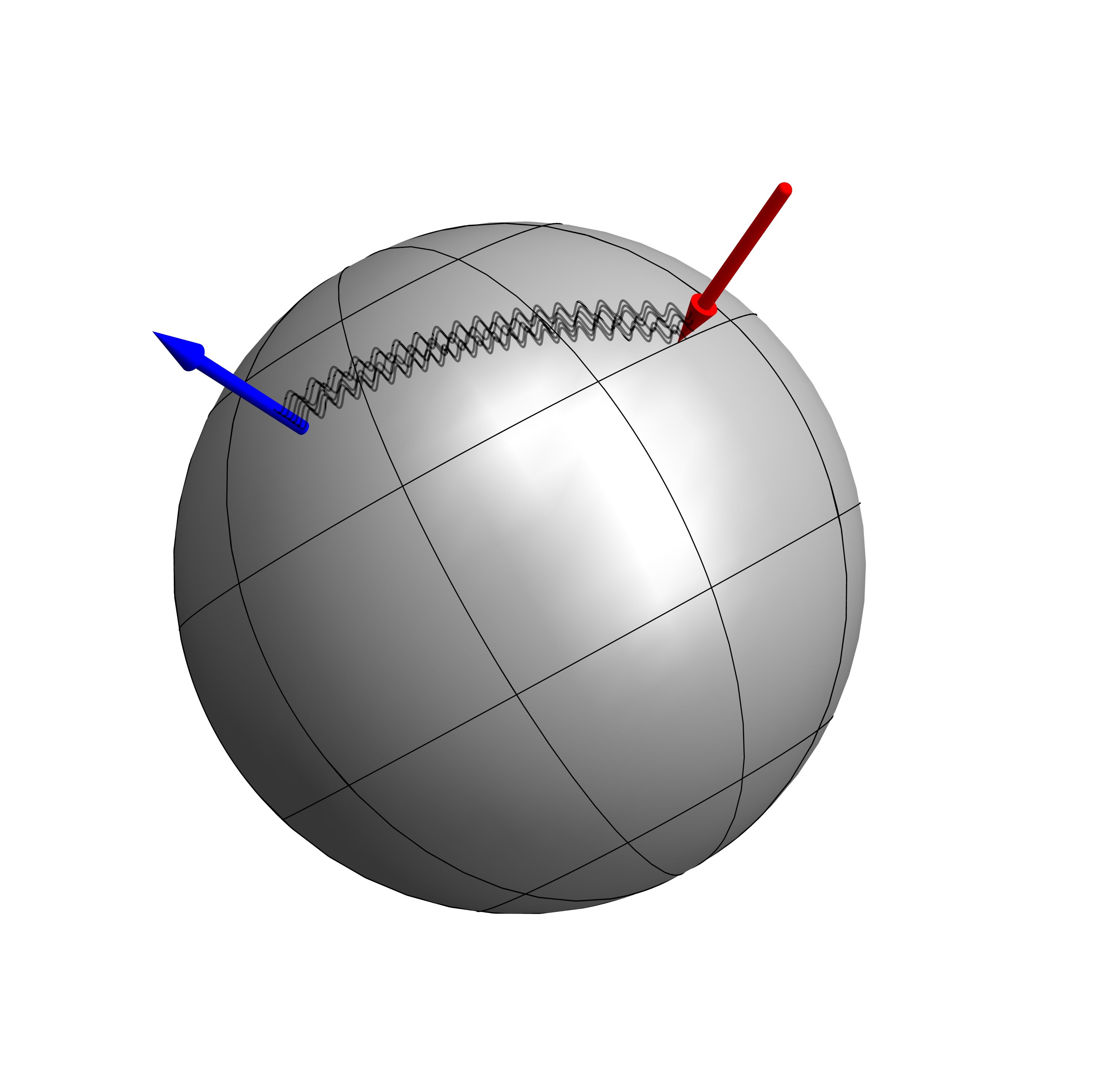}
    \caption{The eikonal scattering between two particles on the horizon. The setup is the same as for tree level, but now the two particles exchange infinitely many soft gauge interactions. In principle all of these interactions live on the horizon but at different times, for this image they are drawn at different radii to show the ladder clearly.}
    \label{fig:eikonal scattering}
\end{figure}

The next step is to approximate the internal momenta to be small, also internally. This means in practice that we approximate for order of magnitudes that $K_i\sim q$ at most. Together with the momentum transformation we then find
\begin{align}
    \frac{1}{(\bar{p}_1+K_i)^2+\mu^2\lambda_{\ell_i}-i\epsilon}&\approx \frac{1}{2p_1\cdot \tilde{K}_i-i\epsilon}\equiv I_i^{eik}\\
    \frac{1}{(\bar{p}_2-\bar{K}_i)^2+\mu^2\lambda_{\bar{\ell}_{\pi(i)}}-i\epsilon}&\approx \frac{1}{-2p_2\cdot \tilde{\bar{K}}_i-i\epsilon}\equiv \bar{I}_i^{eik}.
\end{align}
which is of a similar form as \cite{LevySucher}. The propagators have been simplified as much as possible, and contain no more $\ell_i m_i$ dependence. \\
Next we apply this to the vertex couplings. The vertex couplings consist of two parts: the four momenta of the type $(\bar{p}\pm K\pm B)$ and the graviton propagator. The momenta are polynomial, and so we can easily approximate
\begin{align}
    (\bar{p}_1+K_{i-1}+B_{i-1})^{a_{i-1}}&(\bar{p}_1+K_{i}+B_i)^{b_i}\mathcal{P}^{L_iM_i}_{a_{i-1}b_{i}c_{\pi(i-1)}d_{\pi(i)}}(\tilde{k}_i-b_i)\\
    &\times (\bar{p}_2-\bar{K}_{i-1}-\bar{B}_{i-1})^{c_{\pi(i-1)}}(\bar{p}_2-\bar{K}_{i}-\bar{B}_i)^{d_{\pi(i)}}\nonumber\\
    \approx p_1^{a}p_1^{b}\mathcal{P}^{L_iM_i}_{abcd}(\tilde{k}_i-b_i)p_2^{c}p_2^{d},\nonumber
\end{align}
since $p_1,p_2\sim\sqrt{s}$ whereas $K,B\sim q \sim \frac{1}{\sqrt{s}}$. Here we simplified the indices since they are summation dummy variables, and we can remove any $_i$ or $_{\pi(i)}$ subscripts because the rest of the $i$ dependence drops out. Under the same arguments we changed $\bar{p}_1\to p_1$ removing the mass contribution. Of course technically the terms above that are simplified are vectors, and so some caution is needed, however, when writing out all components into one big scalar, one still finds that the approximation above gives the leading order result, and for brevity we only give the heuristic version here. The only remaining problem is the presence of $b_i$ in the graviton propagator: This still contains dependence on $\ell_i m_i$ and since $b_i$ is of the same order as $\tilde{k}_i$ we cannot neglect it here. Here the specific form of the graviton propagator comes to the rescue: If we write out explicitly the remaining term using our propagators, we find in both gauges to leading order that
\begin{align}
    p_1^{a}p_1^{b}\mathcal{P}^{L_iM_i}_{abcd}(k_i-b_i)p_2^{c}p_2^{d}&\approx-\frac{s^2}{\mu^2\lambda_{L_i}},
\end{align}
where for the eikonal gauge the identity holds exactly. The amplitude becomes:
\begin{align}
    i\mathcal{M}_{n-1}&= -(i\gamma^2)^n\sum\limits_{\pi}\prod\limits_{i=1}^{n-1}\left(\sum\limits_{\ell_i\ell_i}\sum\limits_{\bar{\ell}_i\bar{\ell}_i}\right)\prod\limits_{i=1}^n\left(\sum\limits_{L_iM_i}\int\frac{\ed ^2k_i}{(2\pi^2)}\right)(2\pi)^2\delta^{(2)}(K_{n})\nonumber\\
    &\times \prod\limits_{i=1}^{n-1}\left(I^{eik}_i\bar{I}^{eik}_i\right)\prod\limits_{i=1}^{n}\left(-\frac{s^2}{\mu^2\lambda_{L_i}}\right)\label{eq:harmonicspart}\\
    &\times \prod\limits_{i=1}^n CL(\ell_{i-1} m_{i-1},\ell_{i}m_i,L_iM_i)CL(\bar{\ell}_{i-1}\bar{m}_{i-1},\bar{\ell}_{i}\bar{m}_{i},L_{\pi(i)}M_{\pi(i)}), \nonumber
\end{align}
where we restored some indices because we still need to keep track of all $L_i,M_i$. However, thanks to the replacement of the matter propagators by eikonal ones, the removal of $q_n$ in the delta function, and the small momentum exchange approximation, there are no more $b_i$ present anywhere and thus all dependence on $\ell_i m_i$ has been factorized into the $CL$ functions only. Because they are only present in the $CL$ functions a resummation is now possible. The exact steps are outlined in \appref{app: calculations for resum}, and we find that the amplitude is given by
\begin{align}
    i\mathcal{M}_{n-1}&= 2s\int\ed\Omega\ed\bar{\Omega}Y_{IV}(\Omega,\bar{\Omega})\frac{1}{n!}\left( \frac{i\gamma^2s}{2\mu^2}G_1(\Omega,\bar{\Omega})\right)^n.
\end{align}
Since this is the form of a normal exponential, we can sum over all loops from $n=1$ to $n=\infty$ giving
\begin{align}
    i\mathcal{M}&=2s\int\ed\Omega\ed\bar{\Omega}Y_{IV}(\Omega,\bar{\Omega}) \left(e^{i\chi(\Omega,\bar{\Omega})}-1\right)\\
    &\chi(\Omega,\bar{\Omega})=\frac{i\gamma^2s}{2\mu^2}G_1(\Omega,\bar{\Omega})-iq_{\text{in}}q_{\text{out}}G_0(\Omega_1,\Omega_2).
\end{align}
This gives in principle the full eikonal amplitude including all non-trivial couplings, where the result for electromagnetism has been added. Notably this includes accounting for the presence of mass terms and the presence of spherical harmonics couplings. Notice as expected that the same factor of $2s$ is still in front, because the phase space measure for our $2\to 2$ scattering problem did not change with respect to the Minimal coupling calculation. An illustration of this scattering has been added in \figref{fig:eikonal scattering}.

Remarkably the resummation is to leading order in $s$ still possible. The result is both quantitatively and qualitatively different from that in \cite{LongPaper}. The scattering matrix is still described by a complex exponent, however there is now a dynamical term present depending on position, with integrals in front. This structure is identical to results in flat space, AdS and celestial CFT's \cite{KabatOrtiz,RaclariuEikonalCFT,Cornalba2007two}, and we shall show in \secref{sec:flat space comp} that in flat space we find an exact agreement with \cite{KabatOrtiz}. In generality the final result $\chi$ depends on the (relative) transverse positions measured by the transverse Green's function, which are still to be integrated over together with a set of eigenfunctions. In our case the transverse positions are the angles, expressed in a spherical basis. The Green's function is $G_1$ and the set of eigenfunctions is $Y_{IV}$. This similarity is a remarkable result, although on the other hand this similarity from summation over all $\ell m$ was therefore to be somewhat expected. However, this has never been done in a harmonics basis, and the mathematical structure at the foundation was for that reason quite different.\\ 
It is tempting to think that this amplitude contradicts the results of 't Hooft \cite{tHooft1996}, where the integrals are be inside the exponent. The reason for this difference is the fact that we are still considering $2\to 2$ scattering, and as we show in the next section we need to generalize the amplitude to arbitrarily many particles. In addition we remark that there is no nice limit for the Green's functions $G_{1,0}$ where the amplitude reduces to the results of \cite{LongPaper}.\\
\\
Next we resum over the external $\ell m$ as well, giving
\begin{align}
    S=2s \delta^{(2)}(\Omega_1-\Omega_3)\delta^{(2)}(\Omega_2-\Omega_4)e^{i\frac{i\gamma^2s}{2\mu^2}G_1(\Omega_1,\Omega_2)-iq_{\text{in}}q_{\text{out}}G_0(\Omega_1,\Omega_2)}.
\end{align}
So first of all we see that the in- and out-particles still must have the same angles, similar to tree level. This effect of the small momentum exchange persists through the summation over all loops: The particles on the top line and bottom line keep moving in the same direction.\\
For this reason their only interaction is the familiar phase factor. This phase factor depends on the angular separation between the two particles, and in this way the impact parameter enters into the scattering matrix. This different phase also automatically distinguishes different particles. Since the stress-energy tensor average the energy one might expect two scalars entering the black hole to lose information since they may not be distinguished any more, however this is fixed so long as the particles are at different locations.\\
Finally, we remark that similar to \cite{SQEDpaper} we may remove the $-1$ and $2s$ factor as follows. The canonical commutator for the scalar field is given by
\begin{align}
    [a(p,\Omega),a^{\dagger}(p',\Omega')]=(2\pi)(2p)\delta(p-p')\delta^{(2)}(\Omega-\Omega').
\end{align}
Thus the free field contribution where the scalars do not interact is given by
\begin{align}
    \mathds{1}=2s \delta^{(2)}(\Omega_1-\Omega_3)\delta^{(2)}(\Omega_2-\Omega_4).
\end{align}
The S-matrix may then nicely be written as
\begin{align}
\label{eq:resummed smatrix}
    S_{\text{combined}}=\mathds{1}\text{Exp}\left(\frac{i\kappa^2s}{2}G_1(\Omega_1,\Omega_2)-iq_{\text{in}}q_{\text{out}}G_0(\Omega_1,\Omega_2)\right),
\end{align}
giving the familiar explicit complex phase corresponding to the eikonal approximation. Note that this expression holds for both the real scalar field ($q_{\text{in}/\text{out}}=0$) and the complex one ($q_{\text{in}/\text{out}}=\pm q$).

\subsection{Flat space eikonal comparison}
\label{sec:flat space comp}
\begin{figure}[h!]
    \centering
    \includegraphics[width=\textwidth]{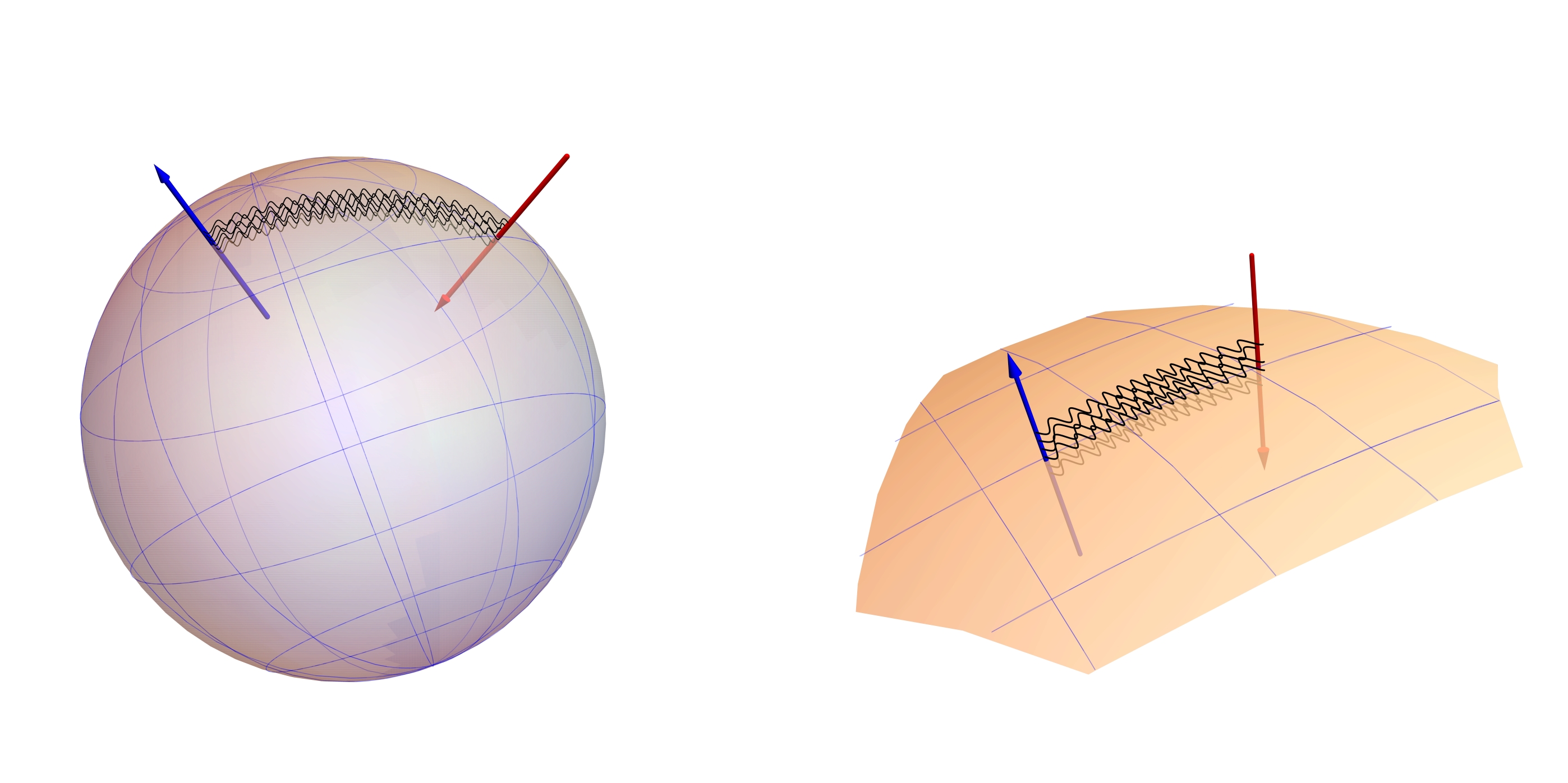}
    \caption{An impression of the scattering in flat space. The interaction takes place at a fixed radius $\rfs$ which holds no special meaning, so now as shown on the left the particles may simply move through the sphere. However when zooming in on a region close to an extremely large sphere $\rfs \to \infty$ the transverse space between the particles becomes flat again. Since in flat space there is translational invariance, one can always shift $\rfs$ to become large, making this a valid limit to take in general.}
    \label{fig:flat space zoom}
\end{figure}
It is worthwhile to try and extend the spherical harmonics basis to flat space as well. The eikonal calculation becomes identical to the one on the black hole since the diagrammatic combinatorics do not change, the only difference is to use our flat-space Feynman rules in \secref{sec: flat space analogue}. We will do this for the gravitational interaction, and compare with literature. The amplitude becomes
\begin{align}
    i\mathcal{M}=2s \delta^{(2)}(\Omega_1-\Omega_3)\delta^{(2)}(\Omega_2-\Omega_4)e^{i\chi_0 G_0(\Omega_1,\Omega_2)},
\end{align}
Using the hard-sphere impact-parameter relation
\begin{align}
    \tfrac{1}{2}-\tfrac{1}{2}\cos\theta=\tfrac{1}{2}\mu^2 b^2,
\end{align}
we write
\begin{align}
    G_{flat}(b)=-\frac{1}{4\pi}\log(\tfrac{e}{2}\mu^2 b^2).
\end{align}
Absorbing the numerical factors into $\bar{\mu}^2=\frac{e}{2}\mu^2$ we can write more compactly
\begin{align}
    G_{flat}(b)=-\frac{1}{2\pi}\log(\bar{\mu} b).
\end{align}
Inserting this into the amplitude gives that
\begin{align}
    i\mathcal{M}=2s \delta^{(2)}(\Omega_1-\Omega_3)\delta^{(2)}(\Omega_2-\Omega_4)e^{-2G i s\log(\bar{\mu} b) },
\end{align}
where the exponent coincides exactly with the result of \cite{KabatOrtiz,ACVScattering} upon using that $Ep=s/4$ \cite{ACVScattering}. In our case however the infrared regulator is provided automatically by $\mu^2$. We can also revert this back to the momentum space formalism with some care. First remark that the S-matrix is given by
\begin{align}
    S=(2\pi)^2 2s \delta^{(2)}(p_{tot,\parallel}) \delta^{(2)}(\Omega_1-\Omega_3)\delta^{(2)}(\Omega_2-\Omega_4)e^{-2iG s\log(\bar{\mu} b)}.
\end{align}
To move back to the original momentum space formulation we Fourier transform the transverse momenta:
\begin{align}
    S=(2\pi)^2 2s &\int\prod\limits_i\left(\ed\Omega_i e^{i p^i_{\perp}\cdot \rfs w^i(\Omega_i)}\right)\delta^{(2)}(p_{tot,\parallel}) \nonumber\\
    &\times \delta^{(2)}(\Omega_1-\Omega_3)\delta^{(2)}(\Omega_2-\Omega_4)e^{-2iG s\log(\bar{\mu} b(\Omega_1,\Omega_2))}.
\end{align}
The integration is with respect to the angles since we are still in spherical coordinates, where the Cartesian inner product in the orthogonal plane has been parametrized using the unit vector $w(\Omega)$ in a direction $\Omega$ with length $\rfs$.\\
Two of the integrations can immediately be removed by the two delta functions. Note that $s$ only contains the parallel momenta. This turns the equation above into
\begin{align}
    S=(2\pi)^2 2s \int \ed^2 \Omega_1 e^{i (p^1_{\perp}-p^3_{\perp})\cdot \rfs w(\Omega_1)}\int \ed^2 \Omega_2 e^{i (p^2_{\perp}-p^4_{\perp})\cdot \rfs w(\Omega_2)}\delta^{(2)}(p_{tot,\parallel}) e^{-2iG s\log(\bar{\mu} b_{12})},
\end{align}
where $b_{12}=\rfs(w(\Omega_1)-w(\Omega_2))$. To rewrite this we may combine the residual exponents as
\begin{align}
   &e^{i (p^1_{\perp}-p^3_{\perp})\cdot \rfs w(\Omega_1)}e^{i (p^2_{\perp}-p^4_{\perp})\cdot \rfs w(\Omega_2)} 
   =e^{i q_{13}\cdot b_{12}}e^{i q_{tot,\perp}\cdot \rfs  w(\Omega_2)}
\end{align}
algebraically, where $q_{13}=p_{\perp}^1-p_{\perp}^3,q_{tot,\perp}=p_{\perp}^1+p_{\perp}^2-p_{\perp}^3-p_{\perp}^4$. We can shift the first integral $\ed \Omega_1\to \ed\Omega_{12}$ to the relative angle between $1$ and $2$ such that
\begin{align}
    S=(2\pi)^2 2s \int \ed\Omega_2 e^{i q_{tot,\perp}\cdot \rfs w(\Omega_2)}\delta^{(2)}(p_{tot,\parallel})\int \ed\Omega_{12} e^{i q_{13}\cdot b_{12}} e^{-2iG s\log(\bar{\mu} b_{12})}.
\end{align}
Finally in this last step we can define for the two integrals separate z-axes. By the assumption that $q$ is small the relevant region of integration is close to the poles, so that we can replace $\ed\Omega_2\to \mu^2\ed^2 x_{\perp}$ and $\ed\Omega_{12}\to \mu^2\ed^2 b_{12}$. Additionally, a very large $\rfs$ which is identical to a large impact parameter $\rfs\sim b$ solidifies this approximation, as for large $\rfs$ any transverse curvature effects may be neglected, as shown in \figref{fig:flat space zoom}. In The result becomes:
\begin{align}
    S=\mu^4(2\pi)^2 2s \int \ed^2 x_{\perp} e^{i q_{tot,\perp}\cdot x_{\perp}}\delta^{(2)}(p_{tot,\parallel})\int \ed^2 b_{12} e^{i q_{13}\cdot b_{12}} e^{-2iG s\log(\bar{\mu} b_{12})}.
\end{align}
Recognizing the integral definition of the delta function and the fact that $\delta^{(2)}(p_{tot,\parallel})\delta^{(2)}(p_{tot,\perp})=\delta^{(4)}(p_{tot})$ we find
\begin{align}
    S=\mu^4(2\pi)^4 \delta^{(4)}(p_{tot}) 2s \int \ed^2 b_{12}e^{i q_{13}\cdot b_{12}} e^{-2iG s\log(\bar{\mu} b_{12})}.
\end{align}
Notice that the $\mu^4$ is present by the original definition of the scalar fields in harmonics (an extra $1/r$). Transforming back finally gives the following momentum space analogue of the amplitude above
\begin{align}
    i\mathcal{M}=2s \int \ed^2 be^{i q\cdot b} e^{-2iG s\log(\bar{\mu} b)},
\end{align}
as in exact agreement with \cite{KabatOrtiz}. This integral can be calculated to find
\begin{align}
    i\mathcal{M}&= \frac{2\pi s}{\bar{\mu}^2}\frac{\Gamma(1-iGs)}{\Gamma(i G s)}\left(\frac{4\bar{\mu}^2}{-t}\right)^{1-iG s},
\end{align}
as in agreement with \cite{KabatOrtiz,HOOFT1987dominance}, upon identification of the emergent scale $\bar{\mu}$ with their infrared regulator, removal of the $2s$ factor, and identification of $t=-q^2=-\tilde{k}^2$. This shows that the calculation of field theory diagrams in a harmonics base is capable of finding familiar results, when compared in limits valid for both. While tempting to perform a similar analysis on the black-hole S-matrix, there are numerous conceptual problems. The first one is that the transverse curvature effects on the black hole background need not be small, as we do not need $\rfs\sim b$ a large radius of curvature, making the last steps difficult. Secondly, the relations used between $\Omega$ and $b$ do not clearly hold on the black hole. Finally, and most importantly, the Fourier transform on the transverse momenta is not defined on the black hole, instead on the black hole we only have the angles and spherical harmonics eigenvalues available, making it impossible to define the S-matrix in a transverse momentum base in the first place. Our results before resumming the $Y_{\ell m}$'s are instead the direct analogue of this.

\clearpage

\section{Many particle eikonal amplitude}
\label{chap: MN to MN}
So far we have fully performed the methods of the familiar flat space eikonal amplitude on the Schwarzschild horizon using a spherical harmonics basis. While the results make sense for $2\to 2$ scattering, in the original paper by 't Hooft \cite{tHooft1996} the semi-classical scattering matrix is derived to be given by
\begin{align}
    S_{\text{'t Hooft}}=e^{i~8\pi G\int\ed\Omega\ed\bar{\Omega}P_{\text{in}}(\Omega)G_1(\Omega,\bar{\Omega})P_{\text{out}},(\bar{\Omega})}
\end{align}
where $P_{\text{in}},P_{\text{out}}$ describe many-particle distributions. This is with $8\pi G=\gamma$ restored. In the case that they describe a single particle $P_{1,2}(\Omega)=p_{1,2} \delta^{(2)}(\Omega-\Omega_{1,2})$ we do immediately find
\begin{align}
    S^{2\to 2}_{\text{'t Hooft}}=e^{i~8\pi G ~p_1p_2G_1(\Omega_1,\Omega_2)},
\end{align}
in agreement with our eikonal result for the shockwave approximation where $s=2p_1p_2$. The general case however does not agree because the integral is inside the exponent. We believed this to be the case because we consider $2\to 2$ scattering, whereas 't Hooft considers $N\to N$ scattering for arbitrary $N$. In this chapter we shall extend our previous eikonal methods into a new diagram, constructed to still obey the important eikonal constraints $s\gg\mu^2,s\gg t$, while being extended to arbitrary many particles. The result is an elastic S-matrix that agrees with the one by 't Hooft. While the diagram, calculation, and S-matrix are all defined with great accuracy, it is difficult to properly analyse if this diagram gives indeed the leading contribution.\\
\\
Let us first define what we mean by a many particle state in this context. For this counting we use the number operator in a Fock space basis valid only locally on the horizon, disregarding spacetime effects for later research. First let us define the on-shell canonical quantization
\begin{align}
    \phi_{\ell m}(x^a)=\int\limits_{\mathbb{R}^+}\frac{\ed p}{(2\pi)(2p)}\left(a_{\ell m}(p)e^{ip_a x^a}+a^{\dagger}_{\ell m}(p)e^{-ip_a x^a}\right)
\end{align}
where $p$ is a component of choice of non-zero momentum, and $p^a$ is fixed by the mass-shell condition. The integral is only over positive momenta because we are looking at lightcone-momenta and the positive sign ensures future directed particles. For a massive scalar field the number operator is defined by by (for a scalar field)
\begin{align}
    \mathcal{N}=\int\frac{\ed p}{(2\pi)(2p)}\sum\limits_{\ell m} a^{\dagger}_{\ell m}(p)a_{\ell m}(p),
\end{align}
since the mass-shell condition ensures we can use either of the two momentum components (so long as we work consistently within that choice). Together with the commutator
\begin{align}
    [a_{\ell m}(p),a^{\dagger}_{\ell'm'}(p')]=(2\pi)(2p)\delta(p-p')\delta_{\ell \ell'}\delta_{mm'},
\end{align}
we can quickly show that $\mathcal{N}$ indeed counts the amount of creation operators on the right.\\
For massless scalars instead we can define a separate number operator for each component, namely one for infalling and one for outgoing modes:
\begin{align}
    \mathcal{N}_{\text{out}}&=\int\frac{\ed p_y}{(2\pi)(2p_y)}\sum\limits_{\ell m} a^{\dagger}_{\ell m}(p_y)a_{\ell m}(p_y),\\
    \mathcal{N}_{\text{in}}&=\int\frac{\ed p_x}{(2\pi)(2p_x)}\sum\limits_{\ell m} a^{\dagger}_{\ell m}(p_x)a_{\ell m}(p_x).
\end{align}
The commutators are given by
\begin{align}
    [a_{\ell m}(p_x),a^{\dagger}_{\ell'm'}(p'_x)]&=(2\pi)(2p_x)\delta(p_x-p'_x)\delta_{\ell \ell'}\delta_{mm'},\\
    [a_{\ell m}(p_y),a^{\dagger}_{\ell'm'}(p'_y)]&=(2\pi)(2p_y)\delta(p_y-p'_y)\delta_{\ell \ell'}\delta_{mm'},\\
    [a_{\ell m}(p_x),a^{\dagger}_{\ell'm'}(p'_y)]&=0.
\end{align}
Under this definition we define a typical $N,M$ state of infalling particles to be given by
\begin{align}
    |\text{in}\rangle=\prod\limits_{i=1}^N a^{\dagger}_{\ell_i m_i}(p_1^i)\prod\limits_{j=1}^M a^{\dagger}_{\ell_j m_j}(p_2^j)|0\rangle.
\end{align}
So notably this is an $M+N$ particle state, even though the creation operators are in spherical harmonics. The classical interpretation that these are distributions does not alter the quantum notion of what a particle is. Notably, for example for a single particle, the $\ell,m$ state us related to the position basis by
\begin{align}
    |\text{in}\rangle= \int \ed \Omega ~ Y_{\ell m}(\Omega)|in,\Omega\rangle=\int \ed \Omega~  Y_{\ell m}(\Omega)a^{\dagger}(p_1,\Omega)|0\rangle.
\end{align}
So the $\ell,m$ creation operators do not correspond to many particle, however they do correspond to a infinite superposition of single particle states. The position space analogue of the general instate is given by
\begin{align}
    |\text{in}\rangle=\prod\limits_{i=1}^N a^{\dagger}(p_1^i,\Omega_1^i)\prod\limits_{j=1}^M a^{\dagger}(p_2^j,\Omega_2^j)|0\rangle,
\end{align}
which contains the same number of particles as in the $\ell,m$ basis since the transformations are linear and do not mix creation and annihilation operators.

\subsection{Angular position space theory}
\label{sec: angular position space}
\begin{figure}[h!]
    \centering
    \includegraphics[width=0.8\textwidth]{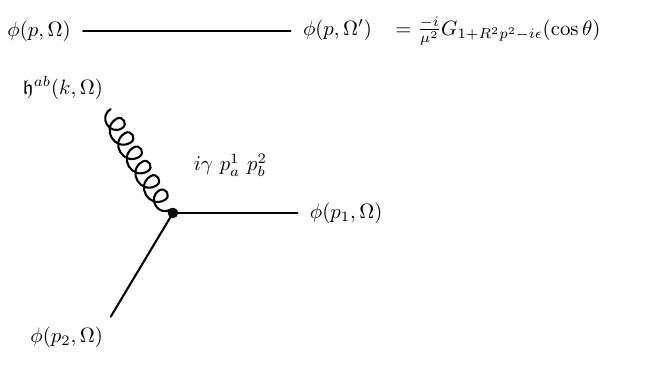}
    \caption{The position space Feynman rules, shown only for a real scalar field and a graviton. For the complex scalar field the propagator and vertex are identical. For the gauge field coupling to a complex scalar the vertex factor changes to $iQ(p_1+p_2)$. The propagator moves a field from an angle $\Omega$ to some $\Omega'$ ($\theta$ is the great circle distance between these two), the vertex is at a fixed $\Omega$.}
    \label{fig:mn to mn feynman rules}
\end{figure}

By using the full resummation over $\ell,m$ in the previous section, we can analogously define a theory using Feynman rules in angular position space. This would place all vertices at a specific angle $\Omega$, while the propagators move between these angles $\mathcal{P}(\Omega,\Omega')$. We shall derive this angular propagator for the scalar to perform the many-particle scattering. The scalar propagator in $\ell,m$ is given by
\begin{align}
    \frac{-i}{p^2+\mu^2\lambda_{\ell}-i\epsilon}=\frac{-i}{\mu^2}\frac{1}{\ell^2+\ell+1+\frac{p^2}{\mu^2}-i\epsilon}.
\end{align}
Then from the Feynman rules we can read off that the propagator is given by
\begin{align}
    \mathcal{P}(p,\Omega;\Omega')&=\sum\limits_{\ell m}\frac{-i}{\mu^2}\frac{1}{\ell^2+\ell+1+R^2p^2-i\epsilon}Y_{\ell m}(\Omega)Y_{\ell m}(\Omega'),\\
    &= \frac{-i}{\mu^2}G_{1+R^2p^2-i\epsilon}(\cos\theta),
\end{align}
where $G_a(\cos\theta)$ is the familiar Green's function, but the value of $a$ is determined by $p^2$. A similar expression may be written down for the graviton and gauge field propagator, but we will not need it. We remind the reader that there is a direct relation between $\cos\theta$ and $\Omega,\Omega'$ given by 
\begin{align}
    \cos\theta=\cos \theta_{\Omega}\cos\theta_{\Omega'}+\sin\theta_{\Omega}\sin\theta_{\Omega'}\cos(\phi_{\Omega}-\phi_{\Omega'}).
\end{align}
From our $\ell,m$-resummed calculation we can observe how to do a calculation in the angular position basis: For all lightcone momenta there is conservation at the vertices and the propagators, where for propagators this means essentially the propagator carries a single momentum. For the position the propagators do not conserve position and generally carry two different positions, however all vertices no matter how high order are at the same position. Additionally while there are loop momenta, there are no loop positions for this reason. The resulting Feynman rules are shown in \figref{fig:mn to mn feynman rules}.

To proceed with our calculation we want to find a description of the Feynman rules in the eikonal phase. In \secref{sec:greens func} we observed that the Green's function $G_a$ contains a pole in $a$. To find a well-defined asymptotic expression, we can extract this pole and calculate the first order residue
\begin{align}
    \lim_{a\to\infty} aG_a(\cos\theta)=\sum\limits_{\ell m}Y_{\ell m}(\Omega)Y_{\ell m}(\Omega' )= \delta^{(2)}(\Omega-\Omega').
\end{align}
This gives the expected limit where the particle does not change direction; the Green's function is infinitely sharply peaked around $\cos\theta=0$. Asymptotically for the scalar we can write the first order Laurent expansion in $\frac{1}{a}$ to be
\begin{align}
    \mathcal{P}_{eik}(p,\Omega;\Omega')&=  \frac{-i}{p^2+\mu^2-i\epsilon}\delta^{(2)}(\Omega-\Omega'),
\end{align}
which is the eikonal (high energy) approximation for the position space theory. This limit alone shows that the black hole eikonal phase matches nicely with the angular position basis. The transverse and longitudinal modes decouple naturally, with high energy modes maintaining their direction of motion unchanged. In the $\ell m$ basis this only became clear after a lengthy resummation. It should be said that the approximation above essentially assumes $p^2\mu^2\gg 1$ which is in principle not correct for on-shell particles. However, in the eikonal phase only the momentum parts contribute, since as we saw the mass terms could be transformed away by appropriate redefinitions of the loop-momenta and/or inclusion of the eikonal momentum exchange $q_-$ exactly. In the following we will use the propagator above as leading order propagator.

\subsection{\texorpdfstring{$M+1\to M+1$}{M+1->M+1} diagram}
\label{sec:m1tom1}
\begin{figure}
\centering
\includegraphics[width=\textwidth]{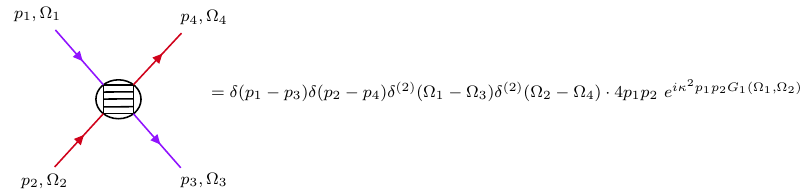}
\caption{\label{fig:eikonal vertex} The new vertex factor to be used, essentially implicitly including the full perturbatively exact S-matrix for $2\to 2$ scattering between the particles. We remark that this is of course not to be understood as a formal vertex but a compact notation for the full scattering in terms of sub-diagrams.}
\end{figure}

\begin{figure} 
\centering
\includegraphics[width=\textwidth]{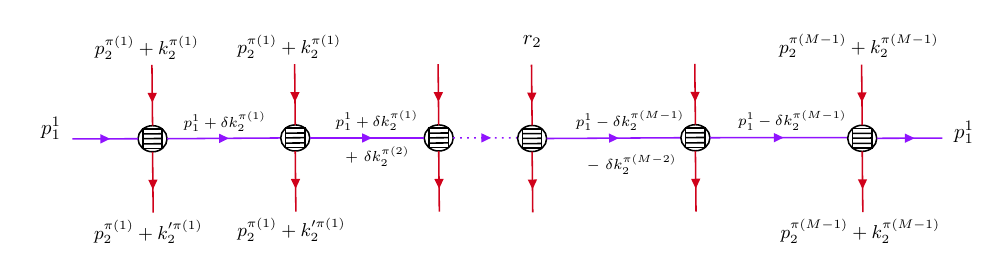}
\caption{\label{fig:flowlineparticle} The kinematical choices for the $M+1\to M+1$ scattering process. The $p_1^i$ momentum is mostly conserved along the blue line, where we assume the momentum differences to be small. All $p_2^j$ particle are added in all different permutations $\pi$. The $r_2$ line corresponds to the insertion of $p_2^M$, before crossing it we add all momentum exchanges $\delta k_2^j$ from the left side, whereas after crossing it we instead subtract all momentum exchanges from the right side.}

\end{figure}

To calculate the specific case that 't Hooft considers we need to look at an eikonal generalization to many particles. As mentioned in the introduction, we consider $K\to K$ scattering for an arbitrary amount of particles $K$, however on the horizon it is natural to split $K=N+M$ into $N$ particles falling into the horizon and $M$ going out. We first calculate the case of one infalling particle $N=1$ with momentum $p_1^i$, and label the $M$ outgoing particles with momenta $p_2^j$ at angle $\Omega_2^j$. We will consider the tree-level diagram where pair of scalar particles interacts once eikonally, with all external $p_2^j$ generally off-shell, so that we can use this diagram as a building block to extend to the full $M+N\to M+N$ diagram. The eikonal interaction we will treat as a $2\to 2$ scalar vertex, and has been drawn in \figref{fig:eikonal vertex}. The corresponding $N=1$ case has been drawn in \figref{fig:flowlineparticle}. In order to keep this general we need to consider also all possible permutations of placing the external legs.

For the $M+1\to M+1$ case there are exactly $M$ vertices and $M-1$ propagators with momentum $p_1^i+\text{transfer}$. The momentum transfers are in principle given by
\begin{align}
q_1^i&=p_1^{\prime i}-p_1^i,\\
\delta k_2^j&=k_2^{\prime j}-k_2^j,
\end{align}
and we assume them to be of order $\mathcal{O}\left(\frac{\mu^2}{s}\right)$. Because we are working in the angular position basis the scalars are now massless, so we require the on-shell particle to obey $(p_1^i)^2=0$. This has important consequences for the momentum exchange $q_1^i$:
\begin{align}
    (q^i_1)^2+2p_1^i\cdot q_1^i=0
\end{align}
has to be true for $(p_1^{\prime i})^2=0$ to hold. Since we assume the momentum transfer to always be of order $\mathcal{O}\left(\frac{\mu^2}{s}\right)$, for the leading order contributions this means
\begin{align}
    p_1^i\cdot q_1^i=0.
\end{align}
Thus when only considering leading order behaviour we may always assume that 
\begin{align}
    q_1^i=\begin{pmatrix}
        q_{1}^i\\ 0
    \end{pmatrix}
\end{align}
so that it is orthogonal (in lightcone coordinates) with $p_1^i=(p_{1}^i,0)$.\\
\\
Combining the small momentum exchange limit with the black hole eikonal phase propagator in \secref{sec: angular position space} we find that the internal scalar propagators are given by
\begin{align}
    \Delta_{\phi}(p_1^i+k)(\Omega,\Omega')\to \frac{-i}{2p_1^i\cdot k-i\epsilon}\delta^{(2)}(\Omega,\Omega').
\end{align}
Here $k$ is a shorthand notation for any respective necessary sum of momenta that we assume to be small of order $\mathcal{O}\left(\frac{\mu^2}{s}\right)$. \\
\\
Let us first regard the behaviour of the angles: Along the entire diagram each vertex will preserve the angles. Thus the $p_2^j$ legs all have the same ingoing and outgoing angle. For the $p_1^i$ particle the first and last vertex are fixed, but the middle ones are in principle to be integrated out. Denoting the internal angles $\bar{\Omega}_n$ this gives:
\begin{align}
    \sim \int \ed\bar{\Omega}_1^i\dots \ed \bar{\Omega}_{M-1} \Delta_{\phi}(p_1+k)(\Omega_1,\bar{\Omega}_1)\dots \Delta_{\phi}(p_1+k' )(\bar{\Omega}_{M-1},\Omega^{\prime i}_1),
\end{align}
where the angle $\Omega_1^{\prime i}$ is the angle of the particle moving out of the diagram, that carries momentum $p_1^{\prime i}$. Using that the propagators are all proportional to delta-functions the entire expression is proportional to
\begin{align}
    \sim \delta^{(2)}(\Omega_1^i-\Omega_1^{\prime i}).
\end{align}
This shows that in the eikonal limit the outgoing particle must still have the same angle as the ingoing particle, just as we saw for $2\to 2$ scattering. The total transverse contribution is then simply given by
\begin{align}
    \delta^{(2)}(\Omega_1^i-\Omega_1^{\prime i}) \times \prod_{i=j}^M \delta^{(2)}(\Omega_2^j-\Omega_2^{\prime j})
\end{align}
Similar for each vertex we can write down the eikonal exponent including the kinematical prefactor. Specifically, the vertex of the interaction with $p_2^j$ is given by
\begin{align}
    (2p_1^i)(2p_2^j)e^{i\chi_0^{ij}G_1(\Omega_1^i,\Omega_2^j)}
\end{align}
where
\begin{align}
    \chi_0^{ij}=\frac{\gamma^2 s_{ij}}{2 \mu^2} \lgap\lgap\lgap
    s_{ij}=-(p_1^i+p_2^j)^2.
\end{align}
This turns the contribution of all vertices into
\begin{align}
\label{eq:vertex contributions}
   (2p_1^i)^M\prod_{j=1}^M(2p_2^j) ~ \text{Exp}\left(\frac{i\gamma^2}{2 \mu ^2}\sum \limits_{j=1}^{j=M}s_{ij}G_1(\Omega_1^i,\Omega_2^j)\right).
\end{align}
We now proceed to adding the contribution of the propagators.

\subsubsection*{Propagator contribution}
We use the momentum configuration in \figref{fig:flowlineparticle}. Essentially all vertical legs have momentum $p_2^j+k^j$ going in, and $k^{\prime j}$ going out. We want to enforce momentum conservation, so we use that
\begin{align}
    \sum\limits_{j=1}^M k_{2}^j= \sum\limits_{j=1}^M k_{2}^{\prime j}+q_1^i.
\end{align}
We insert this explicitly by replacing $k_2^j-k_2^{\prime j}$ with the other momenta at the intersection vertex of $p_2^M$. The location of this vertex $r_2$ is arbitrary, and must be summed over to include all permutations. We sum over the location of $r_2$ separately, such that we split off the remaining permutations subgroup for the rest of the $M-1$ legs explicitly, which we denote by $\pi$. The propagator contribution is then
\begin{align}
    &\frac{-i}{2p_1^i\cdot \left(k_2^{\pi(1)}-k_2^{\prime\pi(1)}\right)-i\epsilon}\times\dots\times  \frac{-i}{2p_1^i\cdot \sum\limits_{n=1}^{r_2-1}\left(k_2^{\pi(n)}-k_2^{\prime\pi(n)}\right)-i\epsilon}\\
     \times &\frac{-i}{-2p_1^i\cdot \sum\limits_{n=r_2}^{M-1}\left(k_2^{\pi(n)}-k_2^{\prime\pi(n)}\right)-i\epsilon}\times\dots\times\frac{-i}{-2p_1^i\cdot \left(k_2^{\pi(M-1)}-k_2^{\prime\pi(M-1)}\right)-i\epsilon}.
\end{align}
In the above in principle we would have needed to add $q_1^i$ as well, but since $p_1^i\cdot q_1^i=0$ this does not contribute. The calculation is outlined in \appref{app:m1tom1}. Essentially the summation over combinatorics automatically ensures that momentum is conserved at every vertex, resulting in a set of delta-functions $\delta(q)$. The eikonal amplitude for the $M+1\to M+1$ diagram may be written in total as
\begin{align}
i\mathcal{M}_{\text{sub}}\left(p_1^i,k_{y,2}^j,k_{y,2}^{\prime j}\right)&=(2p_1^i)(2p_2^ M)\delta^{(2)}(\Omega_1^i-\Omega_1^{\prime i})\nonumber\\
    &\times\prod\limits_{j=1}^{M-1}(2\pi)(2p_2^j)\delta\left(k_{y,2}^j-k_{y,2}^{\prime j}\right)\\
    &\times\text{Exp}\left(\frac{i\gamma^2}{2 \mu ^2}\sum \limits_{j=1}^{j=M}s_{ij}G_1(\Omega_1^i,\Omega_2^j)\right).
\end{align}
The next step is to extend this to $N$ infalling particles as well. Note that we intentionally excluded the transverse delta's $\delta^{(2)}(\Omega_2^j-\Omega_2^{\prime j})$ in the definition to integrate them out early in the next section.

\subsection{\texorpdfstring{$M+N\to M+N$}{M+N->M+N} diagram}
\begin{figure} 
\centering
\includegraphics[width=0.8\textwidth]{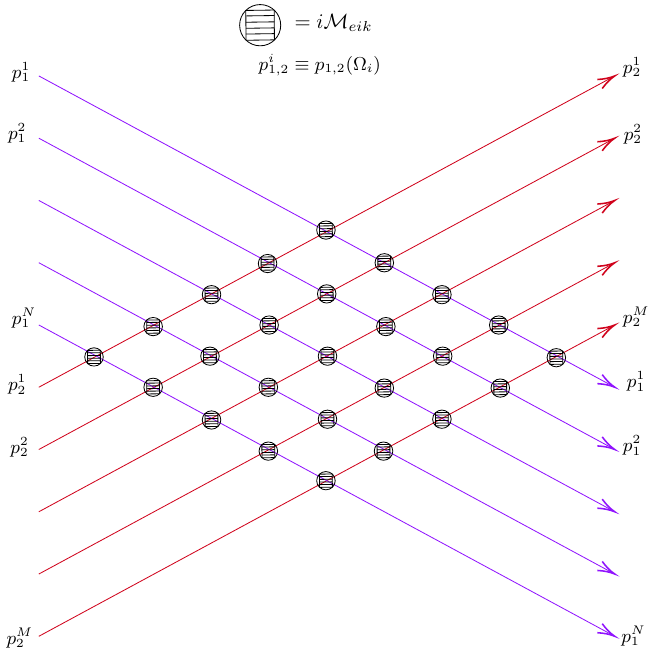}
\caption{\label{fig:manyparticle} The complete elastic eikonal many-particle scattering. All particles with identical momentum direction have been grouped together as either blue or red, these momenta are all different. The key component in the calculation is that all vertices are a complete eikonal interaction, and no other interaction's or couplings take place. For the rest the usual Feynman rules apply. A priori it seems to be possible for the $p_1,p_2$ to spread differently over the grid, but this would automatically result in at least one lower order vertex, and an overall amplitude an order lower in $s$.}

\end{figure}

We now extend this to an arbitrary amount of infalling particles $N$. They still all interact once eikonally. We denote the infalling particles by $p_1^i,\Omega_1^i$ still, but now essentially seek to sum the amplitude of the previous section over all $i$. We will first consider the vertices.

\subsubsection*{Transverse separation}
In this section we outline the calculation for arbitrary $N,M$. We can sum up that there are:
\begin{itemize}
    \item $N$ amount of sub-diagrams $\mathcal{M}_{\text{sub}}$,
    \item $(N-1)(M-1)$ amount of loops to integrate over,
    \item $N(M-1)$ additional matter propagators with $p_1^i$ momenta.
\end{itemize}
We make the same approximation as before, assuming large momenta but small exchanges now also for the $p_2^j$ particles:
\begin{align}
q_1^i&=p_1^{\prime j}-p_1^i,\\
q_2^j&=p_2^{\prime j}-p_2^j,
\end{align}
that are all of order $\mathcal{O}\left(\frac{\mu^2}{\sqrt{s}}\right)$. In the leading order limit the on-shell conditions now restrict
\begin{align}
    q_1^i=\begin{pmatrix}
        q_{1}^i \\0
    \end{pmatrix} \qquad \qquad \qquad \qquad q_2^j=\begin{pmatrix}
        0 \\ q_{2}^ j
    \end{pmatrix}
\end{align}
to ensure that $p_1^i\cdot q_1^i=0,p_2^j\cdot q_2^j=0$. Because of this property we can now observe an important kinematical consequence: The phase space splits explicitly into its $x,y$ component separately. All $p_1$ momenta may only couple to the $y-$component of other momenta, so that it is only non-zero together with $q_2^j$ or the $y$-component of the loop momenta, and vice versa for $p_2$. We can use this phase space separation to neatly write down kinematical choices. Specifically we can add any small momentum exchange in the $x-$direction to a $p_1$ leg without altering the result. For this reason we also write the components of all momenta without $x,y$ subscripts to avoid clutter of notation, $p_1^i$ for example denotes both the vector (when coupled to a dot-product) or the component (when on its own).\\
\\
All vertex factors are automatically included in the sub-diagrams, so we mostly have to look carefully at the new scalar legs. As before we approximate the scalar legs by
\begin{align}
    \Delta_{\phi}(p_2^j+k)(\Omega,\Omega')\to \frac{-i}{2p^j
    _2\cdot k-i\epsilon}\delta^{(2)}(\Omega,\Omega').
\end{align}
Combining the transverse delta's $\delta^{(2)}(\Omega_2^j-\Omega_2^{\prime j})$ with the ones above and integrating all internal angles out result in delta-functions in the transverse space for all $p_2^i$ functions as well:
\begin{align}
    \sim \delta^{(2)}(\Omega^2_j-\Omega_2^{\prime j}).
\end{align}
This shows that in the eikonal limit all outgoing particles will remain at the same angle and all infalling particles as well, indicative of the small transverse exchanges. The total transverse contribution for all particles is thus given by
\begin{align}
\label{eq:vertex angles}
    \prod\limits_{i=1}^N \delta^{(2)}(\Omega_1^i-\Omega_1^{\prime i})\prod\limits_{i=j}^M \delta^{(2)}(\Omega_2^j-\Omega_2^{\prime j}).
\end{align}
Next we look at the momenta to integrate those out.

\subsubsection*{Layers of sub-diagrams}

\begin{figure}
\centering
\includegraphics[width=.9\textwidth]{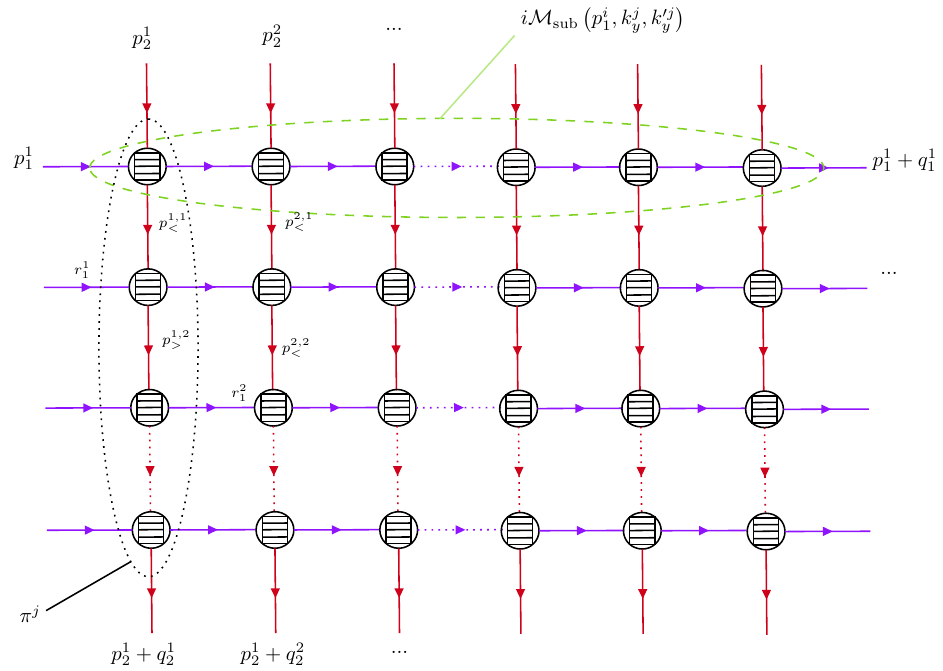}
\caption{\label{fig:manypart momentum2} Schematic notation of the $M+N\to M+N$ diagram where all particles interact only once eikonally. All rows are defined by the sub-diagram calculated before, now patched together over many columns. For each column we allow arbitrary permutations $\pi^j$, and a location $r_1^j$ where the $p_1^N$ momentum enters. We define separate loop-momenta for each row and column, depending on whether we are before $r_1^j$ or a after $r_1^j$.}
\end{figure}

We define the kinematics as shown in \figref{fig:manypart momentum2}. For each row we insert a full sub-diagram that was calculated before. For each column we want to assign the momenta using the same tricks as before. For each column, we define a separate permutation $\pi^j$ on where to attach the $p_1^i$ particle to the $p_2^j$ line, and a separate location $r_1^j$ where the $p_1^N$ particle is attached, to apply overall momentum conservation to the scalar momenta. In \secref{sec:generality} we show that all possible internal leg configurations are now included by these permutations. To be able to write down a calculable amplitude, we define the internal momenta differently in the different directions. On the $j\text{'th}$ row, before the $r_1^j$ intersection we define the $i\text{'th}$ scalar leg by
\begin{align}
    p_<^{i,j}&=\begin{pmatrix}
        \sum\limits_{n=1}^{i} k_{x}^{n,j}\\
        p_2^j+\sum\limits_{n=1}^{i} k_{y}^{n,j}
    \end{pmatrix}.
\end{align}
On the other hand after the $r_1^j$ intersection we define
\begin{align}
p_>^{i,j}&=\begin{pmatrix}
        -\sum\limits_{n=i}^{N-1} k_{x}^{n,j}\\
        p_2^j+\sum\limits_{n=1}^{i} k_{y}^{n,j}
    \end{pmatrix}.
\end{align}
We thus explicitly use the separation of coordinates $x,y$ to make a different choice for the loop momenta in the different directions. We may choose anything convenient for the $y-$direction since it drops out in the inner product, and so we choose the most optimal thing to insert in the sub-diagrams. For the external legs we have to define boundary cases
\begin{align}
    k^{0,j}&=0=k^{i,0},\\
    k^{N,j}&= q_2^j,\\
    k^{i,M}&= q_1^i.
\end{align}
The amplitude for a given permutation is then given by
\begin{align}
\label{eq:ampfirst}
    i \mathcal{M}&=\int\prod_{\substack{i,j=1}}^{\substack{i=N-1\\ j=M-1}}\left(\frac{\ed^2 k^{i,j}}{(2\pi)^2}\right)\prod\limits_{i=1}^N i\mathcal{M}_{\text{sub}}\left(p_1^i,\sum\limits_{n=1}^{i}k_y^{\pi^j(n-1),j},\sum\limits_{n=1}^{i}k_y^{\pi^j(n),j}\right)\\
    \times   \prod\limits_{j=1}^M   &\frac{-i}{2p_2^j\cdot k^{\pi^j(1),j}-i\epsilon}\times\dots\times  \frac{-i}{2p_2^j\cdot \sum\limits_{n=1}^{r^j_1-1}k^{\pi^j(n),j}-i\epsilon}\\
     \times &\frac{-i}{-2p_2^j\cdot \sum\limits_{n=r_1^j}^{N-1}k^{\pi^j(n),j}-i\epsilon}\times\dots\times\frac{-i}{-2p_2^j\cdot k^{\pi^j(N-1),j}-i\epsilon}.
\end{align}
This expression must still be summed over all permutations. The calculation of this expression is performed in \appref{app:mntomn}. The resummation over combinatorics over the column's again ensures that all momentum transfers receive delta functions; the delta-functions over internal momenta remove the loop integrals, while the boundary values ensure no momentum transfer for the external particles. The kinematical choices remain valid for different permutations due to the separation of phase space. To write the final $S$-matrix we first recognize that, because the diagram is elastic, we can write down the free-field contribution using the commutators as
\begin{align}
    S_0\equiv\mathds{1}&=\prod\limits_{i=1}^N\left((2\pi)(2p_1^i)\delta(p_1^i-p_1^{\prime i})\delta^{(2)}(\Omega_1^i-\Omega_1^{\prime i})\right)\nonumber\\
    &\times\prod\limits_{j=1}^M\left((2\pi)(2p_1^j)\delta(p_2^j-p_2^{\prime j})\delta^{(2)}(\Omega_2^j-\Omega_2^{\prime j})\right).
    \end{align}
Then the resulting $S$-matrix simply becomes a pure complex exponent
\begin{align}S&=\mathds{1}\text{Exp}\left(\frac{i\gamma^2}{2 \mu ^2}\sum \limits_{i,j=1}^{\substack{i=N\\j=M}}s_{ij}G_1(\Omega_1^i,\Omega_2^j)\right).
\end{align}

\subsection{S-matrix conclusion}

\begin{figure}[h!]
    \centering
    \includegraphics[width=0.6\linewidth]{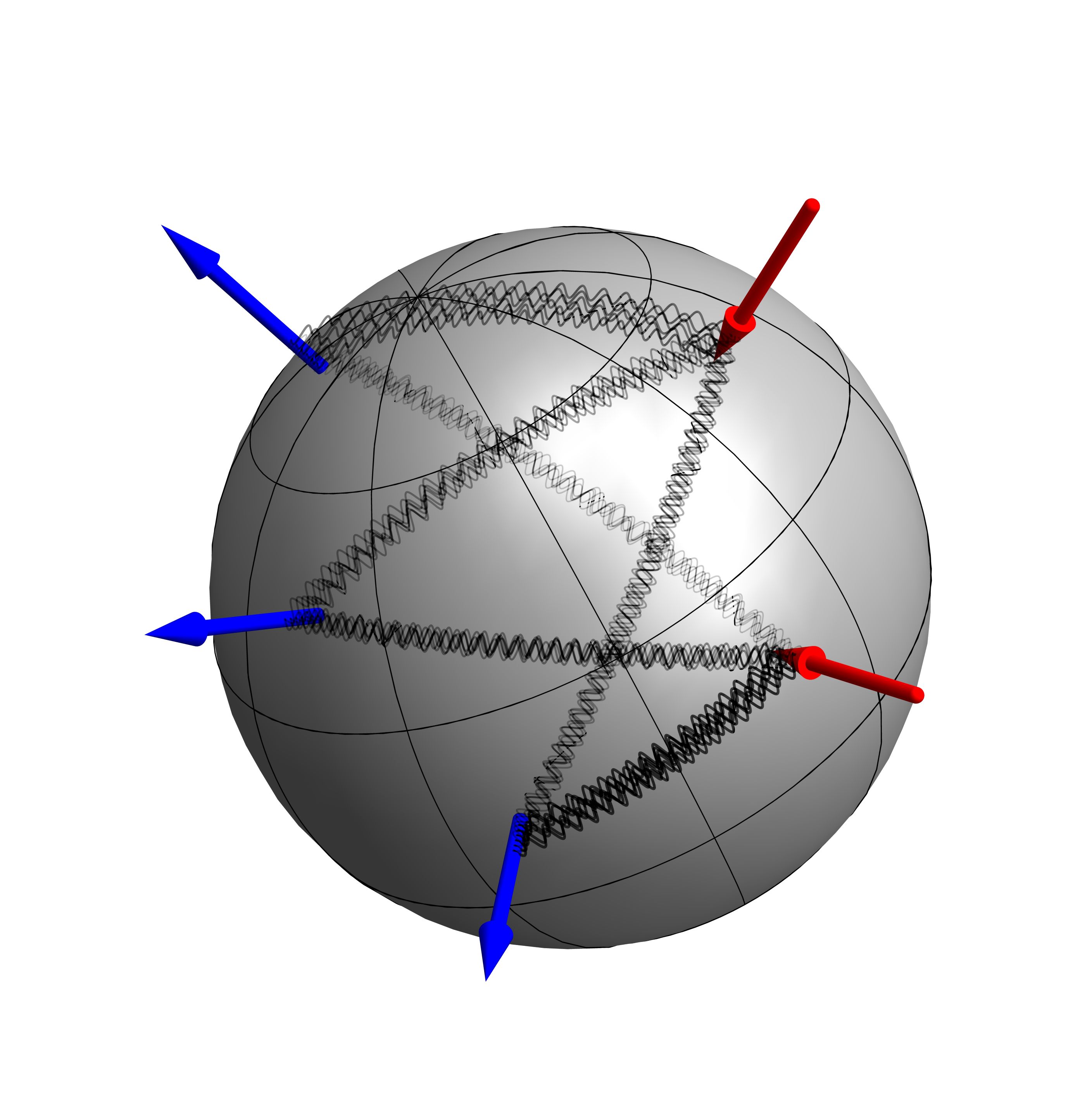}
    \caption{The many-particle scattering setup. Multiple particles $N,M$ are going in and out. These all scatter eikonally, but only the ingoing with the outgoing particles (there are no interactions between the identical coloured arrows). The strength of the gauge interaction depends on the separation. The interactions are drawn as straight lines to avoid clutter.}
    \label{fig:manyparticle scattering}
\end{figure}

The total S-matrix is given by
\begin{align}
\label{eq:mnmnsmatrix}
    S=\mathds{1}\text{Exp}\left(\frac{i\gamma^2}{2 \mu ^2}\sum \limits_{i,j=1}^{\substack{i=N\\j=M}}s_{ij}G_1(\Omega_1^i,\Omega_2^j)-\frac{i}{\mu^2}\sum \limits_{i,j=1}^{\substack{i=N\\j=M}}q_iq_j~G_0(\Omega_1^i,\Omega_2^j)\right),
\end{align}
where the extension to electrodynamics has done by adding the electromagnetic charges and Green's function as well. Here $q_i,q_j$ are the sign-included charges of particle $i,j$ respectively. In \figref{fig:manyparticle scattering} the setup has been illustrated.

We observe that this amplitude is indeed the $M+N\to M+N$ extension of the usual eikonal amplitude; similar to shown in the previous section and \cite{LongPaper,SQEDpaper}, the S-matrix reduces to just a complex phase factor. We can write the phase factor out as follows:
\begin{align}
    \frac{i\gamma^2}{2\mu ^2}\sum \limits_{i,j=1}^{\substack{i=N\\j=M}}s_{ij}G_1(\Omega_1^i,\Omega_2^j)=8\pi G\sum\limits_{i=1}^N\sum\limits_{j=1}^M p_1(\Omega_1^i)p_2(\Omega_2^j)G_1(\Omega_1^i,\Omega_2^j).
\end{align}
To get in touch with 't Hooft's results we manually define distributions using the external particles' eigenvalues:
\begin{align}
    p_1(\Omega_1^i)\to p_{\text{in}}(\Omega)=\sum\limits_{i=1}^N p_1(\Omega_1^i)\delta^{(2)}(\Omega-\Omega_1^i).
\end{align}
Using these distributions our S-matrix may be written as:
\begin{gather}
    S=\mathds{1}\text{Exp}\left[i\int\ed\Omega\ed\Omega'\biggr( 8\pi G ~ p_{\text{in}}(\Omega)G_1(\Omega,\Omega')p_{\text{out}}(\Omega')- Q_{\text{in}}(\Omega)G_0(\Omega,\Omega')Q_{\text{out}}(\Omega')\biggr)\right]
\end{gather}
which is exactly the scattering matrix of 't Hooft \cite{tHooft1996} in \eqref{eqn:shockwaveQMSmatrix}, with the addition of the electromagnetic interaction. This shows that the scattering matrix of 't Hooft is not described by $2\to 2$ scattering, but by any $M+N\to M+N$ scattering, where $N,M$ are free. The scattering must however happen in a generalized eikonal sense, such that each scalar must interact with each other scalar only exactly once in a $2\to 2$ eikonal manner. Naturally in a complete scattering theory this is an extremely specific case, and we can only assume this type of interaction to happen if the angular separation between particles is large enough. In the continuum limit this might be violated without notice: If two particles have too small separation $\Omega_1-\Omega_2$ the eikonal vertex does not hold any more. Since this assumption is embedded in the eikonal phase, we may conclude that within the eikonal phase the $S$-matrix above is valid.\\
This type of diagram provides the most general elastic eikonally resummed amplitude one can construct, and to our knowledge has not been drawn before. Remarkably, the eikonal simplifications still manage to hold out and work in simplifying the diagram, although more general contour integrations were needed. 

\subsubsection{Flat space amplitude}
We note that we calculated the $M+N\to M+N$ diagram in the black hole eikonal phase, however an extension towards flat space is easily done with the flat space Feynman rules. The result is given by \eqref{eq:mnmnsmatrix} upon replacing the momenta by the flat space momenta and shifting the Green's functions accordingly: 
\begin{align}
    S_{\text{flat}}=\mathds{1}\text{Exp}\left(i8\pi G\sum \limits_{i,j=1}^{\substack{i=N\\j=M}}p_i\cdot p_j~ G_0(\Omega_1^i,\Omega_2^j)-i\sum \limits_{i,j=1}^{\substack{i=N\\j=M}}Q_iQ_j~G_0(\Omega_1^i,\Omega_2^j)\right).
\end{align}
The most striking difference is that both Green's functions have the same index: Generally the gravitational interaction can not be distinguished as well from the electromagnetic interaction as for the black hole eikonal phase. Inserting the explicit form for the Green's function and writing $\tfrac{1}{2}-\tfrac{1}{2}\cos\theta=\tfrac{1}{2}\mu^2 b^2$ as described in \secref{sec:flat space comp} gives
\begin{align}
    S_{\text{flat}}=\mathds{1}\text{Exp}\left(i\sum \limits_{i,j=1}^{\substack{i=N\\j=M}}\left(-4G ~ p_i\cdot p_j+\frac{1}{2\pi}Q_iQ_j\right)~ \log\left(\bar{\mu}b_{ij}\right)\right).
\end{align}
This shows the familiar transverse distance-dependent logarithm clearly, and suggests that the amplitude above may be interpreted as the semi-classical description of many particles interacting via shockwaves. The equation above describes the most general possible flat space eikonal amplitude, and we leave it without transforming back to four-dimensional momentum space since there exists no literature solution to compare to.

\subsection{Generality}
\label{sec:generality}

\begin{figure}[h!]
    \centering
    \includegraphics[width=\textwidth]{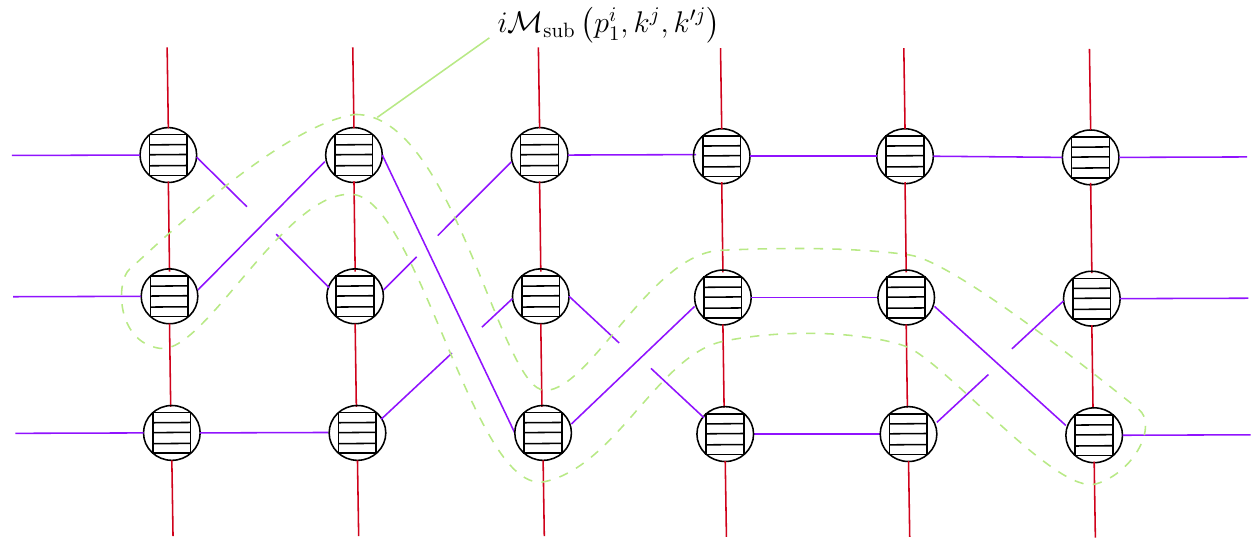}
    \caption{The effect of the permutations on the columns. Generally this will allow for any distribution of the the blue lines, and we always define the sub-diagram to follow the blue lines to ensure that the conditions for the way it was derived are still satisfied.}
    \label{fig:mn to mn column perms}
\end{figure}

\begin{figure}[h!]
    \centering
    \includegraphics[width=\textwidth]{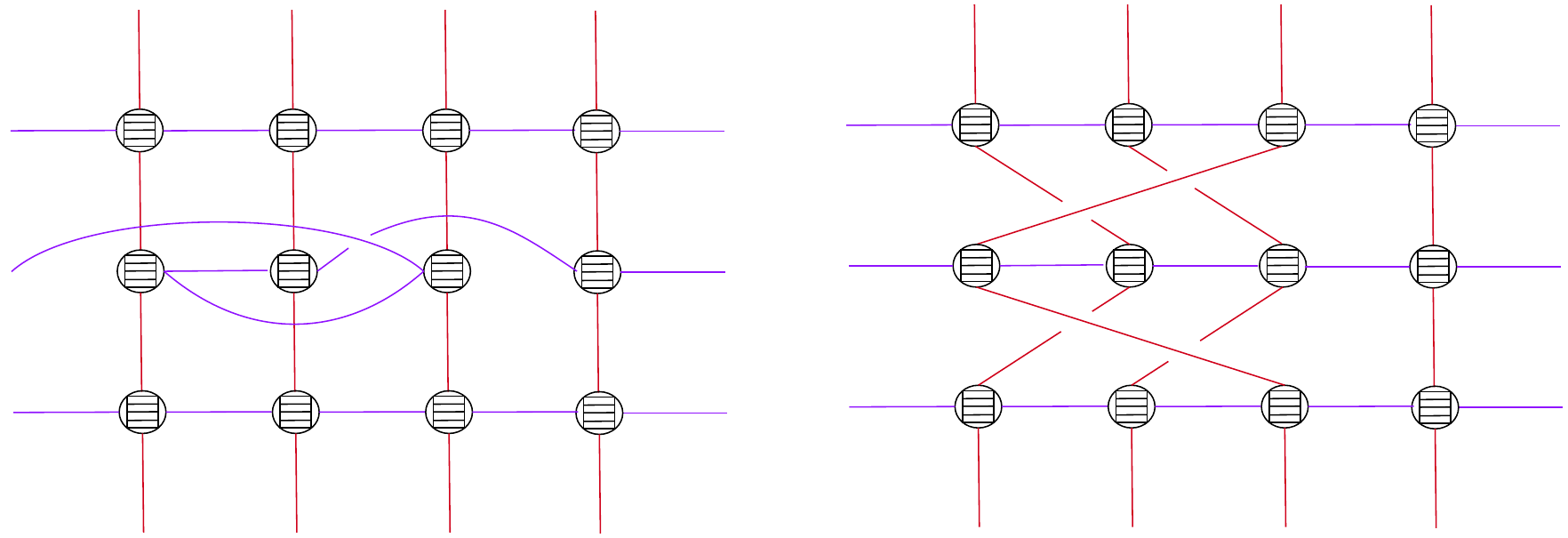}
    \caption{The diagram on the left indicates a possible diagram that does not appear to be included by the column permutations. However it is diagrammatically identical to the diagram on the right, which corresponds to a simple row permutation of the middle row, which is automatically included in $i\mathcal{M}_{\text{sub}}$. Thus by combining the column permutations with the row permutations in $i\mathcal{M}_{\text{sub}}$ we allow all possibilities.  }
    \label{fig:mn to mn row perms}
\end{figure}

In this section we finally comment on the generality to which \eqref{eq:mnmnsmatrix} is the leading order $S$-matrix in the eikonal phase. We first comment on the possible configurations on internal legs, and argue that all possibilities are included. The separate permutations $\pi^j$ for each column ensure that any diagram of the type in \figref{fig:mn to mn column perms} is included. One may then ask well what if the $p_1^i$ legs do not interact with $p_2^1$ first, but somewhere in the middle like on the left in \figref{fig:mn to mn row perms}, but this is identical to a permutation on the row instead (the right picture), which is included in $\mathcal{M}_{\text{sub}}$. Thus by combining both permutations we unsure every possible configuration is included. The only caveat is that in principle these changes in configurations also change what happens with momentum conservation, so that the chosen kinematics in \figref{fig:manypart momentum2} are not necessarily allowed. This is finally where the separation of coordinates are essential. Because $p_{1y}^i=0$ and $p_{2x}^j=0$ for all $i,j$, we may add respectively any loop momentum component $k_x,k_y$ to the $p_1,p_2$ scalar propagators without modifying them. This allows us to fix momentum conservation in the desired way for each row and column in the way described: If a permutation changes the internal momenta, we redefine the internal momenta to the desired choice, and any possible residual change by this redefinition is completely undone by the separation of coordinates. Of course we may use momentum conservation internally for any sub-diagram since all vertices must satisfy momentum conservation. Additionally we make sure to define all sub-diagrams to follow the same momenta $p_1^i$, which is always a possible choice and avoids complicated mixing. We may conclude that all possible permutations of the chequerboard-like diagram are included. The remaining question is if there are other diagrams that are possibly leading or of the same order, of which we argue the only remaining option mixing the eikonal ladders of separate particles instead of treating each eikonal interaction as a factorized vertex.

Let us first comment on the two other scenarios. The first is also allowing interactions between two different $p_1^i$ particles. However since $p_1^i\cdot p_1^{i'}=0$ for any $i,i'$ this vertex will always vanish trivially in the leading order limit. The second is to allow for more than one eikonal interaction. If this is done in repetition for the same two particles this would mean connecting two eikonal ladders of loop by another loop. It is for this diagram that we mentioned that the eikonal vertex is not a formal vertex but a graphical method to write sub-diagrams: Combining two eikonal ladders this way is nonsensical, since, when written out in terms of propagators, it contributes to the same ladder in the end.

\begin{figure}
    \centering
    \includegraphics[width=0.8\textwidth]{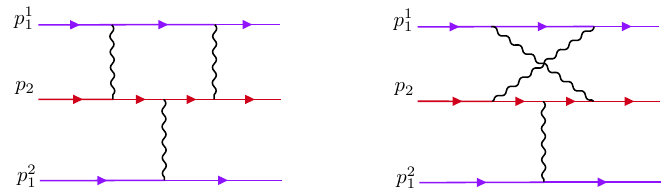}
    \caption{A diagram of $1+2\to 1+2$ scattering, containing the simplest possibility of corrections to the $S$-matrix. All different combinatorial ways of drawing it in the eikonal phase have been shown.}
    \label{fig:eikonal correction}
\end{figure}

Finally, we must consider the remaining option of mixing the eikonal ladders. The most simple diagram of such type one can construct is shown in \figref{fig:eikonal correction}. In the eikonal gauge this amplitude is proportional to
\begin{gather}
\begin{aligned}
    \sim  \gamma^6(p_1^1)^4 (p_2)^6(p_1^2)^2 &\int \ed^2 k \left(\frac{1}{(k-p_1^1)^2-i\epsilon}+\frac{1}{(k+p_1^{\prime 1})^2-i\epsilon}\right)\\
    &\times \left(\frac{1}{(k+p_2)^2-i\epsilon}\frac{1}{(k+p_2+q_1^1)^2-i\epsilon}+\frac{1}{(k-p_2)^2-i\epsilon}\frac{1}{(k-p_2+q_1^1)^2-i\epsilon}\right)
    \end{aligned}
\end{gather}
where we used $q_1^1=-q_1^2$ and $q_1^i=p_1^{\prime i}-p_1^i$. For the first part we can use the eikonal approximation on the propagator terms, neglecting $q_1^1$ to find
\begin{align}
    \frac{1}{(k-p_1^1)^2-i\epsilon}+\frac{1}{(k+p_1^{\prime 1})^2-i\epsilon}\sim \frac{1}{p_1^1}\delta(k_y).
\end{align}
Inserting this gives
\begin{align}
    \sim  \gamma^6(p_1^1)^3 (p_2)^5(p_1^2)^2 &\int \limits_{-\infty}^{\infty}\ed k  \left(\frac{1}{-k-i\epsilon}\frac{1}{-k -q_1^1-i\epsilon}+\frac{1}{k-i\epsilon}\frac{1}{k +q_1^1 -i\epsilon}\right).
\end{align}
We may simply evaluate this integral, since the $i\epsilon$ ensures there are no poles on the real line. We find
\begin{align}
    &\sim  \gamma^6(p_1^1)^3 (p_2)^5(p_1^2)^2 \left(\log\left(\frac{k-q_1^1+i\epsilon}{k+q_1^1-i\epsilon}\right)+\log\left(\frac{k-i\epsilon}{k+i\epsilon}\right)\right)\biggr\rvert^{\infty}_{-\infty}\\
    &= 0.
\end{align}
We thus find that this diagram vanishes exactly in the leading order approximations. Naturally when we consider sub-leading terms the diagram will yield non-zero results, however we may conclude that these are sub-leading in the eikonal phase. Thus the $S$-matrix in \eqref{eq:mnmnsmatrix} is the most general elastic amplitude in the eikonal phase.

\clearpage

\section{Conclusion and Outlook}
In this article we have calculated all possible elastic diagrams with external scalar particles in the (black hole) eikonal phase. To do so we have used the field theory developed in \cite{Toolbox} where scalar particles scatter by exchange of a linear covariant graviton mode $h_{\mu\nu}$. The usual issues with such a theory of quantum gravity are circumvented in the proceeding calculations, by working either at tree level, or within the black hole eikonal phase. The eikonal phase on the black hole is defined by $E\gg\frac{\mpl}{\mbh}\mpl$, so that for large enough black holes the energy conditions are satisfied trivially. This implies that the eikonal phase for black holes should be widely applicable. Additionally, we formulated Feynman rules on flat space in spherical harmonics at a fixed radius $\rfs$, to compare with literature.\\
\\
The important addition of this article compared to previous literature (\cite{ShortPaper,LongPaper,SQEDpaper}) is the fact that we resum over all partial waves $\ell m$. Already at tree level a measure of separation enters into the amplitude, where as expected particles closer to each other start interacting more strongly. Notably, the kinematical phase space trivializes into two configurations, of which one is strongly leading in the black hole eikonal phase.\\
\\
In \secref{chap: eikonal} we extended the resummation over partial waves to the eikonal summation of ladder graphs for $2\to 2$ scattering. The resulting amplitude has the familiar form of an eikonal amplitude, showing that the two particles interact with small longitudinal momentum exchange and without change in transverse separation. The transverse structure matches with eikonal amplitudes in literature \cite{KabatOrtiz,RaclariuEikonalCFT,Cornalba2007two}, where the transverse distance is measured by the Laplacian Green's function and integrated over. \\
\\
We originally expected to match the semi-classical results of 't Hooft for the black hole with this $2\to 2$ scattering graph, however the integration over transverse separation was incorrectly placed. In the next section (\ref{chap: MN to MN}), we concluded by extending the familiar $2\to 2$ eikonal graph to arbitrarily many particles, which completely agreed with 't Hooft's result. The newly developed diagram in \secref{chap: MN to MN} is the most general elastic amplitude one can calculate in the eikonal phase, both on the black hole and in flat space. Since this amplitude agrees with 't Hooft's result, we may conclude that his S-matrix is also as general as may be achieved within the semi-classical regime. Additionally, the S-matrix is found explicitly to be unitary. \\
\\
While this is so by construction, it motivates the idea that information exchange over interactions may be crucial. Furthermore, 't Hooft's original conclusions only considered gravitational interactions, with comments on electrodynamics, we have managed to include these as well. As expected, the contribution of electrodynamics is sub-leading in the energy, but it allows for distinction between otherwise identical particles based on their charge. Within our field theory it is easy to add many types of different particles and interactions as desired.\\
\\
We remark that in the eikonal limit all interactions were mediated by soft gauge fields, avoiding any issues in the problematic UV regime. Within the harmonics basis we also avoided infrared divergences, since the presence of the black hole and the angular momentum $\ell$ introduced natural regulators. This allowed us to perform a resummation in graviton loops over all orders of $G$.\\
\\
Furthermore, we developed an analogous field theory on flat space in spherical harmonics, to compare with literature. Indeed in \secref{sec:flat space comp} we find a match with the familiar $2\to 2$ eikonal amplitude in flat space \cite{KabatOrtiz}. We may conclude that the method of resummation over partial waves within field theory provides correct results at least within the eikonal phase. Most likely this applies to complete generality, however the approximation of constant $r=\rfs$ restricts only to certain types of scattering. Of course for flat space such a harmonics theory is overcomplicated, but for the Schwarzschild black hole it proved to be a great tool to separate the radial curvature from the symmetric angular regions.\\
\\
There are naturally a number of shortcomings. The first one that appears is the horizon approximation. There are numerous arguments for this approximation; particles' energy increases exponentially so interactions are at their strongest, a general interest in behaviour at the horizon because the bulk is well understood and simply an attempt at approximating the system to our region of interest to allow as many calculations as possible. However, we would have preferred to loosen this approximation, or remove it at all. For certain modes (low values of $\ell$, only traceless $\tilde{h}_{ab}$) this may be possible. \\
Furthermore, it is impossible to check if our eikonal amplitudes are truly leading. It has been shown that the eikonal amplitude is not leading for all types of particles \cite{LevySucher,Tiktopoulos1971,Eichten1971}. However because we are considering gravitational interactions it is not possible to calculate other loops to compare.\\
Finally, most results are proofs of concept. We have shown that many amplitudes or systems exhibit the desired or expected behaviour, but without proofs whether the amplitudes are actually leading, or numerical matches. Since our aim was exactly to show that the use of perturbative quantum gravity may still yield a large amount of interesting and consequential results, we did not put our focus on certain details. However, this may be interpreted as that our methods still need further proof or details in order to be validated.\\
\\
The aim of this article was to investigate the application of canonical field theory to interactions on black holes to complete elastic generality. We can now safely say that even for an arbitrary number of particles these amplitudes may be calculated safely without divergent problems. The resulting equations hint that already based on general relativity and quantum field theory alone, we might find interesting behaviour and possibly solutions for fundamental black hole problems by considering the complex system of interactions.\\
\\
Since our elastic amplitude holds for any number of particles, we conclude that we have achieved the most general elastic amplitude possible within the (black hole) eikonal phase. For future work extending inelastic amplitudes (like in \cite{2to2N}) to complete generality, and combining this with properly defined asymptotic states and Hawking radiation, we hope that a resolution to the information paradox may be achieved within the current paradigm.

\section*{Acknowledgements}
We are grateful to Nava Gaddam, Gerard 't Hooft and Fabiano Feleppa for many discussions and insights that contributed to this research. Additionally we thank Mick van Vliet for proofreading this article.

\clearpage
\appendix

\section{Appendix}

\subsection{Conventions}
\label{chap: background}
We will generally work on the Schwarzschild background, defined by
\begin{align}
    \eta_{ab}&=\begin{pmatrix}
    0 & -1 \\ -1 & 0
    \end{pmatrix},\\
    \gamma_{AB}&=\begin{pmatrix}
    1 & 0 \\ 0 & \sin^2\theta
    \end{pmatrix},
\end{align}
in terms of which the full metric is given by
\begin{align}
    \ed s^2=f(r)\eta_{ab}\ed x^a \ed x^b+r^2\gamma_{AB}\ed x^A\ed x^B,
\end{align}
We will use the antisymmetric Levi-Civita tensor. On the angular coordinates we define it by
\begin{align}
\label{eq:epslower}
\epsilon_{AB}= r^2 \sqrt{\gamma}\begin{pmatrix}
0 & 1 \\
-1 & 0
\end{pmatrix},
\end{align}
As mentioned we work in Kruskal-Szekeres coordinates. The main reason for this choice of coordinates is that it describes the entirety of the Schwarzschild Spacetime, and it is regular on the horizon. This last property is important for us to be able to define a stable field theory. We have employed coordinates $x,y$ such that
\begin{align}
    xy&=2R^2\left(1-\frac{r}{R}\right)e^{\frac{r}{R}-1},\\
    x/y&=\text{sgn}\left(1-\frac{r}{R}\right)e^{2\tau} \lgap\lgap \tau=\frac{t}{2R},\\
    f(r)&=\frac{R}{r}e^{1-\frac{r}{R}},
\end{align}
where $R$ is the Schwarzschild radius and $\mu=1/R$ the inverse Schwarzschild radius. In the original Schwarzschild coordinates there was explicit time translation symmetry $t\to t+a$. In the Kruskal-Szekeres coordinates this becomes
\begin{align}
    \text{translation} & : \qquad~  x\to a x & y\to \frac{y}{a}.
\end{align}
This restricts all physical results to have an equal contribution from the $x$ and $y$ coordinates, in order to be translation invariant. This can be used as a tool to check the validity of equations or results, and sometimes for physical reasoning. Raising and lowering is in principle performed with the metric $g_{AB}=r^2\gamma_{AB}$. For commutation on covariant derivatives we use
\begin{align}
    [\nabla_{\mu},\nabla_{\nu}]T^{\rho_1 ... \rho_n}_{\sgap\sgap\sgap \sigma_1 ... \sigma_n}&=\sum_iR^{\rho_i}_{\sgap \bar{\rho}_i \mu\nu}T^{ ...\bar{\rho}_i ...}_{\sgap\sgap\sgap \sigma_1 ... \sigma_n}-\sum_jR^{\bar{\sigma}_j}_{\sgap \sigma_j \mu\nu}T^{\rho_1 ... \rho_n}_{\sgap\sgap\sgap... \bar{\sigma}_j ...},
\end{align}
where the Riemann tensor is defined by
\begin{align}
R^{\rho}_{\gap\mu\sigma\nu} ~ = ~ \p_{\sigma}\Gamma^{\rho}_{\mu\nu}-\p_{\nu}\Gamma^{\rho}_{\mu\sigma}+\Gamma^{\rho}_{\sigma\kappa}\Gamma^{\kappa}_{\mu\nu}-\Gamma^{\rho}_{\mu\kappa}\Gamma^{\kappa}_{\sigma\nu} \, .
\end{align}
The Ricci tensor is then given by
\begin{align}
R_{\mu\nu}=R^{\rho}_{\gap\mu\rho\nu}.
\end{align}

\subsection{Spherical harmonics}
\label{app: spherical harmonics}
Throughout this thesis we will be working in a spherical harmonics basis. In this appendix we define our convention, and write down all useful equations and definitions. For the complex convention we will use
\begin{align}
    Y_{\ell}^m(\theta,\phi)=\sqrt{\frac{2\ell+1}{4\pi}\frac{(\ell-m)!}{(\ell+m)!}}e^{im\phi}P_{\ell}^m(\cos\theta),
\end{align}
where $P_{\ell}^m(x)$ are the Associated Legendre Polynomials. We will denote the argument using the solid angle $\Omega={\theta,\phi}$ for compact notation. To avoid having to distinguish between real and complex fields, and running into problems with the vertices, we will use real spherical harmonics defined by
\begin{align}
    Y_{\ell m}=\begin{cases}
         \frac{i}{\sqrt{2}}\left(Y_{\ell}^m-(-1)^m Y_{\ell}^{-m}\right) &m<0\,, \\ Y_{\ell}^0 & m=0\,,\\
         \frac{1}{\sqrt{2}}\left(Y_{\ell}^{-m}+(-1)^m Y_{\ell}^{m}\right) &m>0\,,
    \end{cases}
\end{align}
where the careful alternation in $m$ is to ensure the same orthogonality conditions hold:
\begin{align}
    \int\ed\Omega ~ Y_{\ell m}(\Omega) Y_{\ell' m'}(\Omega)=\delta_{\ell \ell'}\delta_{mm'}.
\end{align}
In reverse the harmonics obey the following delta-identity
\begin{align}
    \sum\limits_{\ell m} Y_{\ell m}(\Omega)Y_{\ell m}(\Omega')=\delta^{(2)}(\Omega-\Omega')\,,
\end{align}
where $\delta^{(2)}(\Omega-\Omega')$ is defined including the inverse Jacobian. The advantage of using the real definition is subtle. For complex fields using complex harmonics the quadratic action for example would diagonalize over $\ell m$, whereas for real fields using complex harmonics one field would be at $+m$ and one at $-m$. We could then use real harmonics for the real fields only, but this would greatly increase the amount of notation needed for the vertices. The easiest most compact solution was to use the real definition everywhere.
Our harmonics vectors are defined by
\begin{align}
    \eta^+_{\ell m}&=\p_A Y_{\ell m}\,,\\
    \eta^-_{\ell m}&=-\epsilon_{AB}\p^B Y_{\ell m}\,.
\end{align}
The eigenvalue equations are given by
\begin{align}
    \Delta_{\Omega}Y_{\ell m}&= -\ell(\ell+1)Y_{\ell m}\,,\\
    \Delta_{\Omega} \eta^{\pm}_{A,\ell m}&=-(\ell(\ell+1)+1)\eta^{\pm}_{A,\ell m}\,,
\end{align}
where the vector eigenvalue receives a $+1$ due to the commutation of derivatives. Finally we have many different types of higher order couplings (more than two harmonics). These do not decouple. In principle any order higher than three can be written in terms of the third-order coupling, but this is not necessarily useful. The most simple coupling is given by
\begin{align}
    CL[\ell_1m_1,\ell_2m_2,\ell_3m_3]=\int\ed\Omega ~ Y_{\ell_1m_1}Y_{\ell_2m_2}Y_{\ell_3m_3}\,,
\end{align}
which is proportional to Clebsch-Gordan coefficients - hence the name - but not identical. When including derivatives we obtain an even parity coupling symmetric in the first two inputs
\begin{align}
    CL_+[\ell_1m_1,\ell_2m_2;\ell_3m_3]&=\int\ed\Omega ~ \p^AY_{\ell_1m_1}\p_AY_{\ell_2m_2}Y_{\ell_3m_3}\,,\\
    &=-\frac{1}{2}(\ell_3(\ell_3+1)-\ell_1(\ell_1+1)-\ell_2(\ell_2+1))CL[\ell_1m_1,\ell_2m_2,\ell_3m_3]\,,
\end{align}
and an odd parity coupling antisymmetric in the first two inputs
\begin{align}
    CL_-[\ell_1m_1,\ell_2m_2;\ell_3m_3]&=\int\ed\Omega ~ \epsilon^{AB}\p_AY_{\ell_1m_1}\p_BY_{\ell_2m_2}Y_{\ell_3m_3}.
\end{align}
For four derivatives we have the following, also symmetric in the first two inputs:
\begin{align}
    CL_{2+}[\ell_1m_1,\ell_2m_2;\ell_3m_3]&=\int\ed\Omega ~ \p^BY_{\ell_1m_1}\p_AY_{\ell_2m_2}\p^A\p_BY_{\ell_3m_3}\,,\\
    &=\frac{1}{4}((\ell_3(\ell_3+1))^2-(\ell_1(\ell_1+1)+\ell_2(\ell_2+1))^2))CL[\ell_1m_1,\ell_2m_2,\ell_3m_3].
\end{align}
Finally two definitions that are only used for compact notation:
\begin{align}
    CL_{2-}[\ell_2 m_2,\ell_3m_3;\ell_1m_1]&=\int\ed\Omega~ \nabla_A \eta^-_{B,\ell_1m_1}\p^A Y_{\ell_2m_2}\p^B Y_{\ell_3m_3}\,,\\
    CL_G[\ell_2m_2,\ell_3m_3;\ell_1m_1]&=CL_{2+}[\ell_2m_2,\ell_3m_3;\ell_1m_1]-\tfrac{1}{2}\ell_1(\ell_1+1)CL_+[\ell_2m_2,\ell_3m_3;\ell_1m_1]\,.
\end{align}
For the even parity couplings one can simplify higher derivative couplings into just $CL$ using
\begin{align}
   2\p_A \phi_1\p^A\phi_2 =\Delta_{\Omega} (\phi_1\phi_2)-\phi_2\Delta_{\Omega} \phi_1-\phi_1\Delta_{\Omega} \phi_2
\end{align}
and integration by parts.

\subsection{Eikonal calculation for \secref{chap: eikonal}}
\label{app: calculations for resum}
We may arbitrarily redefine the loop momenta. We define
\begin{align}
    k_i&\to \tilde{k}_i+b_i\\
    B_i&=\sum\limits_{j=1}^i b_j,
\end{align}
such that the matter propagator denominators become   
\begin{align}
    &(\bar{p}_1+K_i)^2+\mu^2\lambda_{\ell_i}-i\epsilon\nonumber \\
    &=2\left(p_1+\frac{\mu^2\lambda_1}{2s}p_2+B_i\right)\cdot\tilde{K}_i+ \tilde{K}_i^2 +\left(B_i^2+2\bar{p}_1\cdot B_i+\mu^2\lambda_{\ell_i}-\mu^2\lambda_1\right)-i\epsilon,
\end{align}
and an analogous equation for the bottom row. We seek to remove the mass contribution, which can be done by choosing an appropriate value for $b_i$, such that the last term in the equation above drops out exactly. The equations to solve are given by
\begin{align}
    B_i^2+2\bar{p}_1\cdot B_i+\mu^2\lambda_{\ell_i}-\mu^2\lambda_1&=0\,,\\
    \bar{B}_i^2-2\bar{p}_2\cdot \bar{B}_i+ \mu^2\lambda_{\bar{\ell}_{\pi(i)}}-\mu^2\lambda_2&=0\,.
\end{align}
Let us first write $B_i=p_1 \tilde{B}^i_x+p_2 \tilde{B}^i_y$ to extract the momenta from the parametrization. Then we can write
\begin{align}
    -2s \tilde{B}^i_x\tilde{B}^i_y-2s\tilde{B}^i_y-\mu^2\lambda_1\tilde{B}^i_x +\mu^2\lambda_{\ell_i} -\mu^2\lambda_1&=0\,,\\
    -2s \bar{\tilde{B}}^i_x\bar{\tilde{B}}^i_y+2s\bar{\tilde{B}}^i_x+\mu^2\lambda_2\bar{\tilde{B}}^i_y + \mu^2\lambda_{\bar{\ell}_{\pi(i)}} -\mu^2\lambda_2&=0\,.
\end{align}
Now suppose $\tilde{B}^i_{x,y}$ is of the form 
\begin{align}
    B^i_a=\sum\limits_{n=-\infty}^N B^i_{an} s^n\,,
\end{align}
where we can assume the same upper limit for both coordinates because of the time translation symmetry. Subsequently a quick straightforward order analysis shows that only for $N=0$ non-trivial solutions may be possible ($B_i^n\neq 0$), and specifically for the equation above this only starts to happen at $N=-1$. We can find for the leading order equations that
\begin{align}
    -2s^{-1} \tilde{B}^i_{x,-1}\tilde{B}^i_{y,-1}-2\tilde{B}^i_{y,-1}-\mu^2\lambda_1s^{-1}\tilde{B}^i_{x,-1} +\mu^2\lambda_{\ell_i} -\mu^2\lambda_1&=0\\
    -2s^{-1} \bar{\tilde{B}}^i_{x,-1}\bar{\tilde{B}}^i_{y,-1}+2\bar{\tilde{B}}^i_{x,-1}+\mu^2\lambda_2 s^{-1}\bar{\tilde{B}}^i_{y,-1} + \mu^2\lambda_{\bar{\ell}_{\pi(i)}} -\mu^2\lambda_2&=0.
\end{align}
For both equations, the first and third terms are an order lower in $s$ and so they can be dropped (they will give couplings to the $n=-2$ coefficients). Then the solution can immediately be read off from the remainder
\begin{align}
    \tilde{B}^i_{y,-1}&=\frac{1}{2}\mu^2(\lambda_{\ell_i} -\lambda_1)\,,\\
    \bar{\tilde{B}}^i_{x,-1}&=-\frac{1}{2}\mu^2(\lambda_{\bar{\ell}_{\pi(i)}} -\lambda_2)\,.
\end{align}
Since we will only look at highest order we can write the full solution as given by the leading term
\begin{align}
    \tilde{B}^i_y&=\frac{\mu^2}{2s}(\lambda_{\ell_i} -\lambda_1),\\
    \tilde{B}^i_x&=-\frac{\mu^2}{2s}(\lambda_{\bar{\ell}_{\pi(i)}} -\lambda_2).
\end{align}
Let us now define two vectors that contain all loop momenta scaled by $p_1,p_2$:
\begin{align}
    p_1\vec{\beta}_x=\begin{pmatrix}
         b_1^x \\  b_2^x \\  b_3^x \\ \vdots
    \end{pmatrix}, && p_2\vec{\beta}_y=\begin{pmatrix}
         b_1^y \\  b_2^y \\  b_3^y \\ \vdots
    \end{pmatrix} .
\end{align}
We can write $\tilde{B}$ in terms of these using a lower-triangular matrix $T$ that captures the summation and a permutation matrix $\Pi$ corresponding to $\pi(i)$ such that we find
\begin{align}
    T\vec{\beta}_y&= \frac{\mu^2}{2s}\vec{\Lambda}_,,\\
    T\Pi \vec{\beta}_x&= -\frac{\mu^2}{2s}\Pi\vec{\Lambda}_,.
\end{align}
Here we defined
\begin{align}
    \vec{\Lambda}_1 = \begin{pmatrix}
        \lambda_{\ell_1}-\lambda_1 \\ 
        \lambda_{\ell_2}-\lambda_1 \\ 
        \vdots
    \end{pmatrix}, && \vec{\Lambda}_2 = \begin{pmatrix}
        \lambda_{\bar{\ell}_1}-\lambda_2 \\ 
        \lambda_{\bar{\ell}_2}-\lambda_2 \\ 
        \vdots
    \end{pmatrix}.
\end{align}
These are easily inverted to find the solution for the original transformation functions $b_i$ giving
\begin{align}
    \vec{b}=\frac{\mu^2}{2s}\left(-p_1 \Pi^{-1}T^{-1}\Pi \vec{\Lambda}_2+p_2 T^{-1}\vec{\Lambda}_1\right),
\end{align}
where $\vec{b}=(b_1,b_2,\dots)$ is the vector version of $b_i$. This gives an exact expression for $b_i$, importantly showcasing that it is of the same order as the momentum exchange $q=p_3-p_1$. Using the inverse of the triangular matrix we can write this for $b_i$ as
\begin{align}
    b_i=\frac{\mu^2}{2s}\left(-p_1 (\lambda_{\bar{\ell}_{\pi(i)}}-\lambda_{\bar{\ell}_{\pi(i-1)}})+p_2 (\lambda_{\ell_i}-\lambda_{\ell_{i-1}})\right),
\end{align}
where as initial condition $\lambda_{\ell_0}=\lambda_1,\lambda_{\bar{\ell}_{\pi(0)}}=\lambda_2$. However similarly by extension on the other edge we have that $\lambda_{\ell_n}=\lambda_3,\lambda_{\bar{\ell}_{\pi(n)}}=\lambda_4$ since there are only $n-1$ matter legs and so the $n^{\text{th}}$ takes on the value of the external leg.\\
Summing this up gives in total for $B_i$ that
\begin{align}
    B_i=\frac{\mu^2}{2s}\left(p_1 (\lambda_2-\lambda_{\bar{\ell}_{\pi(i)}})+p_2 (\lambda_{\ell_i}-\lambda_{1})\right),
\end{align}
and in particular
\begin{align}
    B_n=\frac{\mu^2}{2s}\bigr(p_1 (\lambda_{2}-\lambda_{4})+p_2 (\lambda_{3}-\lambda_{1})\bigr)=q.
\end{align}
This value coincides exactly with the momentum exchange as given in \eqref{eq:momentum exchange}, indicating that the transformation simply takes into account the fact that mass has to be exchanged. We expect that exact solutions are in principle also possible but overly complicated, especially for our purpose where we are looking in the eikonal regime.

\subsubsection{Spherical harmonics rewriting}
First rearrange the expression into groups of identical indices:
\begin{gather}
\begin{aligned}
    \int\prod\limits_{i=1}^{n}&\bigr(\ed\Omega_i\ed\bar{\Omega}_i\bigr)Y_{\ell_{0} m_{0}}(\Omega_1) Y_{\ell_{n}m_{n} }(\Omega_{n}) Y_{\bar{\ell}_{0} \bar{m}_{0}}(\bar{\Omega}_1) Y_{\bar{\ell}_{n}\bar{m}_{n}}(\bar{\Omega}_{n})\\
    &\times \prod\limits_{i=1}^{n+1}\biggr( Y_{L_i M_i}(\Omega_i)Y_{L_{\pi(i)} M_{\pi(i)}}(\bar{\Omega}_i) \biggr)\\
    \label{eq:mattersums}
    &\times \prod\limits_{i=1}^n\left(\sum\limits_{\ell_i m_i} Y_{\ell_{i} m_{i}}(\Omega_i) Y_{\ell_{i}m_i }(\Omega_{i+1}) \sum\limits_{\bar{\ell}_i\bar{\ell}_i}Y_{\bar{\ell}_{i} \bar{m}_{i}}(\bar{\Omega}_i) Y_{\bar{\ell}_{i}\bar{m}_i}(\bar{\Omega}_{i+1})\right).
\end{aligned}
\end{gather}
The last line contains separate sums over harmonics with identical indices, so that we can identify these as many different delta functions using the identity in \appref{app: spherical harmonics} giving
\begin{align}
    \prod\limits_{i=1}^{n-1}\left(\sum\limits_{\ell_i m_i} Y_{\ell_{i} m_{i}}(\Omega_i) Y_{\ell_{i}m_i }(\Omega_{i+1}) \sum\limits_{\bar{\ell}_i\bar{\ell}_i}Y_{\bar{\ell}_{i} \bar{m}_{i}}(\bar{\Omega}_i) Y_{\bar{\ell}_{i}\bar{m}_i}(\bar{\Omega}_{i+1})\right) = \prod\limits_{i=1}^{n-1}\biggr(\delta^{(2)}(\Omega_i,\Omega_{i+1})\delta^{(2)}(\bar{\Omega}_i,\bar{\Omega}_{i+1})\biggr).
\end{align}
We see that all $\Omega_i$ can be integrated out exactly, in the sense that first $\delta^{(2)}(\Omega_{n},\Omega_{n-1})$ removes the $\Omega_{n}$ integral setting $\Omega_{n}\to\Omega_{n-1}$, and then the $\delta(\Omega_{n-1},\Omega_{n-2})$ does so again, until in the end only the integral over $\Omega_1$ remains (the product stops at $i=1$). A similar thing happens to all $\bar{\Omega}$'s. This greatly simplifies the harmonics into
\begin{gather}
    \int\ed\Omega_1\ed\bar{\Omega}_1Y_{\ell_{0} m_{0}}(\Omega_1) Y_{\ell_{n}m_{n} }(\Omega_1) Y_{\bar{\ell}_{0} \bar{m}_{0}}(\bar{\Omega}_1) Y_{\bar{\ell}_{n}\bar{m}_{n}}(\bar{\Omega}_1) \times \prod\limits_{i=1}^{n}\biggr( Y_{L_i M_i}(\Omega_1)Y_{L_{\pi(i)} M_{\pi(i)}}(\bar{\Omega}_1) \biggr).
\end{gather}
The first four harmonics depend on the initial values, since the $0^{\text{th}}$ component corresponds to particles $1,2$ and the $n^{\text{th}}$ component corresponds to $3,4$ so we can identify the initial value function in the tree level calculation $Y_{IV}(\Omega,\bar{\Omega}) := Y_{\ell_{1} m_{1}}(\Omega) Y_{\ell_{3}m_{3} }(\Omega) Y_{\ell_{2} m_{2}}(\bar{\Omega})Y_{\ell_4m_4}(\bar{\Omega})
$. Then, upon redefining $\Omega_1\to\Omega,\bar{\Omega}_1\to\bar{\Omega}$, since these are the only ones left, we end up with 
\begin{align}
    \int\ed\Omega\ed\bar{\Omega}Y_{IV}(\Omega,\bar{\Omega}) \times \prod\limits_{i=1}^{n}\biggr( Y_{L_i M_i}(\Omega)Y_{L_{\pi(i)} M_{\pi(i)}}(\bar{\Omega}) \biggr).
\end{align}
Since all $Y_{L_{\pi(i)} M_{\pi(i)}}(\bar{\Omega})$ depend on the same $\bar{\Omega}$ the order is not important, and specifically we can reorder them to remove the permutation as desired:
\begin{align}
    \int\ed\Omega\ed\bar{\Omega}Y_{IV}(\Omega,\bar{\Omega}) \times \prod\limits_{i=1}^{n}\biggr( Y_{L_i M_i}(\Omega)Y_{L_{i} M_{i}}(\bar{\Omega}) \biggr).
\end{align}
This results in the following expression for the amplitude:
\begin{align}
    i\mathcal{M}_{n-1}&= -\left( -\frac{is^2\gamma^2}{\mu^2}\right)^n\int\ed\Omega\ed\bar{\Omega}Y_{IV}(\Omega,\bar{\Omega})\left(\sum\limits_{LM}\frac{1}{\lambda_L}Y_{L M}(\Omega)Y_{L M}(\bar{\Omega})\right)^n\nonumber\\
    &\times \prod\limits_{i=1}^n\left(\int\frac{\ed ^2k_i}{(2\pi^2)}\right)(2\pi)^2\delta^{(2)}(K_n)\sum\limits_{\pi}\prod\limits_{i=1}^{n-1}\left(I^{eik}_i\bar{I}^{eik}_i\right),
\end{align}
where $Y_{IV}$ is the same function as for the tree-level case. Note that we can now recognize the $G_1(\Omega,\bar{\Omega})$ Green's function.\\
Finally we must sum over the remaining permutations, however after all performed rewritings, only the matter propagators still depend on the permutation. This precise summation has been done by \cite{LevySucher}, and we find
\begin{align}
    \prod\limits_{i=1}^n\left(\int\frac{\ed ^2k_i}{(2\pi^2)}\right)&(2\pi)^2\delta^{(2)}(K_n)\sum\limits_{\pi}\prod\limits_{i=1}^{n-1}\left(I^{eik}_i\bar{I}^{eik}_i\right)=\frac{1}{n!}\left(-\frac{1}{2s}\right)^{n-1},
\end{align}
which may be inserted back into the expression.

\subsection{\texorpdfstring{$M+1\to M+1$}{M+1->M+1} calculation for \secref{sec:m1tom1}}
\label{app:m1tom1}
In this appendix we provide the calculation for the propagator contribution to the $M+1\to M+1$ diagram:
\begin{align}
    &\frac{-i}{2p_1^i\cdot \left(k_2^{\pi(1)}-k_2^{\prime\pi(1)}\right)-i\epsilon}\times\dots\times  \frac{-i}{2p_1^i\cdot \sum\limits_{n=1}^{r_2-1}\left(k_2^{\pi(n)}-k_2^{\prime\pi(n)}\right)-i\epsilon}\\
     \times &\frac{-i}{-2p_1^i\cdot \sum\limits_{n=r_2}^{M-1}\left(k_2^{\pi(n)}-k_2^{\prime\pi(n)}\right)-i\epsilon}\times\dots\times\frac{-i}{-2p_1^i\cdot \left(k_2^{\pi(M-1)}-k_2^{\prime\pi(M-1)}\right)-i\epsilon}.
\end{align}
To calculate the expression above we proceed as follows:
\begin{itemize}
    \item The starting point is the arbitrary location of insertion of $p_2^M$ at location $r_2$. This gives $M$ options to sum over.
    \item The set $s_2$ that contains $r_2-1$ elements of all outgoing momenta $p_2^j$. Since there are in total $M-1$ momenta to choose from, the set $s_2$ has 
    $$
    \frac{(M-1)!}{(M-r_2)!(r_2-1)!}
    $$
    different possible unordered options of choosing the momenta, and we must sum over all. We note the elements of $s_2$ by a capital letter $K_2^j$. The complement $\bar{s}_2$ is then automatically fixed as well, as the set of the remaining outgoing momenta that are not in $s_2$, denoted by $\bar{K}_2^j$\footnote{We still index the elements in the range $r_2 \to M-1$ to maintain the same visual structure in the equations as before.}. 
    \item We must still sum over all permutations. Given a certain $s_2$ we define all possible permutations over the set $s_2$ by $\pi_2$. This means that $\pi_2$ contains
    $$
    (r_2-1)!
    $$
    possible permutations. $\pi_2'$ is the permutation for the complement set $\bar{s}_2$ with $(M-r_2)!$ elements. Note that the $\pi_2,\pi'_2$ are the two permutation subgroups that together form $\pi$
\end{itemize}
The next step is to sum over all of these sets, to find the most general result. Since the vertices are permutation independent, we can immediately proceed to sum only the propagators, giving
\begin{gather}
\begin{aligned}
    \sum\limits_{r_2=1}^M\sum\limits_{s_2}&\sum\limits_{\pi_2}\frac{-i}{2p_1^i\cdot \left(K_2^{\pi_2(1)}-K_2^{\prime\pi_2(1)}\right)-i\epsilon}\times\dots\times  \frac{-i}{2p_1^i\cdot \sum\limits_{n=1}^{r_2-1}\left(K_2^{\pi_2(n)}-K_2^{\prime\pi_2(n)}\right)-i\epsilon}\\
    \times&\sum\limits_{\pi_2'}  \frac{-i}{-2p_1^i\cdot \sum\limits_{n=r_2}^{M-1}\left(\bar{K}_2^{\pi_2'(n)}-\bar{K}_2^{\prime\pi_2'(n)}\right)-i\epsilon}\times\dots\times\frac{-i}{-2p_1^i\cdot \left(\bar{K}_2^{\pi_2'(M-1)}-\bar{K}_2^{\prime\pi
    _2'(M-1)}\right)-i\epsilon}.
    \end{aligned}
\end{gather}
Notably there is no sum over the complement set $s_{2}$ since it is automatically determined by the unbarred. We can use the permutation identity defined in \cite{LevySucher}
\begin{align}
\label{eq:perm}
    \sum\limits_{\pi}\frac{1}{A_{\pi(1)}}&\frac{1}{A_{\pi(1)}+A_{\pi(2)}}\frac{1}{A_{\pi(1)}+A_{\pi(2)}+A_{\pi(3)}}...\frac{1}{A_{\pi(1)}+...+A_{\pi(N)}}\\
    &=\frac{1}{A_1A_2A_3...A_N}
\end{align}
to remove all $\pi_2,\pi_2'$ immediately giving just
\begin{align}
    \sum\limits_{r_2=1}^M\sum\limits_{s_2}&\frac{-i}{2p_1^i\cdot \left(K_2^{1}-K_2^{1}\right)-i\epsilon}\times\dots\times  \frac{-i}{2p_1^i\cdot \left(K_2^{r_2-1}-K_2^{r_2-1}\right)-i\epsilon}\\
    \times&\frac{-i}{-2p_1^i\cdot \left(\bar{K}_2^{r_2}-\bar{K}_2^{r_2}\right)-i\epsilon}\times\dots\times\frac{-i}{-2p_1^i\cdot \left(\bar{K}_2^{M-1}-\bar{K}_2^{M-1}\right)-i\epsilon}.
\end{align}
Notably all momenta appear once in a permutation-invariant fashion, however whether they are part of $s_2$ or $\bar{s}_2$ determines the sign with which they appear. To rewrite the expression above, we first define
\begin{align}
    \alpha^{i,j}&=i \left(2p_1^i\cdot \left(K_2^{j}-K_2^{j}\right)-i\epsilon\right),\\
    \beta^{i,j}&= i\left(-2p_1^i\cdot \left(\bar{K}_2^{j}-\bar{K}_2^{j}\right)-i\epsilon\right).
\end{align}
The expression can then be written as
\begin{align}
    &\sum\limits_{r_2=1}^M \sum\limits_{s_2}(\alpha^{i,1}  \dots \alpha^{i,r_2-1})^{-1}(\beta^{i,r_2}\dots \beta^{i,M-1})^{-1}.
\end{align}
We now proceed to sum over $s_2$.

\subsubsection{Combinatorics problem}
This is in principle a general combinatorics problem and we will treat it as such. Given a certain $k_2^j$ for some $j$, we know that it's corresponding momentum must enter either into $s_2$ or $\bar{s}_2$. This means that it must appear either as $\alpha^{i,j}$ or $\beta^{i,j}$, but never both, or never twice. Since the expression involves a summation, this means that whatever it corresponds to must be linear in $(\alpha^{i,j})^{-1},(\beta^{i,j})^{-1}$
\begin{align}
    \sim a \frac{1}{\alpha^{i,j}}+b\frac{1}{\beta^{i,j}}.
\end{align}
The total amount of possibilities where $k_2^j$ is in $s_2$ is given by
\begin{align}
    \frac{(M-2)!}{(r_2-1)!(M-r_2-1)!},
\end{align}
when $1\le r_2< M$, while the number of possibilities where $k_2^j$ is in $\bar{s}_2$ instead is given by
\begin{align}
    \frac{(M-2)!}{(r_2-2)!(M-r_2)!},
\end{align}
when $ 2< r_2\le M $. Summing this over all $r_2$ shows that in both cases the total amount of occurrences is given by
\begin{align}
    2^{M-2},
\end{align}
so both must occur an equal amount of times. This means that we not only require linearity, but also symmetric linearity, restricting to
\begin{align}
    \sim a \left(\frac{1}{\alpha^{i,j}}+\frac{1}{\beta^{i,j}}\right).
\end{align}
Finally, taking the product over all $j$ gives
\begin{align}
    &\sum\limits_{r_2=1}^M \sum\limits_{s_2}(\alpha^{i,1}  \dots \alpha^{i,r_2-1})^{-1}(\beta^{i,r_2+1}\dots \beta^{i,M})^{-1}=\mathcal{N}\prod\limits_{j=1}^{M-1}\left(\frac{1}{\alpha^{i,j}}+\frac{1}{\beta^{i,j}}\right),
\end{align}
where $\mathcal{N}$ is a constant that is not a priori fixed by the requirement of symmetric linearity in all variables. In order to fix it we evaluate the expression when all variables are equal to one, reducing to
\begin{align}
    \sum\limits_{r_2=1}^M \sum\limits_{s_2}=\mathcal{N} 2^{M-1}.
\end{align}
Since we know that $s_2$ has $\frac{(M-1)!}{(M-r_2)!(r_2-1)!}$ options when $r_2>1$, the total contribution on the left hand side is given by
\begin{align}
    \sum\limits_{r_2=1}^M \sum\limits_{s_2}=1+\sum\limits_{r_2=2}^M \frac{(M-1)!}{(M-r_2)!(r_2-1)!}=\sum\limits_{u=0}^M \frac{(M-1)!}{(M-1-u)!u!}=2^{M-1}
\end{align}
as well and we can safely set $\mathcal{N}=1$. The result becomes the following factorization:
\begin{align}   
&\left(\frac{-i}{2p_1^i\cdot \left(K_2^{1}-K_2^{1}\right)-i\epsilon}+\frac{-i}{-2p_1^i\cdot \left(K_2^{1}-K_2^{1}\right)-i\epsilon}\right)\\& \times ...\times \left(\frac{-i}{2p_1^i\cdot \left(K_2^{M-1}-K_2^{M-1}\right)-i\epsilon}+\frac{-i}{-2p_1^i\cdot \left(K_2^{M-1}-K_2^{M-1}\right)-i\epsilon}\right).
\end{align}
Applying the following equation (taken from \cite{KabatOrtiz}):
\begin{align}
    \frac{1}{x+i\epsilon}-\frac{1}{x-i\epsilon}=-2\pi i \delta(x),
\end{align}
this reduces to simple delta-functions:
\begin{align}   
\prod\limits_{j=1}^{M-1}\left(\frac{2\pi}{2p_1^i} \delta\left(k_{y,2}^j-k_{y,2}^{\prime j}\right)\right).
\end{align}
Thus the eikonal amplitude for the $M+1\to M+1$ diagram may be written in total as
\begin{align}
i\mathcal{M}_{\text{sub}}\left(p_1^i,k_{y,2}^j,k_{y,2}^{\prime j}\right)&=(2p_1^i)(2p_2^ M)\delta^{(2)}(\Omega_1^i-\Omega_1^{\prime i})\nonumber\\
    &\times\prod\limits_{j=1}^{M-1}(2\pi)(2p_2^j)\delta\left(k_{y,2}^j-k_{y,2}^{\prime j}\right)\\
    &\times\text{Exp}\left(\frac{i\gamma^2}{2 \mu ^2}\sum \limits_{j=1}^{j=M}s_{ij}G_1(\Omega_1^i,\Omega_2^j)\right).
\end{align}

\subsection{\texorpdfstring{$M+N\to M+N$}{M+N->M+N} calculation for \secref{chap: MN to MN}}
\label{app:mntomn}
In this appendix we calculate the expression given in \eqref{eq:ampfirst}:
\begin{align}
    i \mathcal{M}&=\int\prod_{\substack{i,j=1}}^{\substack{i=N-1\\ j=M-1}}\left(\frac{\ed^2 k^{i,j}}{(2\pi)^2}\right)\prod\limits_{i=1}^N i\mathcal{M}_{\text{sub}}\left(p_1^i,\sum\limits_{n=1}^{i}k_y^{\pi^j(n-1),j},\sum\limits_{n=1}^{i}k_y^{\pi^j(n),j}\right)\\
    \times   \prod\limits_{j=1}^M   &\frac{-i}{2p_2^j\cdot k^{\pi^j(1),j}-i\epsilon}\times\dots\times  \frac{-i}{2p_2^j\cdot \sum\limits_{n=1}^{r^j_1-1}k^{\pi^j(n),j}-i\epsilon}\\
     \times &\frac{-i}{-2p_2^j\cdot \sum\limits_{n=r_1^j}^{N-1}k^{\pi^j(n),j}-i\epsilon}\times\dots\times\frac{-i}{-2p_2^j\cdot k^{\pi^j(N-1),j}-i\epsilon}.
\end{align}
This expression must still be summed over all permutations. Let us first look at the sub-amplitudes:
\begin{align}
\prod\limits_{i=1}^Ni\mathcal{M}_{\text{sub}}&\left(p_1^i,\sum\limits_{n=1}^{i}k_y^{\pi^j(n-1),j},\sum\limits_{n=1}^{i}k_y^{\pi^j(n),j}\right)=\prod\limits_{i=1}^N(2p_1^i)(2p_2^M)\nonumber\\
    &\times\prod\limits_{i,j=1}^{\substack{i =N\mgap\\j=M-1}}(2\pi)(2p_2^j)\delta\left(k_y^{\pi^j(i),j}\right)\times\text{Exp}\left(\frac{i\gamma^2}{2 \mu ^2}\sum \limits_{j=1}^{j=M}s_{ij}G_1(\Omega_1^i,\Omega_2^j)\right)
\end{align}
where we have not written the transverse delta's over the angles, which are kept separate for brevity. All loop-momentum dependence $k^{i,j}$ is now only embedded in the delta-functions, and since it is contained in a product it is independent on $\pi^j$:
\begin{align}
    \prod\limits_{i,j=1}^{\substack{i =N\mgap\\j=M-1}}(2\pi)(2p_2^j)\delta\left(k_y^{\pi^j(i),j}\right)&=\prod\limits_{j=1}^{M-1}(2\pi)(2p_2^j)^N\delta\left(q_2^j\right)\prod\limits_{i,j=1}^{\substack{i =N-1\\j=M-1}}(2\pi )\delta\left(k_y^{i,j}\right)
\end{align}
where we split off the $i=N$ part of the product to isolate the boundary value $k_y^{N,j}=q_2^j$. The remaining $k_y^{i,j}$ are all loop momenta to be integrated, and the delta-functions now trivially set these to zero. Since the rest of the matter propagators in \eqref{eq:ampfirst} only depends on $k_x$, we may define for brevity the separate quantity
\begin{gather}
\begin{aligned}
i\mathcal{M}_{\text{verts}}&=\int\prod_{\substack{i,j=1}}^{\substack{i=N-1\\ j=M-1}}\left(\frac{\ed k_y^{i,j}}{2\pi}\right)\prod\limits_{i=1}^N i\mathcal{M}_{\text{sub}}\left(p_1^i,\sum\limits_{n=1}^{i}k_y^{\pi^j(n-1),j},\sum\limits_{n=1}^{i}k_y^{\pi^j(n),j}\right)\\
    &=\prod\limits_{i=1}^N (2p_1^i)(2p_2^M)\nonumber\times\prod\limits_{j=1}^{M-1}(2\pi)(2p_2^j)^N\delta\left(q_2^j\right)\times\text{Exp}\left(\frac{i\gamma^2}{2 \mu ^2}\sum \limits_{i,j=1}^{\substack{i=N\\j=M}}s_{ij}G_1(\Omega_1^i,\Omega_2^j)\right).
    \end{aligned}
\end{gather}
The amplitude then reduces to 
\begin{align}
\label{eq:ampsec}
    i \mathcal{M}&=i\mathcal{M}_{\text{verts}}\times\int\prod_{\substack{i,j=1}}^{\substack{i=N-1\\ j=M-1}}\left(\frac{\ed k_x^{i,j}}{2\pi}\right)\\
    \times   \prod\limits_{j=1}^M\sum\limits_{r_1^j=1}^M\sum\limits_{s^j_1}\sum\limits_{\pi_1^j}   &\frac{-i}{2p_2^j\cdot K^{1,j}-i\epsilon}\times\dots\times  \frac{-i}{2p_2^j\cdot \sum\limits_{n=1}^{r^j_1-1}K^{n,j}-i\epsilon}\\
     \times &\sum\limits_{\pi_1^{\prime j}}\frac{-i}{-2p_2^j\cdot \sum\limits_{n=r_1^j}^{N-1}\bar{K}^{n,j}-i\epsilon}\times\dots\times\frac{-i}{-2p_2^j\cdot \bar{K}^{N-1,j}-i\epsilon}.
\end{align}
where we have now also inserted the permutations to sum over, and made analogous definitions as before:
\begin{itemize}
    \item The arbitrary location $r_1^j$ for each column.
    \item The set $s^j_1$ that contains $r_1^j-1$ elements of all infalling momenta $p_1^i$, with elements $K^{n,j}$. The complement is $\bar{s}^j_1$ with elements $\bar{K}^{n,j}$. 
    \item The permutations over the set $s_1^j$ denoted by $\pi_1^j$, with $\pi_1^{\prime j}$the permutation for the complement set $\bar{s}_1^j$. Note that $\pi_1^j$ and $\pi_1^{\prime j}
    $ together form $\pi^j$.
\end{itemize}
The remaining summations all factorize over the different $j$ values, and for these individual values the structure is exactly the same as for the individual $M+1\to M+1$ diagram. We can repeat the same combinatorial steps to find for each $j$:
\begin{align}
    \sum\limits_{r_1^j=1}^M\sum\limits_{s^j_1}\sum\limits_{\pi_1^j}   &\frac{-i}{2p_2^j\cdot K^{1,j}-i\epsilon}\times\dots\times  \frac{-i}{2p_2^j\cdot \sum\limits_{n=1}^{r^j_1-1}K^{n,j}-i\epsilon}\\
     \times &\sum\limits_{\pi_1^{\prime j}}\frac{-i}{-2p_2^j\cdot \sum\limits_{n=r_1^j}^{N-1}\bar{K}^{n,j}-i\epsilon}\times\dots\times\frac{-i}{-2p_2^j\cdot \bar{K}^{N-1,j}-i\epsilon}\\
     &=\prod\limits_{i=1}^{N-1}\left(\frac{2\pi}{2p_2^j} \delta\left(k_{x}^{i-1,j}-k_{x}^{i,j}\right)\right).
\end{align}
Writing out the delta's including boundary terms gives
\begin{align}
     \sim \delta\left(-k_{x}^{1,j}\right)\delta\left(k_{x}^{1,j}-k_{x}^{2,j}\right)\delta\left(k_{x}^{2,j}-k_{x}^{3,j}\right) \times \dots\times \delta\left(k_{x}^{N-2,j}-k_{x}^{N-1,j}\right).
\end{align}
Since the first delta-function sets $-k_x^ {1,j}$ to zero, we may iteratively repeat this through all delta-functions setting
\begin{align}
    \sim \delta\left(k_{x}^{1,j}\right)\delta\left(k_{x}^{2,j}\right)\delta\left(k_{x}^{3,j}\right) \times \dots\times \delta\left(k_{x}^{N-1,j}\right)
\end{align}
instead. The amplitude becomes
\begin{align}
    i \mathcal{M}&=i\mathcal{M}_{\text{verts}}\times\int\prod_{\substack{i,j=1}}^{\substack{i=N-1\\ j=M-1}}\left(\frac{\ed k_x^{i,j}}{2\pi}\right)\times   \prod\limits_{j=1}^M\prod\limits_{i=1}^{N-1}\left(\frac{2\pi}{2p_2^j} \delta\left(k_{x}^{i,j}\right)\right)\\
    &=i\mathcal{M}_{\text{verts}}\prod\limits_{j=1}^M\frac{1}{(2p_2^j)^{N-1}} \prod\limits_{i=1}^{N-1}(2\pi)\delta\left(k_{x}^{i,M}\right)
\end{align}
where we integrated out all loop momenta except the boundary term $k_x^{i,M}=q_1^ i$. The complete amplitude thus finally becomes
\begin{align}
\label{eq:ampthird}
    i \mathcal{M}&=(2p_2^M)(2p_1^N)\prod\limits_{i=1}^{N-1}(2\pi) (2p_1^i)\delta\left(q_1^ i\right)\nonumber\times\prod\limits_{j=1}^{M-1}(2\pi)(2p_2^j)\delta\left(q_2^j\right)\\
    &\times \prod\limits_{i=1}^N \delta^{(2)}(\Omega_1^i-\Omega_1^{\prime i})\prod\limits_{i=j}^M \delta^{(2)}(\Omega_2^j-\Omega_2^{\prime j})\\
    &\times\text{Exp}\left(\frac{i\gamma^2}{2 \mu ^2}\sum \limits_{i,j=1}^{\substack{i=N\\j=M}}s_{ij}G_1(\Omega_1^i,\Omega_2^j)\right)
\end{align}
with the transverse delta's restored. Recall that that the transition amplitude is related to the $S$-matrix by a factor
\begin{align}
    (2\pi)^2\delta^{(2)}\left(\sum q\right)&=(2\pi)^2\delta\left(\sum_{i=1}^ N q_1^ i\right)\delta\left(\sum_{j=1}^M q_2^ j\right)\\
    &=(2\pi)\delta\left(q_1^N\right)(2\pi)\delta\left(q_2^ M\right)
\end{align}
where we set all other $q_1^i,q_2^j\to 0$ because of the other delta-functions in \eqref{eq:ampthird}. The $S$-matrix thus becomes
\begin{align}
\label{eq:ampfourth}
    S&=\prod\limits_{i=1}^{N}(2\pi) (2p_1^i)\delta\left(q_1^ i\right)\delta^{(2)}(\Omega_1^i-\Omega_1^{\prime i})\nonumber\times\prod\limits_{j=1}^{M}(2\pi)(2p_2^j)\delta\left(q_2^j\right)\delta^{(2)}(\Omega_2^j-\Omega_2^{\prime j})\\
    &\times\text{Exp}\left(\frac{i\gamma^2}{2 \mu ^2}\sum \limits_{i,j=1}^{\substack{i=N\\j=M}}s_{ij}G_1(\Omega_1^i,\Omega_2^j)\right)\\
\end{align}
Finally we can recognize that, because the diagram is elastic, we can write down the free-field contribution using the commutators as
\begin{align}
    S_0\equiv\mathds{1}&=\prod\limits_{i=1}^N\left((2\pi)(2p_1^i)\delta(p_1^i-p_1^{\prime i})\delta^{(2)}(\Omega_1^i-\Omega_1^{\prime i})\right)\nonumber\\
    &\times\prod\limits_{j=1}^M\left((2\pi)(2p_1^j)\delta(p_2^j-p_2^{\prime j})\delta^{(2)}(\Omega_2^j-\Omega_2^{\prime j})\right)
\end{align}
so that the $S$-matrix simply becomes a pure complex exponent
\begin{align}
\label{eq:ampfifth}
    S&=\mathds{1}\text{Exp}\left(\frac{i\gamma^2}{2 \mu ^2}\sum \limits_{i,j=1}^{\substack{i=N\\j=M}}s_{ij}G_1(\Omega_1^i,\Omega_2^j)\right).
\end{align}

\clearpage

\printbibliography

\end{document}